*Technical report*

# A comprehensive feature comparison study of open-source container orchestration frameworks

**Eddy Truyen \*, Dimitri Van Landuyt, Davy Preuveneers, Bert Lagaisse and Wouter Joosen**

[1] imec-DistriNet, KU Leuven; dimitri.vanlanduyt@cs.kuleuven.be (D.V.L);
davy.preuveneers@cs.kuleuven.be (D.P.); bert.lagaisse@cs.kuleuven.be (B.L);
wouter.joosen@cs.kuleuven.be (W.J.)

\* Correspondence: eddy.truyen@cs.kuleuven.be; Tel.: +32-163-735-85

**Featured Application:** Practitioners and industry adopters can use the descriptive feature comparison as a decision structure for identifying the most suited container orchestration framework for a particular application with respect to different quality attributes such as genericity, maturity and stability. Researchers and entrepreneurs can use it to check if their ideas for innovative products or future research are not already covered in the overall technological domain.

**Abstract:** 1) Background: Container orchestration frameworks provide support for management of complex distributed applications. Different frameworks have emerged only recently, and they have been in constant evolution as new features are being introduced. This reality makes it difficult for practitioners and researchers to maintain a clear view on the technology space. 2) Methods: we present a descriptive feature comparison study of the three most prominent orchestration frameworks: Docker Swarm, Kubernetes and Mesos that can be combined with Marathon, Aurora or DC/OS. This study aims at (i) identifying the common and unique features of all frameworks, (ii) comparing these frameworks qualitatively ánd quantitatively with respect to genericity in terms of supported features, and (iii) investigating the maturity and stability of the frameworks as well as the pioneering nature of each framework by studying the historical evolution of the frameworks on GitHub. 3) Results: (i) we have identified 124 common features and 54 unique features that we divided into a taxonomy of 9 functional aspects and 27 functional sub-aspects. (ii) Kubernetes supports the highest number of accumulated common and unique features for all 9 functional aspects; however no evidence has been found for significant differences in genericity with Docker Swarm and DC/OS. (iii) Very little feature deprecations have been found and 15 out of 27 sub-aspects have been identified as mature and stable. These are pioneered in descending order by Kubernetes, Mesos and Marathon. 4) Conclusion: there is a broad and mature foundation that underpins all container orchestration frameworks. Likely areas for further evolution and innovation include system support for improved cluster security and container security, performance isolation of GPU, disk and network resources and network plugin architectures.

**Keywords:** Container orchestration frameworks; Middleware for cloud-native applications; Commonality and variability analysis; Maturity of features; Feature deprecation risk; Genericity.

# Contents

## 1. Introduction

In recent years, there has been a strong industry adoption of Docker containers due to its easy-to-use approach for distributing and bootstrapping container images. Moreover in comparison to virtual machines, Linux containers have a lower memory footprint and allow for flexible resource allocation to improve server consolidation [1]. The popularity of Docker has also changed the way in which application software can be packaged and deployed: container images are self-contained components that can be tagged with version numbers and are made available for download from private or public Docker registries. Moreover container images are portable across different operating systems and different cloud provider stacks [2].

Container orchestration (CO) frameworks, such as Docker Swarm, Kubernetes and Mesos, build upon and extend container runtimes with additional support for deploying and managing a multi-tiered distributed application as a set of containers on a cluster of nodes [3]. Container orchestration frameworks have also increasingly been used to run production workloads as for example demonstrated in the annual OpenStack user survey [4]– [6].

We have used the OpenStack user survey as the main inspiration for selecting popular open-source CO frameworks as OpenStack itself is a cloud provider company that is fully rooted in the open-source culture and is a rather neutral with respect to promoting a specific CO framework. Figure 1 gives an overview of the most popular PaaS platforms in OpenStack deployments according to the last two surveys of October 2016 and November 2017. It shows that Kubernetes, OpenShift, Docker Swarm and Mesos are the most used container orchestration frameworks for running production-grade services. Note that OpenShift 3.0 [7] has been completely built on top of Kubernetes and Cloud Foundry [8] also provides support for Kubernetes. Moreover, as OpenShift and Cloud Foundry are not pure container orchestration frameworks, but also offer additional PaaS development services, we choose to focus on Docker Swarm, Kubernetes and Mesos for deriving a base of common and unique features.

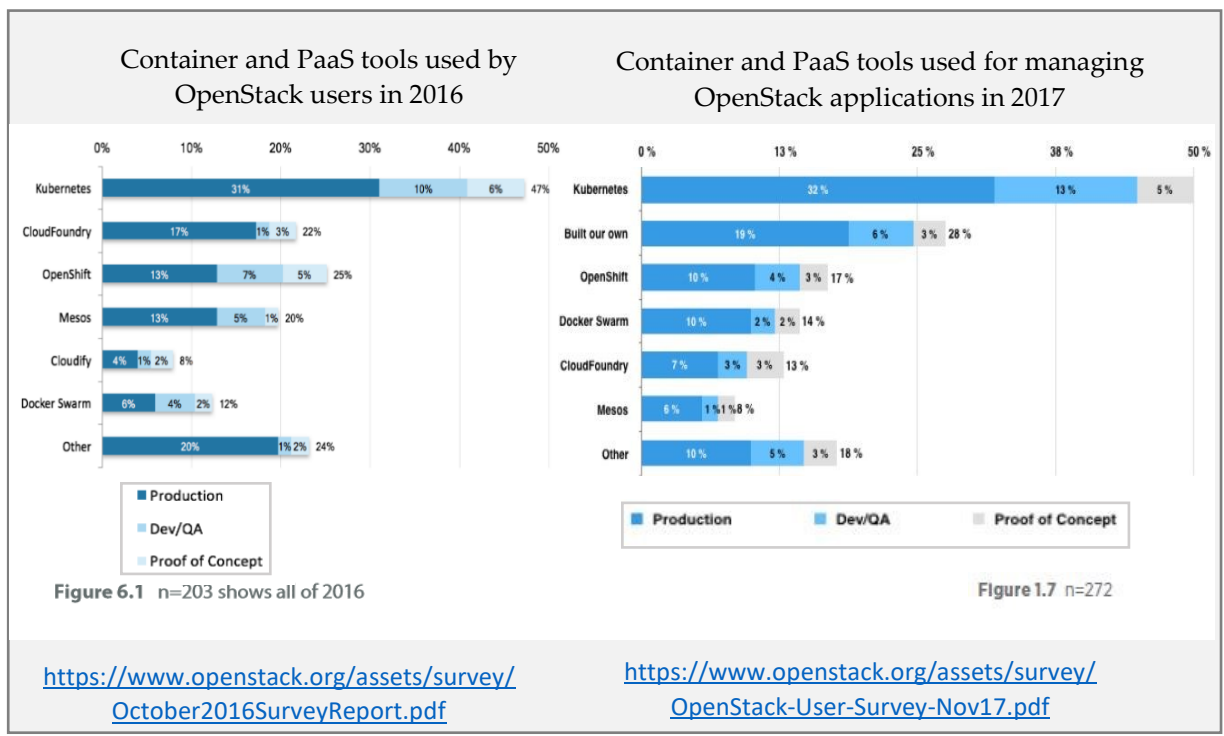

**Figure 1.** The two most recent annual OpenStack public user surveys show that Kubernetes, OpenShift, Docker Swarm and Mesos are the most popular container orchestration frameworks in OpenStack deployments. Cloud Foundry has decreased in popularity.

Note that Docker Swarm and Mesos actually cover different frameworks. As such, we compare in total 7 CO frameworks:

1. Kubernetes [9] supports deploying and managing both service- and job-oriented workloads as sets of containers.

Docker Swarm actually comes with two different distributions:

2. Docker Swarm stand-alone [10] manages a set of nodes as a single virtual host that serves the standard Docker Engine API. Any tool that already communicates with a Docker daemon can thus use this framework to transparently scale to multiple nodes. This framework is minimal but also the most flexible because almost the entire API of the Docker daemon is available. As such it is mostly relevant for platform developers that like to build a custom framework on top of Docker.

3. The newer Docker Swarm integrated mode [11] departs from the stand-alone model by re-positioning Docker as a complete container management platform that consists of several architectural components, one of which is Docker Swarm.

4. Apache Mesos [12], [13] supports fine-grained allocation of resources of a cluster of physical or virtual machines to multiple higher-level *scheduler frameworks*. Such higher-level scheduler frameworks do not only include container orchestration frameworks but also more traditional non-containerized job schedulers such as Hadoop.

Currently, the following three Mesos-based CO frameworks are the most popular:

5. Aurora[14] (by Twitter) supports deploying long-running jobs and services. These workloads can optionally started inside containers.

6. Marathon [15] supports deploying groups of applications together and managing their mutual dependencies. Applications can optionally be composed and managed as a set of containers.

7. DC/OS [16] is an easy-to-install distribution of Mesos and Marathon that extends Mesos and Marathon with additional features.

### 1.1 *Motivation*

There has been several high paces of feature additions among the most popular CO frameworks as illustrated by Figure 2, which shows the number of feature additions over the course of time between June 2013 and June 2018. As shown, there was a first peak of feature additions between June 2014 and January 2015 because Mesos v0.20.0 [17] and Marathon v0.8.0 [18] added support for Docker containers and Google open-sourced Kubernetes v0.4.0 [19] that from its inception offered support for Docker containers. Moreover, Kubernetes v0.6.0 included several innovating features such as container IP and service IP networking [20], pods [21] and persistent volumes [22]. This caused a ripple effect of feature additions across the other CO frameworks. For example, support for persistent volumes has been added to Docker v1.7 [23] in June 2015. By August 2016, Docker's architecture for persistent volumes has also been supported by Mesos v1.0.0 [24], Marathon v1.3.0 [25] and DC/OS v1.8 [26]. As another example, support for container IP networking has been added to Mesos v0.25.0 [27], Marathon v0.14.0 [28] and Docker Swarm stand-alone v1.0.0 [29] by January 2016.

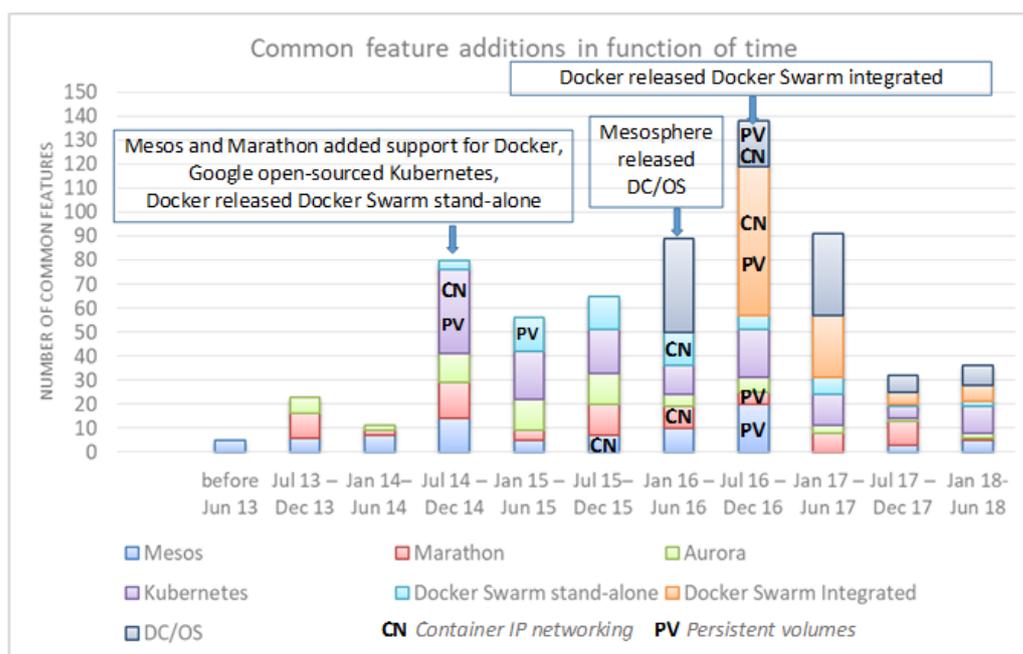

**Figure 2**. Density of feature additions over time (common features only).

This high pace of feature additions has been a challenge for companies to (a) keep track with understanding what constitutes the conceptual foundation of the overall domain and to (b) determine which CO framework matches most closely with their requirements and to (c) determine which framework is most mature with respect to these requirements. This is both a risk for companies who start using container orchestration technology and companies who consider migrating from one CO framework to another framework. They are also faced with (d) feature deprecation risks, i.e. there is strong dependence on a feature that will not be supported anymore by future versions of the employed CO framework. Finally, (e) academic researchers and entrepreneurs are also faced with the challenge that innovative functionality may become obsolete when a new version of the CO framework has been released.

An illustration of these challenges from the entrepreneur side is the story of ClusterHQ, a company that pioneered in 2014 with the container data management service Flocker [30]. Flocker initially gained a lot of traction and the company raised 12$ million in 2015 [31] and there was a well-working integration [32] with Kubernetes, Mesos and Docker Swarm. However, by the end of 2016, the company stopped all its activities because of reportedly "self-inflicted wounds" [33]. Actually, by that time all major CO frameworks provided also built-in support for external persistent volumes.

A final challenge is to (f) keep track and interpret ongoing standardization efforts in this space. For example, the Cloud Native Computing Foundation has pushed Docker's containerd [34] architecture and the associated OCI specification [35] as the de-facto standard for container runtimes [36] and has pushed Kubernetes as the de-facto standard for container orchestration [37]. Indeed Kubernetes has been the most popular framework for several years now [4], [5], [38] and has also the largest community on GitHub [39]. Moreover, DC/OS offers besides Marathon also support for Kubernetes [40] and Docker Enterprise Edition (Docker EE) also supports Kubernetes as an alternative orchestrator for Docker Swarm [41]. Even Amazon Web Services provides support for Kubernetes [42]. Nonetheless, the development of the other CO frameworks remains to continue and they also push other incompatible standards or architectures for networking and persistent volumes. This raises therefore the question what are the relevant standardization initiatives to which different CO frameworks align.

### 1.2 *Contribution statement*

To help address these challenges, we have performed a systematic assessment of the documentation of the aforementioned 7 CO frameworks on GitHub with respect to three main software qualities: genericity, maturity and stability. When the documentation appears inconclusive, we rely on experience drawn from earlier run-time experiments with CO frameworks or we have just tested out the specific feature.

A CO framework is defined as more generic than another when it supports more features than another framework. After all, the more features are supported, the more application and cluster configurations can be supported by a CO framework. The first aim of the systematic assessment is to determine a mapping from CO frameworks to commonly supported features and unique features. In order to provide an easy-to-navigate structure and draw higher-level insights from the results of this systematic assessment, we logically group the found features into 9 functional aspects and 27 sub-aspects that each cover a specific coherent set of related use cases (see Table 1). A functional aspect is defined as a set of related use cases that are of concern to the same type of stakeholder, whereas a functional sub-aspect is defined as an aspect of which the related use cases all represent interactions with the same architectural component or logical substrate of functionality of CO frameworks. We conduct not only a qualitative discussion of the identified aspects and CO frameworks, but also present a quantitative analysis of the number of supported features in each aspect and CO framework.

We also assess the maturity and stability of the different CO frameworks by studying the historical evolution of these CO frameworks in terms of subsequent releases on GitHub. More specifically, we have inspected all versions that are shown in Figure 3. The aim is to rank CO frameworks with respect to the time when they have released support for a particular feature for the first time. We also study the rate of feature deprecations in the development history to gather a more complete insight in the overall stability of the technological domain and we project this history of feature deprecation to an estimate of feature deprecation risks in the future.

This systematic assessment with respect to genericity, maturity and stability provides thus answers on the following 10 research questions:

With respect to genericity:

*RQ1. What are the common features of CO frameworks and what are the different implementation strategies for realizing the common features?*

*RQ2. How can common features be organized in functional (sub)-aspects?*

*RQ3. What are the unique features of CO frameworks?*

*RQ4. How are functional (sub)-aspects ranked in terms of number of common and unique features?*

*RQ5. How are CO frameworks ranked in terms of number of common and unique features?*

*RQ6. (a) Which functional (sub)-aspects are best supported by a CO framework in terms of highest number of common features? (b) What if unique features are taken into account?*

With respect to maturity:

*RQ7. What is the maturity of a CO framework with respect to a common feature or a functional (sub)-aspect?*

*RQ8. Which functional sub-aspects are mature enough to consider them as part of the stable foundation of the overall domain? Which CO frameworks have pioneered in what sub-aspect?*

With respect to stability:

*RQ9. What are the relevant standardization initiatives and which CO frameworks align with these initiatives?*

*RQ10. What is the risk that common or unique features might become deprecated in the future?*

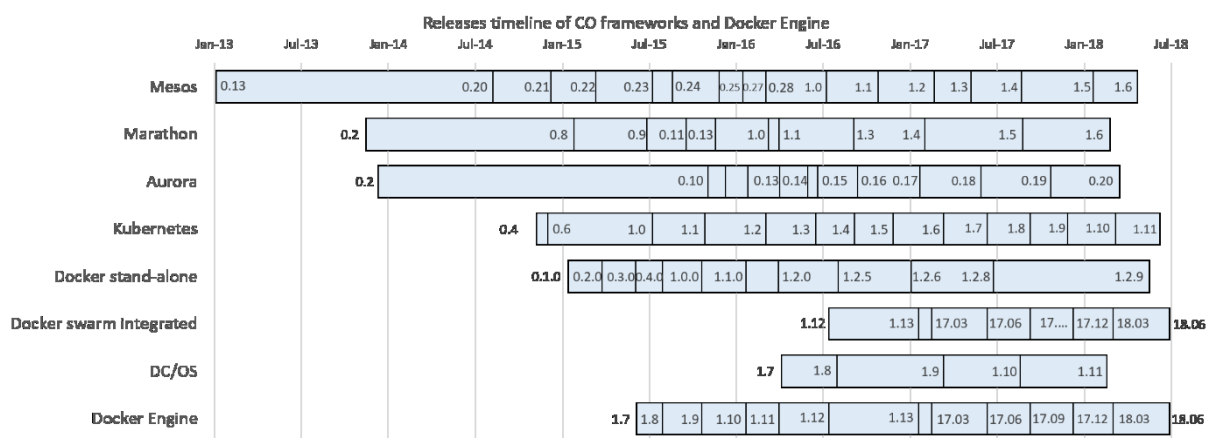

**Figure 3.** Timeline of when successive versions of CO frameworks have been released (until sept 2018).

The contributions of this article are thus as follows: with respect to genericity, it will enable industry practitioners and researchers to

1. compare CO frameworks on a per feature-basis (thereby avoiding comparing apples with oranges),
2. quickly grasp what constitutes the overall functionality that is commonly supported by the CO frameworks by inspecting the 9 functional aspects and 27 sub-aspects,
3. understand what are the unique features of CO frameworks,
4. determine which functional aspects are most generic in terms of common features,
5. identify those CO frameworks that support the most common and unique features across all (sub)-aspects,
6. identify the most generic CO framework for a specific functional (sub)-aspect.

With respect to maturity and stability, it will enable industry practitioners and researchers to

7. identify and understand the impact of relevant standardization efforts,
8. compare the maturity of CO frameworks with respect to a specific common feature,
9. understand which features have a higher risk of being halted or deprecated, and

10. determine those (sub)-aspects that can be considered as mature and well-understood and therefore shape the stable foundation of the technological domain; moreover, academic researchers and entrepreneurs are guided to invest their time and energy in adding innovative functional or non-functional aspects that have not yet been well supported.

We have started this research around the beginning of 2017 as the pace of feature additions has clearly slowed down after June 2017 (see Figure 2). As such, we believe that the insights resulting from this systematic assessment will not become obsolete when new versions of CO frameworks have been released.

**Table 1**. Overview of functional aspects and sub-aspects and their number of common and unique features.

| Functional aspects | Functional sub-aspects | #common features | #unique features |
|---|---|---|---|
| **Cluster architecture and setup** | | **13** | **2** |
| | Configuration management approach | 1 | 0 |
| | Architectural patterns | 5 | 0 |
| | Installation methods and deployment tools | 7 | 2 |
| **CO system customization** | | **6** | **9** |
| | Unified container runtime architecture | 3 | 0 |
| | Framework design of orchestration engine | 3 | 9 |
| **Container networks** | | **20** | **8** |
| | Services networking | 8 | 2 |
| | Host ports conflict management | 2 | 0 |
| | Plugin architecture for network services | 4 | 0 |
| | Service discovery and external access | 6 | 6 |
| **Application configuration and deployment** | | **29** | **10** |
| | Supported workload types | 7 | 1 |
| | Persistent volumes | 9 | 6 |
| | Reusable container configuration | 5 | 2 |
| | Service upgrades | 6 | 1 |
| **Resource quota management** | | **4** | **1** |
| **Container QoS Management** | | **15** | **6** |
| | Container CPU and mem allocation with support for over-subscription | 5 | 1 |
| | Allocation of other resources | 2 | 4 |
| | Controlling scheduling behavior by means of placement constraints | 3 | 0 |
| | Controlling preemptive scheduling and re-scheduling behavior | 5 | 1 |
| **Securing clusters** | | **9** | **4** |
| | User identity and access management | 3 | 1 |
| | Cluster network security | 6 | 3 |
| **Securing containers** | | **7** | **3** |
| | Protection of sensitive data and proprietary software | 2 | 0 |
| | Improved security isolation between containers and OS | 5 | 3 |
| **Application and cluster management** | | **21** | **10** |
| | Creation, management and inspection of cluster and applications | 4 | 1 |
| | Monitoring resource usage and health | 4 | 3 |
| | Logging and debugging of CO framework and containers | 3 | 1 |
| | Cluster maintenance | 5 | 2 |
| | Multi-cloud deployments | 5 | 3 |
| | | **124** | **54** |

1.3 *Structure of the article*

The remainder of this article is structured as follows. First, Section 2 overviews related surveys and research articles that provide an overview of CO frameworks. Then, Section 3 presents our research method to perform the systematic assessment. Thereafter, Section 4 presents a qualitative assessment of the genericity of the CO frameworks, i.e., research questions RQ1-RQ3. Subsequently, Section 5 presents the quantitative analysis with respect to the genericity requirement, i.e., research questions RQ4-RQ6. Thereafter, Section 6 presents an assessment of the maturity and pioneering nature of the CO frameworks, i.e., research questions RQ7-RQ8. Then, Section 7 presents the assessment of the stability of the CO frameworks, i.e., research questions RQ9-RQ10. Finally, Section 8 discusses the threats to validity and concludes with the main lessons learned. All the collected data including hyperlinks to relevant documentation pages of CO frameworks at GitHub is available at Zenodo[1].

## 2. Related work

There are a number of papers that mainly focus on describing (and evaluating) the common features of the Linux container technology, i.e. system virtualization, and/or specific features of Docker [43]–[48]. However, these works provide little to no overview of the common functions of state-of-the-art container orchestration frameworks.

Heidari et al. [49] presents a survey of seven container orchestration frameworks that were identified as most promising: Apache Mesos, Mesos Marathon, Apache Aurora, Kubernetes, Docker Swarm and Fleet. This survey concisely and clearly describes the architecture of these frameworks and zooms into a number of features of these platforms. However, it does not present a systematic assessment of commonality and variability. Moreover, it does not study the maturity of these frameworks and the risks of feature deprecation.

Jennings et al. [50] and Costache et al. [51] present classifications of resource management techniques in cloud platforms. More specifically, Jennings et al. provides a review of the literature in cloud resource management, while Costache et al. focuses on complete Platform-as-a-Service (PaaS) platforms, including commercial and research solutions. The latter work by Costache et al. studies commercial solutions include Mesos [12] and Borg [52], the predecessor of Kubernetes. Costache et al. also presents a list of opportunities for further research, which includes the use of container orchestration frameworks to support (i) generic resource management for any type of workload and (ii) provisioning of cloud resources from multiple IaaS clouds. However these works do not study the resource management concepts of container orchestration frameworks in detail, such as support for oversubscription and neither includes an assessment of other functional aspects such as cluster setup tools, virtual networking, customizability, security and multi-cloud support.

Pahl et al. [53] analyses required container orchestration functions for facilitating deployment and management of distributed applications across multiple clouds and how these functions can be integrated in PaaS platforms and relevant standards for portable orchestration of cloud applications. However, these functions are presented at a high level.

Kratzke et al. [3], [54] define a reference model of container orchestration frameworks, i.e. these works identify common functionalities of existing container orchestration frameworks such as scheduling, networking, monitoring and securing clusters as well as their inter-dependencies. These common functionalities are similar with the found commonalities of our study but these functionalities are described shortly at a high-level while our work decomposes each functionality into a detailed set of individual features.

In another paper, Kratzke et al. also present a domain-specific language (DSL) for specifying portable, multi-cloud application descriptors that can be translated to application descriptors for multiple container orchestration frameworks such as Docker Swarm and Kubernetes. This DSL is mainly concerned with expressing common concerns that are of interest to an application manager, i.e. specifying units of deployments and configuring their allocated resources and replica levels, customizing scheduling decisions, auto-scaling rules. Additionally, Kratzke et al. [55] studies concerns that are of interest to a cluster administrator in order to build a middleware platform to transfer container clusters from one cloud provider to another cloud provider. As one of the

---

requirements of the DSL and the middleware platform is to favour pragmatism over expressiveness [56], this DSL and middleware platform supports concepts that are supported by Kubernetes, Docker Swarm and Mesos.

We confirm by large extent the common functionalities of container orchestration frameworks as presented by Kratzke et al. However, we also extend the findings of these works in several dimensions. Firstly, we relax the definition of what is a common feature, i.e. a common feature is supported by at least two CO frameworks. Secondly, we also determine unique features that are only supported by one CO framework. Thirdly, we give a systematic and exhaustive overview of all common and unique features whereas the work of Kratzke et al. presents meta-models of configuration languages that encompass concepts to support expressing cluster or application configurations that are commonly supported by all CO frameworks; in other words, our work is complementary as it can be used to refine and update the meta-models with support for common features that have not been discovered by Kratzke et al. Finally, we do not only study common features but we also study the maturity of these common features to distinguish between stable features and those features that are relatively immature and subject to change; additionally we also discuss the risks of feature deprecation.

In summary, to our knowledge, this is the first work that presents a detailed and exhaustive commonality and variability analysis among popular container orchestration frameworks and that studies the maturity frameworks as well as the risks of feature deprecation.

## 3. Research method

This section presents how we have been working towards studying the genericity, maturity and stability of the CO frameworks. The reader can skip this section if she or he is not interested in these methodological aspects of our work.

Before starting the research for this article, we have already acquired plenty of experience with container orchestration frameworks in the context of the DeCOMAdS research project [57] that aims to design advanced deployment and configuration middleware for adaptive multi-tenant SaaS applications. At the beginning of this project we performed a technical SWOT[2] analysis of containers and container orchestration framework that helps a SaaS provider to make a cost-benefit analysis to move their applications to a container orchestration framework [2]. Subsequently, we have also compared the performance of Docker Swarm and Kubernetes for NoSQL databases [58] and we have built a tool for comparing different auto-scalers for container-orchestrated services in Kubernetes [59].

The following six subsections explain our method for (1) the qualitative assessment of the genericity requirement, (2) the quantitative analysis with respect to the genericity requirement, (3) the qualitative assessment of maturity, (4) the quality assessment of stability, (5) gathering feedback from industry to improve the coherency and correctness of our research findings and (6) dealing with the continuous evolution of the CO frameworks during the course of the research work.

### 3.1 Qualitative assessement of genericity requirement

The following three subsections explain how (1) features of CO frameworks have been identified, (2) how common and unique features across CO frameworks have been discovered and modelled, (3) how features have been organized in functional aspects and sub-aspects.

### 3.1.1 Identifying features in documentation of CO frameworks

For all CO frameworks, except DC/OS, we have manually processed their documentation on GitHub because this platform has been used to manage the editing of the documentation as well as the versioning of documentation. To manage the documentation of different versions of a CO framework, *git tags* are typically used. These git tags allow us to dynamically browse through different versions of the documentation. This was essential for us in order to discover the addition and removal of features across versions (see Section 3.4).

---

[2] Strengths, Weaknesses, Opportunities and Threats

For DC/OS, however, git tags have not been used for versioning the documentation. Instead, the documentation of different versions of DC/OS are stored in different directories on GitHub. This makes it tedious to browse across versions on GitHub. Fortunately, it is possible to easily browse through different versions of the documentation on the official website of DC/OS. Therefore we have processed the documentation of the official DC/OS website.

Our method for identifying and modeling common and unique features among different CO framework is widely inspired on feature modeling [60], [61] that is the commonly accepted method in product line engineering for modeling the commonalities and variabilities of a family of frameworks. A feature is defined as a characteristic of a framework that is visible to the end-user or as a distinguishable characteristic that is relevant to some stakeholder [61].

We have first derived **an initial list of features** for each CO framework separately by inspecting the release notes, change logs and feature planning documents of the latest version. We have then refined this initial list of features with additional features by reviewing the full documentation of the latest version of each CO framework. We also found multiple GitHub documentation pages that explained the same feature with different audiences or purposes in mind. We have **grouped these pages** so we could later study them together to fully understand the implementation strategy for the feature or discover additional related features.

### 3.1.2 Discovering common and unique features

We have identified **common features and unique features** by comparing the feature lists of all CO frameworks pair-wise. We define a common feature as appearing in the documentation of two or more container orchestration frameworks (or related incubation projects) and having passed the beta stage in at least one of the frameworks.

We first identify all common features. The question whether two documentation pages from different CO frameworks describe the same feature is concurred based on our previously acquired research experience in using and evaluating CO frameworks [2], [58], [59].

We then determine all unique features for each CO framework. We define a unique feature as a feature that has been documented by only one framework and other CO frameworks have no related incubation projects or design proposals on GitHub. By striking through all documentation pages of common features, we withhold documentation pages of possible unique features.

This work resulted into one table with common features and one table with unique features. The former table has row and column names that correspond respectively with common features and CO frameworks, while the latter table has only column names that correspond with CO frameworks. Both tables order the aforementioned grouped documentation pages (see Section 3.1.1) per CO framework and per specific feature. A table cell in these tables therefore contains all relevant information about a specific feature of a specific framework.

### 3.1.3 Organizing features in functional aspects and sub-aspects

We have organized the common and unique features in **functional aspects** because the number of discovered features was too large to be comprehendingly presented as a flat list. We have used the principles of card sorting[62] as the method for grouping features in usable aspects and naming these aspects. We decided that two features belong to the same aspect when they relate to similar use cases or requirements and have the same stakeholder in common.

Our first pass through the aforementioned two tables with features resulted into grouping the features into 8 functional aspects into a Google Docs document [63]. Based on the feedback from industry (see Section 3.5), we concluded that it takes too much time to process the volume of the presented information in these tables. As such, we have refined the functional aspects into **functional sub-aspects** because the lists of features in some functional aspects were still too large in order to be comprehensively grasped from a helicopter view. We decided that two features of a functional aspect belong to the same functional sub-aspect when they concern the same architectural component or logical substrate of functionality that is found in many CO frameworks.

We have then written an exhaustive inventory of common features by carefully reading the documentation pages of the CO frameworks. This helped us for a given common feature to (i) determine **differences in feature implementation strategy** among CO frameworks and (ii) to discover **new features** that are also distinguishable in other CO frameworks. Moreover, (iii) we discovered one new functional aspect and many sub-aspects; finally we have identified **9 functional aspects and 27 functional sub-aspects** (see Section 4 and Tables 2 to 10).

We also classified the found unique features in the different found sub-aspects. It was possible to perform this task without introducing new functional sub-aspects which increased confidence that the set of identified sub-aspects covered the whole technological domain of container orchestration. This work resulted in an extension of Section 4 with a short description of the unique features and a summarizing Table 28.

### 3.2 *Quantitative analysis with respect to genericity*

The results of the qualitative assessment of genericity allowed us to quantify rankings between (sub)-aspects in terms of number of supported common and unique features. Similarly it possible to determine rankings between CO.

To find evidence for overall significant differences between the CO frameworks with respect to the number of supported features across all 27 sub-aspects, we have used the statistical tests for checking the overall ranking of multiple CO frameworks with respect to different sub-aspects. The goal is to identify if there are significant differences in genericity between different CO frameworks, i.e., although a CO framework may support a higher number of features for several sub-aspects, the difference with other CO frameworks may still be just one or two features and therefore not significant. We have used the Friedman and Nemenyi tests that are designed with this goal in mind, but for un-replicated experimental designs [135]: un-replicated experiments take for each metric only one sample of the performance of a system, but many different metrics are evaluated; in the context of this study, metrics correspond with the 27 sub-aspects.

### 3.3 *Study of maturity*

Initially we have established an **historical timeline of the versions of each CO framework** by storing the date when each version of a CO framework has been released. We have extracted this information from official release notes.

Then we have determined a **historical timeline for each common feature**. The historical timeline of a common feature starts with a *feature addition* event, then has zero or more *feature update* events and optionally ends with a *feature removal/deprecation* event. We annotate these events with the version of the CO framework during which the events have occurred. These timelines have been defined per CO framework by using the following pseudo-algorithm:

1. We first open in our browser the latest version of the root directory of the CO framework's documentation on GitHub.
2. We then iteratively trace back to the preceding versions of the root directory by modifying the version tag in the URL naming scheme of GitHub.
3. When we discover that the documentation page of a specific feature disappears from the list of files in the directory when tracing back from version *x.y* to version *x.y-1* of a CO framework, and there is no evidence that the documentation page has been renamed, we concur that this feature has been officially added in version *x.y*, unless specified otherwise in the documentation page of version x.y itself. On the contrary, when the documentation page has been renamed in version x.y, we concur that the implementation strategy of the feature has been updated. In both cases, we record the GitHub URL of version *x.y* of that documentation page as the seed of that feature addition or update.

4. When we observe that a new file appears in the directory when tracing back from version *x.y* to version *x.y-1*, and the new file describes features that are not yet in our list of features, we concur that the features described in this documentation page have been removed in version *x.y* of the CO framework, unless deprecation information is specified in the documentation page itself.

The obtained timelines of different CO frameworks are then merged per common feature in order to understand which CO framework pioneered in which feature and which functional sub-aspects. The presentation of these merged timelines are structured according to the 9 functional aspects (see Tables 18-26, Section 6, RQ7).

Finally, an overall assessment of the maturity of the sub-aspects has been conducted (see Section 6, RQ8). We define a sub-aspect as mature and well-understood if it meets the following three criteria: (i) the sub-aspect has been consolidated by the pioneering framework at least two traditional release cycles of 18 months [130] ago, (ii) the corresponding feature implementation strategies of the pioneering framework have at least reached beta-stage in the meantime and (iii) there are no deprecation or removal events of important features in the latest traditional release cycle.

### 3.4 *Assessment of stability*

The existing standardization initiatives in the container orchestration space are an important indicator for the stability of the platform development artifacts of the leading CO frameworks. We have already identified the existing initiatives and the mapping towards adopting CO frameworks during the commonality analysis. As such, we could easily derive a compact table from this work to assess the overall state of these standardization initiatives (see Section 7, RQ9).

We have performed the assessment of feature deprecation risks during the last part of the writing. The risks of feature deprecation have only been assessed for the unique features because an analysis of the historical evolution of common features has shown that there were very few deprecations of common feature implementation strategies by any CO framework (see Section 7, RQ10).

### 3.5 *Involvement and feedback from industry*

We have asked three senior platform developers to provide feedback on the grouping into functional aspects based on the aforementioned Google Docs document [63]. All three platform developers have worked and still work for companies who aim to create commercial platforms and tools for container orchestration in cloud computing environments. Moreover they lead the development of installation tools and network plugins for setting up container clusters in Docker Swarm, Kubernetes and Mesos. They did not provide any substantial feedback however. This made us doubt about whether there is any interest in comparisons between CO frameworks from platform industry. When asked for the reasons of providing no feedback, it was because of lack of time.

We have also asked to review the current form of this article by a senior developer from a software services company who has used DC/OS and Kubernetes for running their application services. We have received detailed feedback that enabled us to improve the clarity and correctness of the feature descriptions in this article.

### 3.6 *Dealing with continuous evolution of CO frameworks during the research*

We have performed the above research from April 2017 till December 2017. After that period, new versions of CO frameworks have of course been continuously released. We have kept the collected information up-to-date as follows. Each time a new version of a CO framework has been released, we reviewed the release notes and change logs of that new version in order to discover feature additions, feature updates and feature deprecations. As a result, new common features have been discovered when a unique feature of a CO framework becomes also supported by another framework; if so, timeline information was also updated.

As the article reached completion, we decide to take into account only versions released before 1 July 2018. Currently, Kubernetes versions 1.12 has been released. We have not thus not taken into account changes introduced by that version.

## 4. Qualitative assessment with respect to genericity

We present in this section answers to research questions RQ1-RQ3

*RQ1. What are the common features of CO frameworks and what are the different implementation strategies for realizing the common features?*

*RQ2. How can common features be organized in functional (sub)-aspects?*

*RQ3. What are the unique features of CO frameworks?*

In summary, we have identified 124 common features and 54 unique features. A common feature is supported by at least two CO frameworks or related incubation projects and has not been released in the latest version of at least one of the frameworks, whereas a unique feature is supported by only one CO framework and has not been released in the latest version of the framework.

As stated above, common and unique features are grouped in 9 functional aspects that cover a set of related functionalities that are of concern to a single type of stakeholder. For reasons of simplicity we distinguish between two high-level stakeholders that each may subsume different user types:

- **Application Manager:** A person who develops, deploys, configures, controls or monitors an application that runs in a container cluster. An application manager can be an application developer, application architect, release architect or site reliability engineer.
- **Cluster administrator:** A person who installs, configures, controls and monitors container clusters. A cluster administrator can be a framework administrator, a site reliability engineer, an application manager who manages a dedicated container cluster for his application, a framework developer who customizes the CO framework implementation to fit the requirements of his or her project.

A particular stakeholder, after reading the features of a particular functional aspect, will have a clear understanding of how CO frameworks work and how they must be operated with respect to that functional aspect.

In total we distinguish between the following 9 aspects:

1. cluster architecture and setup tools relevant by a cluster administrator
2. customization of container orchestration framework components by a cluster administrator
3. container networking, i.e. setup of inter-container networks by a cluster administrator
4. application configuration and deployment by an application manager
5. resource quota management by a cluster administrator
6. container QoS management by an application manager
7. securing clusters by a cluster administrator
8. securing containers by an application manager
9. application and cluster management
    a. cluster management by an cluster administrator
    b. application management by an application manager

This section is thus structured as follows. Sections 4.1-4.9 present the common and unique features for each of the above 9 aspects. For each aspect, sub-aspects are indicated with a **bold paragraph heading.** For each sub-aspect, a common feature is indicated in an *italic paragraph heading*. Finally, for each common feature different feature implementation strategies for different container orchestration frameworks are qualitatively compared based on relevant documentation webpages of the frameworks. The URLs to these documentation pages are represented as bibliographic references.

Direct hyperlinks are also available in table format as part of the supplementary material of this article.

This section also includes for each functional aspect Tables 2-10 that map common features to their corresponding implementation strategies of CO frameworks. The mapping includes also structured information about (a) whether a common feature is fully or partially supported by that CO framework, (b) whether it is available in the open-source distribution or only in the commercial version of that CO framework, and (c) whether any standards related to the feature are implemented by that particular CO framework.

As argued above, we refer to the links to the GitHub documentation instead of the official documentation site because it enables many advantages in comparison to the official documentation websites. The only disadvantage of presenting the documentation on GitHub is that dynamic scripted content is not readable. Such dynamic content included scripts for downloading and displaying a source file, for rendering graphical UI elements, and for generating reference documentation of HTTP APIs and command-line commands. Therefore when the documentation is unreadable we refer to a link to the official documentation site.

### 4.1 Cluster architecture and setup

This aspect represents common architectural patterns and features of CO frameworks that a cluster administrator must understand in order to be able to setup a running container cluster on top of a particular operating system and/or cloud provider infrastructure.

#### 4.1.1 Common features

**Configuration management approach**. All container orchestration (CO) frameworks follow a *declarative configuration management* approach instead of an imperative configuration management approach [64]. Declarative configuration management implies that an application manager describes or generates a declarative specification of the desired state of the distributed application. The CO framework then continuously adapts the deployment and configuration of containers until the actual state of the distributed application matches the described desired state. The configuration language that is used for describing the desired state varies among CO frameworks. Docker Swarm stand-alone [65], Docker Swarm integrated mode [66] and Kubernetes [67] support the YAML mark-up language. Kubernetes also support the JSON mark-up language but recommends YAML. Aurora [68] uses the Python programming. Marathon [69] and DC/OS [70] use the JSON mark-up language.

**Architectural patterns.** The core architectural pattern underlying a container cluster is very similar: it is based on the *Master-Workers architecture* where a Master node controls that running applications are always in their desired state by scheduling containers to the Worker nodes and by monitoring the actual run-time state of nodes and containers. Masters use a distributed data store (e.g., *etcd, Consul, or Zookeeper)* for storing the actual configuration state about all deployed containers and services. The specific naming of master and worker nodes differs among CO frameworks:

- Docker Swarm [71] refers to Managers and Workers
- Kubernetes [72] refers to Masters and Nodes
- Mesos [73] refers to Masters and Agents.
- Aurora [74], Marathon [75] and DC/OS [76], which run on top of Mesos, refer to Schedulers and Executors.

The scheduler of Mesos supports fine-grained sharing of the resources of a cluster of machines across multiple frameworks, such as Aurora and Marathon. The overall scheduling architecture of Mesos can be described as follows:

1. To deal with the differences between frameworks (e.g. some frameworks execute applications in containers, while other frameworks don't), Mesos uses the generic concept of Task for launching both containerized and non-containerized processes.

2. Mesos consists of a two-level scheduler architecture [73], i.e. the Mesos master and multiple framework schedulers. The protocol between a framework scheduler, the central Mesos master and multiple agent nodes to achieve the scheduling of a container on an agent node are as follows:

   a. Each agent node notifies the Mesos master when it has resources available.

   b. The Mesos master then uses the Dominant Resource Fairness [77] algorithm to determine to which framework to offer these available resources. The Mesos master sends the resource offer then to the Scheduler component of the selected framework.

   c. The selected framework can then accept the offer by reserving a subset of the offered resources on that agent [78]. The framework can also reject the offer because it does not fit with data locality constraints for instance [12]; Mesos will then send the resource offer to another framework

   d. Once a subset of resources is reserved by a framework, the scheduler of that framework can schedule tasks using these resources by sending the tasks to the Mesos master [79].

   e. The Mesos master then sends the tasks to the Mesos agent from which the resource offer originates.

   f. The Mesos agent delegates the execution of the tasks to the co-located Executor [80] component of the framework.

   g. The Mesos master continues to offer the reserved resources to the framework that has performed the reservation. This is because the framework can respond by unreserving [81] the resources.

3. Since the state of a task is stored by both the Mesos master and the framework scheduler, this task state needs to be kept synchronized. Mesos' architecture supports at-most-once [82] unreliable message delivery between the Mesos master and the frameworks. Therefore, when a framework's scheduler has requested the master to start a task, but doesn't receive an update from the Mesos master, the framework scheduler needs to perform task reconciliation [83].

*Highly-Available (HA) master design*. To ensure high-availability of the cluster, Masters can be replicated in all CO frameworks (see Table 2).

*Generic and automated setup of HA masters*. A fully automated and portable framework for setting up replicated Masters in different execution infrastructures is supported in Docker Swarm integrated mode [84], Aurora [85], Marathon [82] and DC/OS [86]. The procedure for Docker Swarm standalone [87] assumes that a distributed key-value store has been setup in advance. A fully automated HA framework for Kubernetes does not exist in the open-source distribution. However a large number of public cloud provider services (e.g. Google Compute Engine (GCE) [88], Amazon Elastic Container Service for Kubernetes (EKS) [42] and Google Kubernetes Engine (GKE) [89]) and a number of tools for installing and managing Kubernetes clusters (e.g. juju [90] and tectonic [91]) include support for an automated HA setup procedure.

*Versioned HTTP API and Client API libraries*. All CO frameworks except Aurora offer a versioned API that defines the concepts for specifying the desired state of the cluster and distributed applications (see Table 2). In the remainder of this article, we refer to an atomic element in such desired state specification as an object. For example, a request to the Master API for registering a new worker node will lead to the creation of Node object, which is specified in YAML and stored in the distributed data store of Master nodes. Mesos also offers an HTTP API for the Scheduler [92] and Executor [80] interfaces of frameworks, but these implement the interactions described as part of the above described two-level scheduler architecture of Mesos.

To support evolution of the API, a specific versioning schema is devised for each CO framework. In general, a specific version of the API corresponds with a certain version of the CO framework. The version schema also allows demarcating stable parts of the API from those parts that are still beta.

An HTTP API becomes only usable if there are client libraries available for one or more programming languages. Kubernetes [93], and Marathon [94] both provide several robust client libraries, while Mesos provides a client library for writing frameworks on top of Mesos' Scheduler API [92] and Executor API [80]. Docker supports the Docker SDK [95] for the Engine API. Finally, as DC/OS [96] extends Mesos and Marathon with additional components, it offers additional REST-based APIs for these components. However, client libraries for theses APIs do not yet exist.

*Simple and policy-rich scheduling algorithm*. An important element of every CO framework is the scheduling algorithm that is responsible for computing on which node a container should be placed. All CO frameworks, except Mesos, have a simple yet highly customizable scheduling algorithm. This is an interesting difference with schedulers for traditional clusters like Hadoop which must compute job placements at massive scale in a time-efficient manner such that node resources are utilized well and resources are fairly distributed across different users  [77], [97]–[100]. Container clusters, on the other hand, need to run dozens of small services that need to be organized and networked to optimize how they share data and computational power [101].

Docker Swarm stand-alone [102] supports three scheduling strategies: spread, binpack and a trivial random strategy. The spread strategy places a new container on the node with the least number of containers, while the binpack strategy places a new container on the node which is most packed, but can still fit the container. Docker Swarm integrated mode [103] supports two distinct spread strategies for respectively replicated services and global services.

Kubernetes [104] offers a generic scheduler component that performs the following three steps for computing a placement for a container: (1) filter the nodes using a set of predicates, (2) prioritize the filtered list of nodes using priority functions and (3) select the best fit node. The default scheduling algorithm is an instantiation of this generic scheduler with a set of default predicates and default priority functions [105]. The ensuing default scheduling algorithm guarantees for instance that replicated containers of the same application are always spread on different nodes, and that nodes with conflicting hardware states (such as ports already in use by other containers) are filtered out.

The scheduling algorithms of Aurora [106], Marathon [107] (and by inclusion DC/OS) are also simple. They randomly select the first Mesos agent with a reserved resource offer that fits the task, but the placement decision can be restricted by means of different kinds of constraints (see Section 4.6.1).

**Installation methods and tools for setting up a cluster**. In order to simplify the installation procedure, a number of deployment methods and associated tools or platforms exist (see Table 2 for a detailed overview):

- Methods that install the CO software itself as a set of Docker containers.
- Methods that use VM images with the CO software installed for local development.
- Methods that install the CO software from a traditional Linux package.
- Methods that use configuration management tools such as Puppet or Chef.
- Cloud provider owned tools and APIs
- Cloud provider independent orchestration tools that come with specific deployment bundles for installing a container cluster on one or multiple public cloud providers.
- Container orchestration-as-a-Service platforms
- Setup-tools for Microsoft Windows or Windows Server

In our experience with these tools, we have found that methods, which install the CO software from a Linux package, are easy-to-use and can be universally applied on any type of virtual machine or cloud provider. Moreover, the Linux package comes typically with a CLI-based setup tool for automating the setup of a secure cluster based on TLS certificates and authentication tokens. For example, Docker Swarm integrated mode [108] is automatically installed when installing docker-engine. A master node can then be created by running the docker swarm init command and a worker node can be created using the docker swarm join command. Kubeadm [109] is a deployment tool that installs the kubeadm CLI for setting up Kubernetes clusters and the kubelet agent from a Linux package. Similar to Docker Swarm, a new master node can be created by running the kubeadm init command. When this command is executed, the rest of the Kubernetes software is installed as containers. DC/OS [110] can also be deployed from a linux package that comes with a CLI-based setup tool.

Proprietary tools and APIs of cloud providers [3] or Container-Orchestration-as-a-Service platforms are more easy-to-use than Linux packaging tools because clusters are automatically setup and several management aspects such as cluster software upgrades and HA masters are automatically handled by the cloud provider. The disadvantage of these methods is that one has to pay for this management automation and one is bound to using a provider-specific API for configuring the management aspects. The cluster administrator also gives up some control over how particular functionalities are implemented (e.g. upgrading the CO software is implemented by the cloud provider in a specific way that may be in conflict with application-specific SLAs; workarounds for handling known open issues[4] are predetermined by the cloud provider.

---

[3] e.g. gcloud[623] for setting up a Kubernetes cluster on top of Google Computer Engine (GCE) and Microsoft Azure Container Service Engine[624] for setting up a Swarm mode, Kubernetes or DC/OS cluster on Microsoft Azure

[4] e.g., lack of N+1 fault tolerance guarantees in Kubernetes when rebooting VMs[403]

**Table 2.** Common features of the "cluster architecture and setup" aspect.

<u>Column Legend:</u>
- **Sa**: Docker Swarm stand-alone
- **Si**: Docker Swarm integrated
- **Ku**: Kubernetes
- **Me**: Mesos
- **Au**: Mesos+Aurora
- **Ma**: Mesos+Marathon
- **Dc**: DC/OS

<u>Cell Legend:</u>
- ✓: The feature is fully supported by the open-source distribution of the platform. The URL to the corresponding documentation is included.
- externalComponent: Support for the feature is not included in the open-source distribution of the CO framework, but the feature is supported by a third party component or platform. The name of the URL refers to the name of the component. The URL to the corresponding documentation is included.
- DC/OS Legend:
  - ○ Dlgt (Delegate): The feature is implemented by Mesos+Marathon and DC/OS relies on it completely
  - ○ Extnd (Extend): The feature is implemented by Mesos+Marathon, DC/OS relies on it but also extends it with additional functionality
  - ○ Sprsd (Supersede): The feature is supported by Mesos+Marathon, but its implementation is superseded by a new component of DC/OS
  - ○ Add (Add): The feature is not supported by Mesos+Marathon, but DC/OS adds support for it.

| Cluster architecture and setup sub-aspects | Features | Swarm stand-alone | Swarm integrated | Kubernetes | Mesos | Mesos + Aurora | Mesos +Marathon | DC/OS |
|---|---|---|---|---|---|---|---|---|
| | | Sa | Si | Ku | Me | Au | Ma | Dc |
| Configuration management Architectural patterns | *Declarative configuration management* | ✓ | ✓ | ✓ | n/a | ✓ | ✓ | Dlgt |
| | *Master-Worker architecture* | ✓ | ✓ | ✓ | ✓ | ✓ | ✓ | Dlgt |
| | *Highly-available (HA) master design* | ✓ | ✓ | ✓ | ✓ | ✓ | ✓ | Dlgt |
| | *Generic, automated setup of HA masters* | ✓ | ✓ | GCE juju tectonic | | ✓ | ✓ | Dlgt |
| | *Versioned HTTP API and client libraries* | ✓ | ✓ | ✓ | | | ✓ | Extnd |
| | *Simple, policy-rich scheduling algorithm* | ✓ | ✓ | ✓ | n/a | ✓ | ✓ | Dlgt |
| Installation methods and tools for setting up a cluster | *Dockerized CO software* | ✓ | | ✓ | ✓ | | ✓ | |
| | *VM images with CO software for local dev* | | ✓ | ✓ | ✓ | ✓ | ✓ | Extnd |
| | *Linux packages + CLI for cluster setup* | | ✓ | ✓ | ✓ | ✓ | ✓ | Extnd |
| | *Configuration management tools* | | | ✓ | ✓ | | | |
| | *Cloud-provider tool or platform* | MsAz | MsAz | ✓ | | | | MsAz |
| | *Cloud-provider independent tools* | | ✓ | ✓ | | | | Add |
| | *Microsoft Windows or Windows Server* | | ✓ | ✓ | ✓ | | | |

Table 2 presents the common features of the "cluster architecture and setup" aspect, organized according to the three above sub-aspects. The first column references the name of the sub-aspect, while the second column references the name of the common feature. CO frameworks with a ✓ provide full support for the feature. If only a particular deployment tool or platform provides support for the feature, the abbreviated names of the tools/platforms are shown. Finally as DC/OS builds upon and extends Mesos+Marathon, we characterize the nature of how DC/OS supports a feature as follows:

- Delegate (Dlgt): The feature is implemented by Marathon+Mesos and DC/OS relies on it completely
- Extend (Extnd): The feature is implemented by Marathon+Mesos and DC/OS relies on it and extends it
- Supersede (Sprsd): The feature is supported by Marathon +Mesos, but its implementation is superseded by a new component of DC/OS
- Add (Add): The feature is not supported in Mesos+Marathon and DC/OS adds support for it.

### 4.1.2 Unique features

The following CO frameworks have also unique features for the sub-aspect "installation methods and tools":

Kubernetes:

- Kubernetes-as-a-Service [111]: Microsoft Azure [112], Google Kubernetes Engine [89], AWS [42] and other cloud providers offer public Kubernetes-as-a-Service offerings with the highest-level of automation and ease-of-use in comparison to other cloud provider specific tools and APIs.

DC/OS:

- A GUI installer [113] provides a simple graphical user interface that guides the cluster administrator during the installation of DC/OS.

### 4.2 *CO framework customization*

This aspect corresponds with features of CO frameworks that a cluster administrator must understand in order to create a customized version of the CO framework.

### 4.2.1 Common features

**Unified container runtime architecture**. All CO frameworks provide support for a *unified container runtime architecture such that multiple container runtimes can be plugged in,* and optionally different container image formats can be supported. Docker launched the *containerd* container runtime architecture [34]. Kubernetes has defined the *Container Runtime Interface* (CRI) [114] for this purpose. There is also a CRI plugin for containerd [115]. Mesos [116] defines its own *Universal Container Runtime* (UCR) that supports different image formats: the Docker image specification [117] and the Appc specification [118].

*Support for Open Container Initiative specifications.* The Open Container Initiative (OCI) [119] defines a specification for container runtimes and a specification for container images. Containerd [34] supports both specifications. Kubernetes [120] has an OCI-based implementation of its Container Runtime Interface. Mesos-based frameworks will provide support for the OCI image specification in the future [121].

*Other supported container runtimes.* As a consequence of the unified container runtime architectures, each CO framework supports besides Docker Engine also other container runtimes: Docker Swarm supports runC [122] that runs containers according to the OCI specification. Kubernetes supports the rkt container runtime, runC and any other OCI-based container runtime [123]. Mesos-based frameworks support besides the Docker containerizer also the Mesos containerizer [124]. The Mesos containerizer is composable, i.e. a cluster administrator can setup a customized and more light-weight container runtime by selecting from an extensive list of existing isolator modules [125] that each implement a particular aspect of how the execution environment for a Mesos task (or container) is constructed. Isolators implement features such as resource isolation, monitoring, networking and security. For example, the cgroups/devices [126] and linux/devices [127] isolators enable cluster administrators to control access of containers to linux devices under the /dev directory.

**Framework design of core orchestration engine.** All CO frameworks except Aurora support *an external plugin architecture for customizing multiple cluster operations* (see Table 3 for a detailed overview). The following cluster operations can be typically customized by means of a plugin: container networking, persistent volume operations, and Identity and Access Management (IAM). Network plugin architectures are presented in Section 4.3, volume plugin architectures are discussed in Section 4.4, and security plugin architectures are discussed in Section 4.7.

*Plugin-architecture for schedulers.* It is also possible to plug-in a custom scheduler in Kubernetes [128], Mesos [129], Aurora [130](see scheduler configuration [131], parameter -offer_set_module), Marathon [132] (and by inclusion DC/OS). In Kubernetes it is even possible to plug-in multiple schedulers in parallel [133].

*Modular interceptors for functional extension of the orchestration engine.* Modular interceptors encapsulate specific extensions of existing CO components. Different kinds of modular interceptors are supported by Kubernetes, Mesos and Aurora. Kubernetes [134] supports three kinds of interceptors:

- Admission controllers [134] are run in sequence before each authorized request to the Master API for creating a Kubernetes object. They can accept, reject or mutate the request. They can also update the state of other Kubernetes objects. Admission controllers are only applied to API requests that have been successfully authenticated and authorized. Admission controllers are used for implementing various functionalities such as resource quota management (see section 4.5) and Pod Security Policies (see section 4.7). A disadvantage of admission controllers is that new admission controllers cannot be loaded into a running cluster as a rebuild of the cluster software is required.
- Two types of run-time pluggable admission controllers that deal with the aforementioned disadvantage of admissions controllers:
  - Initializers [135] are useful for cluster administrators to force policies or inject defaults in a running cluster.
  - Validating or mutating admission web hooks [136] are HTTP call backs that can respectively reject or change the contents of an API request for creating a new Kubernetes object.

Mesos [137] supports module hooks that allow framework developers to tie into internal components of Mesos. Aurora also supports two kinds of interceptors:

- Client hooks [138] but these are limited to pre- and post-hooks around API client methods when they are called by Aurora CLI commands.
- Thrift interceptors [139] that are able to intercept Thrift method calls from the Aurora scheduler [131]. Apache Thrift [140] is a cross-language service development framework for representing structured data in client/server RPC protocols as well as for internal data structure.

**Table 3.** Commonly supported features for the "CO framework customization" aspect.

<table>
<tr><td colspan="2">Cell legend:<br>• <u>future</u>: The feature is not yet part of the open-source distribution of the CO framework. It has however been planned according to the documentation, or there is a separate incubation project. The URL to relevant roadmap documentation is included.</td></tr>
</table>

| CO framework customization sub-aspects | Features | Swarm stand-alone | Swarm integrated | Kubernetes | Mesos | Mesos+ Aurora | Mesos + Marathon | DC/OS |
|---|---|---|---|---|---|---|---|---|
| | | Sa | Si | Ku | Me | Au | Ma | Dc |
| Unified container runtime architecture | *Unified container runtime architecture* | ✓ | ✓ | ✓ | ✓ | ✓ | ✓ | Dlgt |
| | *Support for OCI specifications* | ✓ | ✓ | ✓ | future | | | |
| | *Other supported container runtimes* | ✓ | ✓ | ✓ | ✓ | ✓ | ✓ | Dlgt |
| Framework design of orchestration engine | *External plugin architecture* | ✓ | ✓ | ✓ | ✓ | | ✓ | Dlgt |
| | *Plugin-architecture for schedulers* | | | ✓ | ✓ | ✓ | ✓ | Dlgt |
| | *Modular interceptors* | | | ✓ | ✓ | ✓ | | Dlgt |

### 4.2.2 Unique features

The following CO frameworks have unique features for the sub-aspect "framework design of core orchestration engine":

Docker Swarm integrated mode:

- It is possible to implement new types of plugins as global services [141].

Kubernetes [142] is highly extensible:

- Cloud-provider specific functionality is encapsulated in a separate CloudController plugin [143]. This plugin supports several functions, e.g. configuration of external load balancers when a new service is created and automatic labeling of nodes and persistent volumes to ensure pods are scheduled in the availability zone where the persistent volume is located.
- The Kubernetes API can be extended with custom Kubernetes API objects [144] and associated Controller plugins. Such custom objects can be versioned [145] and can have custom status and scale sub-objects [146].
- Additional APIs [147] can also be aggregated in the overall Kubernetes API
- Attaching arbitrary metadata to Kubernetes objects is possible via annotations [148]. This metadata can serve many purposes. The most common use case is to introduce alpha features that are not yet supported by the Kubernetes API. For example, the PodSecurityPolicy API does not yet support enforcing AppArmor profiles. Therefore, annotations to a PodSecurityPolicy [149] specify the desired enforcement.
- Support for dynamically reconfiguring the Kubelet agent of a running cluster via dynamic kubelet configuration [150].
- A device plugin architecture [151] for writing plugins that discover hardware resources of a specific type of device. This feature is useful when Kubernetes orchestrates virtual network functions in NFV architectures where certain network functions can only run on nodes with specific hardware features. This architecture does not allow that a single instance of a device can be shared among containers.

Mesos:

- A resource provider abstraction [152] for easily extending and customizing how a Mesos agent synchronizes with the Mesos master about available resources the agent's node and handling operations on these resources.

Aurora:

- Aurora can be configured to use a custom Executor [153] instead of the default Thermos executor.

### 4.3 *Container networking*

This aspect corresponds with features of CO frameworks that a cluster administrator must understand in order to customize how containers are networked, load balanced and discovered.

### 4.3.1 Common features

**Services networking.** A container exposes a certain service at a well-defined container port. In order to support availability and fault-tolerance, multiple replicas of the container need to be started across multiple nodes and health checked. In order to support connectivity to such container-based, replicated services the following elements are necessary: (i) a stable service name or IP address that is unique to this service irrespective of the state of the pool of containers of that service, (ii) a network to connect to the containers via a unique network address, and (iii) a service proxy that enables to lookup service network addresses and translate them to container replica network addresses; the service proxy may also encompass a load balancer to spread the workload of a service across the different replica's.

There are three different approaches to enable these three elements of services networking. We consider them as parent features that can be decomposed into a number of child features.

*Routing mesh for global service ports.* Here, (i) every service is identified by means of a unique port that is opened at each node of the cluster where a container replica runs, (ii) a container is thus addressed using the IP address of its local cluster node and the unique service port, (iii) at one or more nodes of the cluster a load-balancer serves requests to a service port by forwarding the requests to the cluster nodes where the containers of that service are running.

Load balancers (LBs) can be classified as according to the following sub-features:

1. Whether the LB is *automatically distributed* on every node of the cluster vs. *centrally installed* on a few nodes *by the cluster administrator*. In the latter case, sending a request to a service port requires a multi-hop network routing to an instance of the LB.
2. Whether the LB supports *Layer 4 (i.e. TCP/UDP) vs Layer 7 (i.e. HTTPS)* load balancing. Layer 7 load balancing allows implementing application-specific load-balancing policies.
3. Whether the L4 LB implementation is *based on the ipvs load balancing module of the Linux kernel* [154]. This ipvs module is known as a highly-performing load balancer implementations
4. Whether containers can run in *bridged* or in *virtual network* mode. In the former mode containers can only be accessed via a host port of the local node; the host port is mapped to the container port via a local virtual bridge. In the latter case, remote network connections to a container can be served.

Docker Swarm integrated mode, Kubernetes, Marathon and DC/OS provide full support for a routing mesh:

- Docker Swarm integrated mode [155] and Kubernetes [156] allow exposing a service via so called NodePort. Moreover, the load balancer of these frameworks can be characterized according to the four above features:
    1. Requests to NodePorts are automatically load balanced via *a distributed* service *proxy* that is automatically installed at every node of the cluster, but can also be load balanced by a *centralized load balancer* for exposing services to external clients outside the cluster. The central load balancer is not automatically installed, but must be manually activated and configured in both Docker Enterprise Edition (Docker EE) [157] and Kubernetes [158].
    2. The distributed service proxy operates a Layer 4 (L4), while the centralized load balancer operates at Layer 7 (L7).
    3. Docker Swarm's L4 service proxy is *by default based on ipvs* [159], while in Kubernetes v1.8+, the L4 service proxy can be *optionally configured to use ipvs* [160].
    4. In Docker Swarm, containers can run in either *bridged or virtual network mode*, while in Kubernetes containers must always run in *virtual network mode*.

- Mesos+Marathon [161] supports *service ports* that are served via Marathon's load-balancer, named marathon-lb [162]. This load balancer can be characterized as follows:
    1. Instances of marathon-lb must be *centrally installed* by the cluster administrator. Different types of load balancer tiers can be setup by the cluster administrator, typically an internal and external load balancing tier.
    2. Both *L4 and L7 load balancing* is supported by marathon-lb [163].
    3. The *L4* implementation is *not based on ipvs*.
    4. Containers must run in *bridged mode* in order to assign a service port to the encompassing Marathon application [164].

    Note that Mesos provides an important building block, named port mapping isolator [165], which manages the range of allowed service ports and isolates the network traffic on a per-container-basis.

- Finally, DC/OS' support for *global service ports is identical to Mesos+Marathon* as described above [166], but DC/OS also supports a richer L7, centrally-deployed load balancer, named Edge-LB [163] that provides support for multi-tenancy and load balancing non-container-orchestrated services.

*Virtual IP network for containers*. Here, (i) each service is either identified by means of a stable DNS name or a stable service IP address, (ii) containers run in virtual network mode, i.e. each container has a unique IP address that can be remotely connected via a virtual network; this virtual network is supported by overlay network software that preferably supports IPv6 network addresses in order to allow for a massive amount of containers in a single cluster, (iii) Service IPs are load balanced by an automatically distributed Layer 4, ipvs-based load balancer, (iv) DNS names are served by an internal DNS service that is automatically installed at one or multiple nodes of the cluster. For load-balanced services with a Service IP, the DNS service resolves to the Service IP by default; otherwise the DNS services resolves to the list of IP addresses of the containers behind that service. The DNS service of different CO frameworks can be classified according to several features which are described in the "service discovery and external access" sub-aspect.

All CO frameworks, except Aurora, support this approach, but some CO frameworks provide only partial support:

- Docker Swarm stand-alone [167] partially supports this approach: only an IP address and DNS name for containers is assigned and containers are not automatically replicated across multiple nodes for ensuring fault tolerance; moreover, no unique Service IP address or DNS name for addressing a load-balanced service as a whole is generated. As such clients need to implement container replication and load balancing themselves. Support for IPV6 network addresses is an optional feature [168]. In addition, a distributed DNS server is automatically installed when installing Docker engine [169].

- Docker Swarm integrated mode [155] and Kubernetes [156] allow to expose a service via a unique Service IP address and DNS name. Requests to Service IPs are load balanced via a distributed service proxy that runs at every node of the cluster.
  - As stated above, Docker Swarm offers support for IPV6 addresses [168] and offers a distributed DNS server [169].
  - In opposition, Kubernetes' DNS service is centrally deployed and, therefore, a DNS lookup requires multi-hop routing. Moreover, support for IPv6 addresses is only supported by some deployment tools (such as kubeadm [170]) or network plugins (such as Calico [171]).
- Mesos+Marathon [172] only partially supports this approach: containers of a Marathon application are automatically replicated. However, only individual containers are associated with an IP address; in other words, there is no stable Service IP address that is served by an L4 load balancer. As such clients need to implement a load-balancer themselves. IPv6 addressing is supported since Marathon v1.6.0, but only for Docker containers.
- DC/OS extends Marathon's approach: Load-balanced services can be manually created by defining a name-based Virtual IP for a Marathon application using the DC/OS GUI [173]. This name-based VIP is served by a distributed service proxy that runs at every node of the cluster and forwards to the container IPs in a round-robin fashion [174]. DC/OS [175] also exposes this VIP as a stable DNS name [176] that is served by a distributed DNS server, named dcos-dns (aka Spartan) [177]. DC/OS v1.11 also supports IPv6 addresses for Docker containers [178].
- Marathon applications without a VIP can also be addressed via a stable DNS name [179]: the list of the container IP addresses for a Marathon application are stored as a list of DNS A records in dcos-dns [180]. Additionally, when containers expose a service at a named port, a list of SRV records [181] is registered in the central DNS server of Mesos, named mesos-dns [182]. An SRV record of a specific container consists of the unique DNS name of each container, the container IP address and the container port at which the service of the container is exposed [183].

*Host port networking*. Here, (i) services are identified by means of *a stable DNS name*; (ii) container ports are exposed via a host port – here containers can run either in *host mode* (i.e. share the network stack of the underlying host) or in *bridge mode* which is less performant but more secure than host mode because of the intermediate virtual bridge; (iii) in the internal DNS service, the IP addresses of the nodes on which a container of the service is deployed are registered as a list of A records and these records are returned according to the *DNS round robin* scheme; (iv) it may also be possible to register exposed host ports as SRV records.

- Docker Swarm stand-alone [184] partially supports this approach: containers are not automatically replicated across multiple nodes. Moreover, there is no unique DNS name for addressing the service as a whole. Containers can run either in bridged [185] or host mode [186]. The advantage of host mode is that there is no performance overhead of using a virtual bridge for mapping container ports to host ports. Its disadvantage is that the container has access to the network namespace of the underlying host, which is an important security vulnerability. The good practice of scanning a container's image for malicious code via a Trusted Docker Registry [187] must in this case certainly be applied.
- Kubernetes [188] does not recommend using host ports for the reason mentioned above. Actually, it partially supports this approach: the number of container replicas of the service can be scaled up or down, but no unique DNS name for addressing the service as a whole is generated. Kubernetes only supports bridged mode and the host-mapper plugin [189] from the Container Network Interface (CNI) project (see feature *Support for Container Network Interface specification* below) must be separately installed on each node of the cluster.
- Docker Swarm integrated mode [190], Mesos+Aurora [191], Mesos+Marathon [192] and DC/OS [166] fully support this third approach:
  - the number of container replicas can be dynamically scaled up or down,
  - services are identified by a unique DSN name,

- o Docker Swarm integrated mode [186], Mesos [193], Marathon [194] and DC/OS [195] document that containers can run in bridged or host mode.
- o In Docker Swarm integrated mode [190], IP addresses of the nodes on which the containers run are registered as a list of A records in the internal DNS service and are returned using DNS round-robin.
- o In Mesos-based frameworks [181], in opposition, the set of replicated containers of a service are registered as a list of SRV records in the centrally deployed mesos-dns [191]. DC/OS [166] additionally also allows to register the containers as a list of A records in the distributed dcos-dns service [179].

**Host ports conflict management**. A common problem with host ports networking is that containers with the same host port cannot be scheduled on the same node and therefore the number of scheduled containers with the same host port is limited to the number of nodes in the cluster. For these reasons, host mode is not recommended by Kubernetes [188] except for very specific use cases such as running plugins as global containers (see Section 4.4) on every node of the cluster.

*Dynamic allocation of host ports*. To deal with host port conflicts at the same node, host ports for a container are preferably dynamically allocated so that every allocated host port is guaranteed to be unique within the cluster. Such dynamic allocation can be requested in the desired state specification of a container in Docker Swarm integrated mode [196] by only specifying the container port and not specifying the host port, in Aurora [197] by invoking the underlying Thermos executor (remember, configuration management files in Aurora are written in Python – see Section 4.1) and in Marathon [198] by setting the host port equal to 0.

Note that dynamic allocation of host ports also requires that the containerized application is reconfigured via a custom Docker entry point so that the default port of the application is changed to the dynamically allocated host port (see Section 4.4).

*Management of statically specified host port conflicts on the same node.* For those applications where dynamically changing the default port is not possible or too cumbersome, or those CO frameworks that do not support dynamic host port allocation, it is still possible in Docker Swarm integrated mode and Kubernetes to statically reserve a particular port with support for resolution of port conflicts:

- in Docker Swarm integrated mode [199], host ports are centrally managed at the service level such that requests for creating a new service with an already allocated host port is a priori rejected
- in Kubernetes [200], the default scheduler policy (see Section 4.1) ensures that containers are automatically scheduled on nodes where the requested host port is still available.

There is no specific support for host port conflicts the other CO frameworks. As a workaround, in Docker Swarm stand-alone [201], Aurora [106] and Marathon [202], however, scheduling constraints (see Section 4.6) can be specified per container in order to ensure that containers with the same host port are scheduled on different nodes. Moreover, in Aurora and Marathon, Mesos agents must be configured to offer port ranges that include the requested static port [203].

**Plugin architecture for network services**.

*Network plugin architecture.* In order to support different network implementations, all CO frameworks support a common interface and composition framework for network plugins. What network plugin is preferred by an application depends on various contextual parameters such as the underlying cloud provider network, the desired performance, desired routing topology, etc. The implementation of routing mesh and/or virtual network can be customized to accommodate performance requirements of the containerized applications. The involved customizations include the implementation of the local virtual bridge, the virtual overlay network software, and the distributed load balancer. We refer to Table 4 for the relevant documentation pages of each CO framework's network plugin architecture.

*Support for the Container Network Interface specification.* A noteworthy standardization initiative for network plugins is the Container Network interface (CNI) project [204], which consists of a specification and a library for writing network plugins as well as a set of helper plugins. Currently, Kubernetes [205], Mesos [206] and the DC/OS distribution of Mesos and Marathon [207] support CNI. The CNI specification also allows for multiple networks to exist simultaneously. Mainstream Kubernetes deployment tools currently do not support the creation of multiple co-existing networks however. Instead a single network must be installed for the entire cluster when bootstrapping the master node. As such exhaustion of the number of available subnet IP addresses is an issue. Another limitation is that most CNI plugins do not yet support hairpin mode which allows containers to reach themselves via their Service IPs [208].

*Support for Docker Swarm's libnetwork.* Docker Swarm uses its own networking plugin architecture, libnetwork [209]. The advantage of this architecture in comparison to CNI is that multiple networking plugins can be dynamically installed/removed and co-exist in an already running cluster, using the docker plugin command-line interface. Mesos v1.0.0+ [210] and DC/OS v1.9+ [211] also support Docker's libnetwork architecture. Due to Mesos' architecture, it is however not possible to add or remove virtual networks at run-time [212], nor is it possible to connect a container to multiple Docker networks [213].

*Separation of data and control traffic.* Docker v17.12 [214] can be configured to to use separate network interfaces for handling data traffic and swarm control traffic. For CNI-based networks, a specific Kubernetes network plugin, named Multus [215], also supports separating data and control traffic by means of distinct container network interfaces.

**Service discovery and external access.**

*Internal DNS service.* All CO frameworks support an internal DNS service for mapping service DNS names to IP addresses. Two approaches for deploying the DNS service exists:

- Docker Swarm integrated mode [169] and DC/OS [175] support a *distributed DNS server* that it automatically installed on every node of the cluster. As such DNS lookups can be resolved locally. Moreover DC/OS' DNS server, named dcos-dns, can work with any type of container mode, i.e. it resolves to virtual IP addresses when containers are attached to a virtual network, or it resolves to IP addresses of the nodes when containers run in host or bridged mode.
- Kubernetes' DNS service [216] and mesos-dns [181], which is used by Aurora, Marathon and DC/OS, are deployed in a *central fashion*. As such DNS lookups always require multi-hop network routing. Note that mesos-dns is used in DC/OS as central back-end for the distributed dcos-dns servers: when services have been launched in Mesos, the Mesos master synchronizes with mesos-dns to add the appropriate DNS records; thereafter the dcos-dns synchronizes its state with mesos-dns.

*DNS SRV records.* It is also possible to lookup named ports as SRV records [183] in Kubernetes [217], Aurora [191] and DC/OS [218]. Note, in DC/OS, SRV records are only supported by the central mesos-dns.

*Bypassing the L4 service load balancer.* Kubernetes and Docker Swarm allow to bypass the built-in L4 load balancer of respectively the virtual network layer and routing mesh by means of round-robin DNS. In Kubernetes [219], this feature is supported as Headless Services, which don't have a Service IP address. Instead the IP addresses of the containers are stored as a DNS record in the internal DNS service. Of course, clients need to implement the load-balancing itself. In Docker Swarm integrated mode [190] it is only possible to bypass the L4 load balancer if the service is exposed as a global service port via the routing mesh. A DNS lookup for the service name returns then a list of IP addresses for the nodes running the containers behind that service.

*Support for access to services from outside the cluster via the routing mesh.* In order to support access to services from external clients that run outside the cluster, an external load balancer solution must be used. In CO frameworks with a routing mesh, the built-in L4 load balancer can play the role of such

**Table 4**. Commonly supported features for the "container networking" aspect.

Cell legend:
- $..$: Support for the feature is not included in the open-source distribution of the CO framework, but is included in a commercial product or cloud service of the CO framework. The URL to the corresponding documentation is included. The name of the URL refers to the name of the product or service.
- externalComponent: Support for the feature is not included in the open-source distribution of the CO framework, but the feature is supported by a third party component or platform. The URL to the corresponding documentation is included. The name of the URL refers to the name of the component.
- *partial support*: the CO framework offers partial support for the feature. The URL to a relevant documentation page is included. The name of the URL refers to the essence of what is being supported.

| Container networking sub-aspects | Features | | Swarm stand-alone (Sa) | Swarm integrated (Si) | Kubernetes (Ku) | Mesos (Me) | Mesos+ Aurora (Au) | Mesos+ Marathon (Ma) | DC/OS (Dc) |
|---|---|---|---|---|---|---|---|---|---|
| Services networking | *Routing mesh for global service ports* | L4, ipvs-based LB distributed on all nodes | | ✓ | ✓ | | | | |
| | | central L4-L7 LB (without ipvs) | | $Docker EE$ | ✓ | port mapping isolator | | ✓ | Extnd |
| | *Virtual IP network for containers* | L4 distributed LB (with ipvs) | | ✓ | ✓ | | | | Add |
| | | with stable DNS name for services | | ✓ | ✓ | | | | Add |
| | | IP per container | ✓ | ✓ | ✓ | ✓ | | ✓ | Dlgt |
| | *Host ports networking* | Mapping container port to host port | ✓ | ✓ | ✓ | ✓ | ✓ | ✓ | Extnd |
| | | with stable DNS name for service | | ✓ | | ✓ | ✓ | ✓ | Extnd |
| | | Host mode networking | ✓ | ✓ | | ✓ | | ✓ | Dlgt |
| Host ports conflict management | *Dynamic allocation of host ports* | | ✓ | ✓ | | port mapping isolator | ✓ | ✓ | Dlgt |
| | *Management of host port conflicts* | | | ✓ | ✓ | | | | |
| Plugin architecture for network services | *Network plugin architecture* | | ✓ | ✓ | ✓ | ✓ | | ✓ | Dlgt |
| | *Support for CNI specification* | | | | ✓ | ✓ | | ✓ | Dlgt |
| | *Support for Docker's libnetwork* | | ✓ | ✓ | | ✓ | | ✓ | Dlgt |
| | *Separation of data and control traffic* | | ✓ | ✓ | Multus plugin | | | | |
| Service discovery and external access | *Internal DNS service* | Distributed DNS server on all nodes | ✓ | ✓ | | | | | Extnd |
| | | Central DNS server | | | ✓ | ✓ | ✓ | ✓ | Dlgt |
| | *DNS SRV records (only in central DNS)* | | | | ✓ | ✓ | ✓ | ✓ | Dlgt |
| | *Bypassing the L4 service load balancer* | | | ✓ | ✓ | | | | |

load balancer if the public or private IP addresses of one or more nodes of the cluster and the global service port of the service are reachable for external clients. DC/OS' Edge-LB load balancer is specifically designed for this purpose [220] and also allows to load balance non-container orchestrated services of DC/OS [221].

It is also possible to let a cloud provider's load balancing service forward client requests to the L4 load balancer. However, none of the CO frameworks, except Kubernetes [222], provide automated support for provisioning of a cloud-provided load balancer (see Section 4.3.2).

Note, in CO frameworks without a routing mesh, host ports of containers may also be accessible for external clients. Of course, clients have to implement then their own service load balancing.

*Co-existence of service IPs and service ports for a single service.* Docker Swarm integrated mode [223] and Kubernetes [156] allow assigning to a single service both a global service port and a service IP. After all, both network addresses are served by the same distributed L4 load balancer.

Note that this is not possible in Marathon or DC/OS [161]: global service ports and virtual container networking cannot be combined for the same application: global service ports can only be assigned to Marathon applications of which the containers do not run in container network mode and thus these containers cannot be reached via a name-based VIP (which is served by DC/OS' distributed L4 load balancer [174]). As a consequence, internal clients are required to send their requests to the centrally deployed L4-L7 marathon load balancer. Vice versa, Marathon applications that do run in container network mode cannot be accessed via the marathon-lb and thus are not externally accessible. As a work-around, DC/OS containers that run in host or bridged mode can be assigned a global service port and an internal DNS name (which is resolved by the distributed DNS service of DC/OS to a cluster node IP [175]), but internal clients need to implement their own load balancing.

### 4.3.2 Unique features

Docker Swarm integrated mode and DC/OS have the following unique feature for the sub-aspect "services networking":

Docker Swarm integrated mode:

- Support for SCTP port mapping for stand-alone containers [224]. The Stream Control Transmission Protocol (SCTP) is widely used in cellular networks as a transport protocol. One of the popular application of SCTP is Diameter[225] . This feature is limited to running these applications in stand-alone Docker containers [226] and connecting their SCTP port to an existing virtual overlay network of Docker Swarm [227].

DC/OS

- Support for load-balancing of non-container orchestrated DC/OS services [228].

Kubernetes has the following unique features for the sub-aspect "service discovery and external access":

- Automated integration with external load balancers [222] of cloud providers via provider-specific libraries [229]. Services of type LoadBalancer [230] are automatically provisioned with an external load balancer of the underlying cloud provider if the Kubernetes cluster has been installed with the cloud-provider-specific package.
- External DNS [231] synchronizes exposed Services and Ingresses with external DNS providers such as Google Cloud DNS, AWS Route 53, etc.
- IP masquerading [232] is a form of NAT that can be used for hiding a Pod's virtual IP address behind the IP address of its Node. This feature is typically used when a Pod sends network traffic to destinations outside the cluster's Pod CIDR range.
- Support for adding host-aliases (similar to entries in /etc/hosts) of Pods in order to override DNS lookup [233].

- Support for using another name server for DNS lookup in a Pod by enabling the CustomPodDNS feature gate and setting the Pod's DNS policy to "None" and specifying all DNS settings in the dnsConfig field in the Pod specification [234].
- Kubernetes 1.9+ also offers the possibility to replace the implementation of the default internal DNS service kube-dns by another DNS service implementation, CoreDNS [235], which is moving to become the new default.

## 4.4 *Application configuration and deployment*

This aspect covers features of CO frameworks that an application manager must understand in order to configure, compose, deploy, scale and upgrade containerized software services and applications.

### 4.4.1 Common features

**Supported workload types.** All CO framework offer support for running different types of workloads: user-facing latency-sensitive, elastically scalable, stateless services; throughput-sensitive job processing; and stateful applications. In this sub-aspect we zoom into the former two types of workloads while the next sub-aspect focusses on the support for stateful applications.

*Smallest unit of deployment.* Docker Swarm integrated mode [236] and all Mesos-based frameworks [79] propose the concept of Task, which is the smallest atomic unit of deployment. In Docker Swarm, a task encapsulates always a single container. In Mesos-based frameworks, a task encapsulates at most one container, but a task can also run non-containerized processes.

In opposition, in Kubernetes, the smallest unit of deployment is a Pod [237], which is a set of co-located containers that logically belong together and therefore are always deployed together on the same node.

*Pods.* The abovementioned Kubernetes concept of Pod has also be adopted in Mesos [238], Marathon [239] and DC/OS [240]. Here multiple containers can be launched atomically as part of a *task group* and these containers are all started inside an underlying Executor container. Such nested containers are only supported in the Mesos containerizer runtime [241].

*Container-based jobs.* Batch-oriented workloads where one or more jobs run in parallel are supported by Kubernetes [242] and Aurora [243]. In Aurora, a single job instance runs a task which consists of one or more processes that are executed sequentially or in parallel. Optionally, this task can be run inside a container. In Kubernetes jobs are always started inside a container and sequential or parallel processing of jobs is also possible. Cron jobs, which run at predetermined time intervals, are also supported by Kubernetes [244] and Aurora [245].

*Container-based services.* As already stated in section 4.3, all CO frameworks, except Docker Swarm stand-alone, offer a Service concept for exposing a replicated set of containers to customers as a stable service endpoint that is served by the distributed load balancer or the internal DNS service. Such stable service endpoint can be represented by one or more forms: a virtual IP address, a global service port, a fully qualified DNS name, or just a unique service name.

In Docker Swarm integrated mode [223] the following items must be declared by the application manager in the configuration file of a Service specification: (i) the container image to be deployed, (ii) optionally a virtual network to which the container must be connected – this virtual network must be created in advance [155] -- and (iii) optionally one or more global service ports or host ports. If a virtual network is specified, Docker Swarm will then also automatically generate a virtual IP address for the service. Moreover it will configure the L4 load balancer of its routing mesh to serve the declared global service ports.

Kubernetes expects that the application manager specifies in a Service configuration file one or more target ports at which the service must be exposed within the virtual IP cluster network; for each target port, it is also possible to expose the service within the routing mesh using a global service port [246]. However, no information about the virtual network must be declared; after all exactly one virtual network is mandatory installed in advance for the entire cluster. Moreover the application manager must not specify information about the container image. Instead, Kubernetes introduces the concept of ReplicaSet [247] for deploying a container image, exposing one or more container ports and host ports, for managing the number of container replicas and for attaching labels [248] to the running containers. The Service configuration file then only contains a so called label selector [249], which selects containers based on their attached labels. Labels and label selectors allow thus defining multiple services that load balance containers from different replica sets.

In order to expose a set of containers as a global service port in Marathon [250] and DC/OS [251], the cluster administrator must first install one or more instances of the marathon-lb or edge-lb load balancer [252]. To deploy an application, the application manager must then declare the following information in a so called Application specification [251]: (i) the container image, (ii) port mappings which are tuples of (host port, container port, global service port), (iii) the load balancer that should process the requests for the application and (iv) optionally the fully qualified domain name of the application when such DNS name is managed by an underlying cloud provider [253].

As DC/OS also supports virtual IP addresses per container [173], Marathon applications can alternatively be discovered via dcos-dns [176] and load balanced via a layer-4 load balancer [174]. The cluster administrator must first install one or more virtual networks [254]. The application manager must then specify in the Application specification (i) what virtual network to use [255] and (ii) a so-called name-based virtual IP [173] at which the service of the application must be exposed.

Finally, in Aurora a service is defined as a job of type Service that runs a container image in daemon mode [256]. As Aurora only supports host ports, DNS load balancing is used via mesos-DNS [191]: the containers of that service are registered in mesos-DNS as a list of SRV records (i.e. IP address and named ports see Section 4.3).

*Elastic scaling of services.* In all CO frameworks, the containers behind a service can be replicated across one or more nodes. The number of container replicas can be increased or decreased and for those CO frameworks with a service proxy, the service proxy will automatically reconfigure itself to take into account the change. Finally, although Docker Swarm stand-alone [257] does not offer the Service concept, it is still possible to replicate a specific container via Docker Compose [258], which is a tool for deploying composite applications across different environments in a portable way.

*Auto-scaling of services.* Kubernetes [259] also supports an auto-scaler functionality that automatically adapts the number of replicas depending on one or more threshold values with respect to a performance or resource consumption metric. In order for the auto-scaling to work, resource monitoring features (see Section 4.8.2) must be enabled. The DC/OS [260] distribution of Marathon also supports auto-scaling of Marathon applications by running a Python implementation inside a separate Docker container. It also includes third party documentation with other approaches for auto-scaling services.

*Global containers.* For some applications or framework support services it is necessary that a particular container image is running at every node of the cluster. This concept is supported in Docker Swarm integrated mode [103] and Kubernetes [261] as respectively global services and daemon sets.

*Composite applications.* Docker Swarm integrated mode and Marathon provide support for deploying multiple tiers of a distributed application in such an order that the dependencies between the tiers are respected. In Docker Swarm integrated mode [262], different service configurations and their mutual dependencies can be specified as part of a ComposeV3 file [66]. The docker stack deploy command takes as input such ComposeV3 file and deploys all the specified services together as one group while respecting the interdependencies between the services. Marathon supports a similar concept called Application groups [263].

Kubernetes does not support a similar concept natively, but several tools exist. First, Helm [264] is a command-line interface and run-time configuration management server for creating and managing the Helm charts. A Helm chart is a highly-configurable deployment package that encapsulates inter-dependent Kubernetes objects such as services, configuration setting or authentication credentials. Second, the Kompose tool [265] takes as input a Docker ComposeV3 file and translates this file to Kubernetes configurations such as multiple Services, ReplicaSets, Persistent volume configurations or Helm charts.

**Persistent volumes**. In all container orchestration frameworks, containers are stateless; when a container dies, any internal state is lost. Therefore, so called persistent volumes, which store the persistent state, can be attached to containers. Persistent volumes can be implemented by various mechanisms: services for attaching block storage devices to virtual machines such as Cinder, distributed file frameworks such as NFS, cloud services such as Google Persistent Disk, or local volumes that reserve a subset of the local disk resources. Persistent volume mechanisms can be categorized according to the following 9 features:

*Local volumes* that are comprised of disk resources of a container's local host node are supported by all CO frameworks. Mesos also enables frameworks to configure local volumes that are composed of multiple disk resources [266].

*Automatic (re)scheduling of containers to local volumes.* Containers that are configured to use a specific local volume are automatically (re)scheduled to the node where that local volume resides. Kubernetes [267], Mesos [268], Marathon and DC/OS [269] and Aurora [270] include full support for local persistent volumes with automatic scheduling support. Docker Swarm integrated mode [271] supports local volumes, but a manual scheduling constraint (see Section 4.6) must be specified to ensure that containers are always scheduled to that node only.

*Shareable volumes between containers* are supported by Docker Swarm integrated mode and stand-alone [272], Kubernetes [273] and Mesos [274]. Such shareable volumes can be used as an asynchronous data communication channel between containers. Note however, that in general not all types of persistent volumes support sharing.

*External persistent volumes* are supported by all CO frameworks. Such external volumes support managing data sizes that exceed a node's disk capacity and also allow for state recovery in case of node failures. Docker Swarm stand-alone [275], Docker Swarm integrated mode [276], Mesos [277], Aurora [278], Marathon [279] and DC/OS [280] provide support for mounting external Docker volumes by relying on a specific Docker volume plugin implementation [281]. Kubernetes also offers support for various persistent volume implementations but uses its own library of persistent volume implementations. Mesos and Kubernetes also support the Container Storage Interface (CSI) specification [282].

*Volume plugin architecture*. All CO frameworks except Aurora support a unified interface for different volume implementations. Overall there are two different architecture that are adopted by multiple CO frameworks: the Docker Engine plugin framework and the CSI-based plugins. These standardization efforts will be explained as part of the following paragraphs.

*Support for Docker volume plugin architecture.* The Docker Engine plugin framework [283], which offers a unified interface between the container runtime and various volume plugins, is adopted by Mesos [277], Marathon [279] and DC/OS [284] in order to support external persistent volumes. In Mesos-based frameworks, Docker volume plugins must be integrated via a separate Mesos module, named dvdi [285], which requires writing plugin-specific glue code. As such a limited number of Docker volume plugins are currently supported in Mesos.

*Support for the Common Storage Interface (CSI) specification.* The Common Storage Interface specification [282] aims to provide a common interface for volume plugins so that each volume plugin needs to written only once and can be used in any container orchestration framework. The specification also supports run-time installation of volume plugins. Typically, CSI can be implemented in any CO framework as a normal volume plugin, which itself is capable interacting with multiple external CSI-based volume plugins. Currently, CSI has been adopted by Kubernetes [286], Mesos [287] and DC/OS [288].

*Support for run-time installation of volume plugins* has been supported by the Docker Engine plugin framework[281] since Docker engine v1.12 and therefore also supported by Docker Swarm. As Mesos, Marathon and DC/OS have adopted the Docker volume plugin framework, these frameworks in principle also support run-time installation. Kubernetes also supports a unified interface for different volume implementations, but these are packaged in the source code of the Kubernetes releases [289]. As such, they cannot be dynamically installed in a running cluster. However, Kubernetes v1.9+ [286] and Mesos v1.5+ [287] support the CSI specification [282] that allows run-time installation of external volume plugins.

*Dynamic provisioning of persistent volumes* is supported by most CO frameworks. This feature entails that volumes must not be manually created by the application manager in advance, but instead volumes are automatically created or re-provisioned when a new container is started.

- In Docker Swarm integrated mode [290], volumes are linked to a service by means of the --mount option in the docker service create command. Here the volume plugin, known as volume driver, must be specified. When Docker schedules a task of a service on a specific node, and the volumes of the service are not present or linkable on that node, Docker Swarm tries to create a new one using the specified volume driver.
- Kubernetes [291] uses a more elaborate approach. Volumes are associated with Pods. Pods declare the required persistent volume type as a persistent volume claim [292] that requests a specific StorageClass [293] and specific data size quota and access modes. When none of the statically created, unmounted persistent volumes match that claim, Kubernetes dynamically provisions a volume based on the requested StorageClass.
- In Mesos-based frameworks, pinned persistent volumes can be dynamically provisioned thanks to Mesos' scheduler architecture that supports dynamic reservation and un-reservation of any type of resource [294]. Mesos also inherits the capability of the Docker Engine plugin framework to automatically provision Docker volumes, but this is not recommended in the documentation of Mesos [295].

**Reusable container configuration.** There are a number of commonly supported features related to supporting generic yet configurable container images.

*Passing environment variables to a container.* First, all CO frameworks allow to pass environment variables to a container, which is a common way for configuring the software that is running inside the containers (see Table 5).

*Self-inspection API.* Kubernetes [296] and Marathon[297] enable a container to retrieve information about itself via a so called downward API. Therefore, this information must not be specified as part of the container configuration or container image.

*Storing non-sensitive information (such as configuration files) outside a container's image.* Docker Swarm integrated mode [298] and Kubernetes [299] additionally support separating configuration data from images in order to keep containerized applications portable.

*Configuring a custom ENTRYPOINT and CMD.* All CO frameworks allow customizing the default ENTRYPOINT and CMD entries of a Docker image at run-time. ENTRYPOINT specifies the command that must be run when starting the container (e.g. /bin/sh –c opens a shell), while CMD specifies the arguments for the entrypoint's command (e.g. cassandra –f starts the Cassandra program of the official Cassandra container image).

Docker engine's docker run command allows to customize both the ENTRYPOINT and CMD at run-time. When only a custom CMD is specified, the default ENTRYPOINT is ran with the custom CMD. When a custom ENTRYPOINT is specified, the default CMD is cleared [300].

Docker Swarm stand-alone [301] supports the same customization scheme as the Docker engine because Docker Swarm stand-alone manages a set of nodes as a single virtual host that serves the standard Docker Engine API. Docker Swarm integrated mode [302] also supports the same scheme via the docker service create command.

Kubernetes [303], Mesos' Docker containerizer [304] and Mesos containerizer [305] as well as Marathon [306] also support the same customization scheme as Docker engine. Aurora requires that Python is installed inside a Docker container. However it supports passing a number of parameters to the Docker Engine that may include a customized ENTRYPOINT and CMD parameter.

**Service upgrades.** All CO frameworks support *rolling upgrades of services* by means of restarting the containers of the service with a new image. In this way the old version of the service gets gradually replaced with a new version. The status of the rolling upgrade can be monitored. Health or readiness checks can be configured in order to monitor the health or readiness of new container replicas. In case of failures, the upgrade can be paused, resumed or rolled back. The upgrade process itself can also be customized with respect to the desired availability of the old and new version of the service during the upgrade. Note Aurora supports both rolling upgrades of jobs and services [307].

*Monitoring the progress of a rolling upgrade.* Docker Swarm integrated mode [308] allows to monitor the progress of a rolling upgrade via the docker service inspect command. Aurora [309] allows to check the health of new tasks by means of heartbeat mechanism. A lost heartbeat pulse will block the update. Kubernetes [310] and Marathon [311] introduce the concept of Deployment for monitoring the progress of a rolling upgrade. A blocked update must be explicitly unblocked. Kubernetes provides the most extensive support for detecting a failed upgrade.

*Configuration of custom readiness checks.* It is also possible to configure custom readiness checks in Kubernetes [312] and Marathon [313]. These checks control when a newly started container is ready to process requests.

*Customizing the enactment of the rolling upgrade.* Docker Swarm integrated mode [258], Kubernetes [314], Marathon [315] and DC/OS offer various options to customize how the rolling upgrade process is executed/enacted. A common enactment customization is controlling how many instances of the old and new version of the service should always be running during the upgrade. In Docker Swarm integrated mode [316], the maximum number of containers that can be upgraded in parallel can be specified. Kubernetes [317] allows to specify the maximum number of unavailable pods during the upgrade and to specify the maximum surge [318], which is the number of pods that can be created over the desired number of pods. Marathon [315] allows specifying the minimum health capacity of the old version as a percentage of old containers for which a new container must be deployed side-by-side, after which the new version is scaled to 100% and the old version is stopped. Aurora [319] only supports side-by-side replacements, but the number of tasks that can be updated in parallel can be configured.

*Roll back.* Docker Swarm integrated mode [320], Kubernetes [321], Aurora [319] and the DC/OS [322] distribution of Marathon support rolling back an upgrade. Aurora does not offer a command for rolling back an upgrade but can be configured to automatically rollback in case of a failure. Note that recovering from a failed upgrade is a more complicated problem than what a roll back can resolve. In most case, it is better to roll forward by upgrading to a resolved application state.

*Canary deployments.* A variant of rolling upgrades, named blue-green deployments or canary deployments, intents the same effect as a rolling upgrade but allows for more manual control over the upgrade. The application manager will deploy a completely new service next to the existing service and the application manager can manually control when to redirect users from the old to the new service. Typically this redirection is only performed after testing the health and readiness of the new service. Moreover, users are redirected in a gradual way so the old service is gradually scaled down while the new service is gradually scaled up. Kubernetes [323], Aurora [319] and DC/OS [324] support performing such canary deployments.

*In-place updates of application configurations.* Several CO frameworks allow narrow updates to application configuration files such as changing the value of a field. Two different implementation strategies exist:

- Kubernetes [325] and DC/OS [326] support updating the configuration file of a Pod with a new version of that configuration file. Kubernetes does not only support updating Pods but also any other API objects such as Deployments and Secrets; moreover different kubectl commands exist for updating the API configuration files: patching [327], applying [328] or editing [329].
- Docker Swarm stand-alone [330], Docker Swarm integrated mode [331] and DC/OS [332] offer an update command that allows changing one or more properties in an application configuration directly without creating a new configuration file.

**Table 5.** Commonly supported features for the "application configuration and deployment" aspect.

Cell legend:
- *Partially supported feature*: The feature is only partially supported
- externalComponent: Support for the feature is not included in the open-source distribution of the CO framework, but the feature is supported by a third party component or platform. The URL to the corresponding documentation is included. The name of the URL refers to the name of the component.
- tutorial: The feature is not directly supported by the framework,     but a set of tutorials how to add auto-scaling capabilities using third-party components has been provided as part of documentation.

| Application configuration and deployment sub-aspects | Features | Swarm stand-alone | Swarm integrated | Kubernetes | Mesos | Mesos + Aurora | Mesos + Marathon | DC/OS |
|---|---|---|---|---|---|---|---|---|
| | | Sa | Si | Ku | Me | Au | Ma | Dc |
| Supported workload types | *Pods* | | | ✓ | ✓ | | ✓ | Dlgt |
| | *Container-based jobs* | | | ✓ | | ✓ | | Add |
| | *Container-based services* | | ✓ | ✓ | | ✓ | ✓ | Dlgt |
| | *Elastic scaling of services* | ✓ | ✓ | ✓ | | ✓ | ✓ | Dlgt |
| | *Auto-scaling of services* | | | ✓ | | | marathon-autoscale | |
| | *Global containers* | | ✓ | ✓ | | | | |
| | *Composite applications* | ✓ | ✓ | Helm Kompose | | | ✓ | Dlgt |
| Persistent volumes | *Local volumes* | ✓ | ✓ | ✓ | ✓ | ✓ | ✓ | Dlgt |
| | *Automatic (re)scheduling* | | | ✓ | ✓ | ✓ | ✓ | Dlgt |
| | *Shareable volumes between containers* | ✓ | ✓ | ✓ | ✓ | | | |
| | *External volumes* | ✓ | ✓ | ✓ | ✓ | ✓ | ✓ | Dlgt |
| | *Volume plugin architecture* | ✓ | ✓ | ✓ | ✓ | | ✓ | Dlgt |
| | *Run-time installation of volume plugins* | ✓ | ✓ | *CSI* | ✓ | | ✓ | Dlgt |
| | *Docker volume plugin system support* | ✓ | ✓ | | ✓ | | ✓ | Dlgt |
| | *Common Storage Interface (CSI) support* | | | ✓ | ✓ | | | Dlgt |
| | *Dynamic provisioning of volumes* | ✓ | ✓ | ✓ | *Supported for local volumes but not recommended for Docker volumes* | | | |
| Reusable container configuration | *Pass environment variable to container* | ✓ | ✓ | ✓ | ✓ | ✓ | ✓ | Dlgt |
| | *Self-inspection API* | | | ✓ | | | ✓ | Dlgt |
| | *Separate configuration data from image* | | ✓ | ✓ | | | | |
| | Custom ENTRYPOINT | ✓ | ✓ | ✓ | ✓ | ✓ | ✓ | Dlgt |
| | Custom CMD | ✓ | ✓ | ✓ | ✓ | ✓ | ✓ | Dlgt |
| Service upgrades | Rolling upgrades of services | | ✓ | ✓ | | ✓ | ✓ | Dlgt |
| | Monitoring of a rolling upgrade | | ✓ | ✓ | | ✓ | ✓ | Dlgt |
| | Roll back | | ✓ | ✓ | | ✓ | | Add |
| | Configuration of custom readiness checks | | | ✓ | | | ✓ | Dlgt |
| | Customizing the rolling upgrade process | | ✓ | ✓ | | ✓ | ✓ | Dlgt |
| | Canary deployments | | | ✓ | | ✓ | | Add |
| | In-place updates of app configurations | ✓ | ✓ | ✓ | | | | Add |
| | Non-disruptive, in-place updates | ✓ | ✓ | ✓ | | | | |

*Non-disruptive, In-place updates without restarting containers*. CO frameworks differ in whether the aforementioned in-place updates can be performed with or without restarting containers.

- In Docker Swarm stand-alone [330], the docker update command allows performing changes to containers without restarting them. The set of possible properties than can be updated is extensive, but application managers should be aware that some properties such as resource limits, should be updated carefully in order to prevent service outages.
- In Docker Swarm integrated mode [331], the docker service update command performs every in-place update by means of a rolling upgrade but containers are not always restarted in order for the update to take effect.
- In Kubernetes [327], in-place updates of a Pod API object are always performed using a rolling upgrade and Pods are always restarted regardless of the property. As a result, Pods may also be rescheduled on another node. Note, however, that in-place updates of other API objects (e.g. editing the upgrade strategy of a Deployment object) can be performed without restarting the related Pods.
- In DC/OS [332], this feature is not supported: every in-place update is performed by means of a rolling upgrade and containers are always restarted regardless of the properties.

### 4.4.2 Unique features

Several CO frameworks have unique features in several sub-aspects.

**Supported workload types.** For this sub-aspect, Kubernetes offers the following unique features:

- Init containers [333] are specialized containers that run before application containers and can contain utilities or setup scripts not present in the application image.
- Vertical Pod Autoscaler [334] is an infrastructure service that automatically sets resource allocation policies of Pods and dynamically adjusts them at runtime, based on analysis of historical resource utilization, amount of resources available in the cluster and real-time events, such as out-of-memory events. Adjusting resource allocation policies requires that Pods are killed and new Pods will be recreated with adjusted policies set.

**Persistent volumes.** For this sub-aspect Kubernetes, Mesos, and DC/OS offer the following unique features

Kubernetes:

- Higher-level, automated support for deploying stateful services such as database clusters is provided via the StatefulSet [335] concept. The realization of this concept depends on the *automated provisioning of persistent volumes* feature and two container networking features: *bypassing the L4 load balancer* and *the internal DNS for service discovery* (see Section 4.3).
- Support for managing raw block storage inside containers [336] without the abstraction of a file system. This allows for higher performance of containerized databases [337].

- Kubernetes v1.8+ [338] enables to resize existing persistent volumes. Kubernetes v1.11+ [339] allows resizing of persistent volumes without having to restart the Pods that refer to these persistent volumes.
- Support for dynamic maximum volume count [340] enables volume plugins to specify a limit on the maximum number of volumes that can be attached to a node and allows this limit to be configured per type of node.

Mesos:

- A shared local volume can be shared by tasks of different frameworks [274].
- Mesos v1.6+ [341] extends the Framework API with operations for growing or shrinking persistent volumes, but these new operations have not yet been used by any Mesos-based CO framework.

DC/OS

- DC/OS commons [342] is a collection of tools, libraries, and documentation for easy integration and automation of stateful services, such as databases, message brokers, and caching services. It comes with pre-configured packages for deploying such stateful services in DC/OS. These services do not run on top of DC/OS's container orchestration framework however. This might give a performance gain in comparison to other CO frameworks because the substantial performance overhead of a virtual network layer is avoided.

**Reusable container configuration.** For this sub-aspect, Kubernetes and Docker Swarm integrated mode offer the following unique features:

Kubernetes:

- Podpresets [343] can be used to inject volume mounts, secrets or environment variables into a Pod at creation-time. It helps application developers to avoid rewriting the same Pod configuration specification across multiple Pods. It also enables separation of concerns: developers of containers consuming a specific service do not need to know all the details about that service

Docker Swarm:

- An option can be set for running a simple service initialization system inside containers. Applications that fork child process typically rely on a service initialization system for reaping these child processes to prevent resource leaks and zombie processes. As existing service init systems for Linux such as systemd or upstart are overkill for use in containers, a simplified service init system called tini [344] can be set as the ENTRYPOINT of a container image (see Common feature *"configuring a custom ENTRYPOINT and CMD")*. Such reconfiguration of the entry point must in principle be applied for any application that forks and haven't been written with child reaping in mind as normally they would leave this up to the init system. A typical case is the java Jenkins applications.

  Docker Swarm provides an init option [345] to automatically apply this reconfiguration. When creating a service and the init option is set, Docker Swarm will automatically set the ENTRYPOINT to tini and passes the CMD to it or whatever is specified in the command-line. This option is possible for Docker Swarm stand-alone [346] and Docker Swarm integrated mode [345].

**Service upgrades.** For this sub-aspect, Docker Swarm integrated mode offers the following unique feature:

- Docker Swarm integrated mode allows customizing the enactment of a roll back [320] of a service. Several options can be specified.

4.5 *Resource quota management*

This aspect covers features of CO frameworks that a cluster administrator must understand in order to organize the hardware resources of a cluster among different teams or organizations.

4.5.1 Common features

*Concept for partitioning API objects into logically named user groups.* All container orchestration frameworks offer a concept for partitioning one or more types of API objects (e.g. services, volumes) into a logically named user group that corresponds with a specific organization or tenant that is able to contract resources from the cluster. Docker EE [347] names this concept Collections [348], Kubernetes calls it Namespaces [349], Apache Aurora uses Job roles [350] that directly refer to a Unix user account. Mesos Marathon does not support this concept, but the extended DC/OS distribution of Marathon supports Service Groups [351]. The typical use case of user groups is to reserve a subset of resources for a tenant of the cluster.

Mesos does not offer the exact concept of user groups. Instead it offers a similar concept, named *framework roles* [352], for dividing hardware resources across multiple scheduler frameworks. A specific Mesos framework is authorized by a cluster administrator (see Section 4.7) to run tasks using the resources of one or more roles. When a framework reserves a set of resources, it must specify a role so that the Mesos master can account for the total resource usage of that role.

**Table 6**. Commonly supported features of the "resource quota management" aspect.

Cell legend:
- **$EnterpriseEdition$**: Support for the feature is not included in the open-source distribution of the CO framework, but is included in the commercial enterprise edition. The URL to the corresponding documentation is included. The name of the URL refers to the name of the enterprise edition.
- *partial support*: the CO framework offers partial support for the feature. The URL to a relevant documentation page is included. The name of the URL refers to what's essentially supported.

| Resource quota management sub-aspects | Features | Swarm stand-alone | Swarm integrated | Kubernetes | Mesos | Mesos + Aurora | Mesos + Marathon | DC/OS |
|---|---|---|---|---|---|---|---|---|
| | | Sa | Si | Ku | Me | Au | Ma | Dc |
| Resource quota management | *Partitioning API objects in user groups* | | $Docker EE$ | ✓ | ✓ | ✓ | | Add |
| | *CPU, mem and disk quota per user group* | | | ✓ | ✓ | ✓ | | |
| | *Object count quota limits per user group* | | | ✓ | *ports* | | | |
| | *Reserving resources for the CO framework* | | | ✓ | | | ✓ | Dlgt |

*Declaring a minimum guarantee and/or maximum limit on CPU, memory and disk quota per user group.* Kubernetes, Mesos and Aurora provide support for declaring a minimum and a maximum quota of CPU, memory and disk resources per user group. More specifically:

- Kubernetes supports attaching to Namespaces minimal guarantees and maximum limits for CPU and memory quota [353] and maximum limits for disk quota per storage class [354].
- Mesos supports attaching to framework roles minimal guarantees [355] for CPU, mem and disk quota [356] for local volumes as well as weights [357] for dividing resources across roles. Apache Aurora allows attaching to job roles quota for memory and disk [358] via the aurora_admin set_quota command.

*Declaring an object count quota limit for the number of API objects per user group.* Kubernetes and Mesos allow assigning to user groups a maximum number of API objects such as the number of nodes, containers, services, etc. More specifically, in Kubernetes [359], object count quota can be declared by expressing a maximum quantity for different kinds of Kubernetes API objects. In Mesos [356], port ranges can be associated to framework roles. In Docker EE [347] high-level resources such as nodes [360], volumes [361] and services can be organized in collections, but there is no declaration of a maximum limit. The DC/OS distribution of Marathon [362] also allows organizing services into service groups without enforcing a limit on the number of services for a service group.

*Reserving resources for the CO framework.* The available set of resources on a node is automatically computed via the operating system in all CO frameworks. Additionally, Kubernetes [363] and Marathon [364] can be configured to reserve a subset of the node resources for the framework's operation and local daemons.

### 4.5.2 Unique features

Mesos offers the following unique feature for the aspect **Resource quota management**. This feature contributes to improved performance isolation between Mesos frameworks:

- Framework rate limiting [365] aims to protect high-SLA frameworks (e.g., production, service) by setting limits to the request rate to the Mesos Master. Frameworks that violate the request rate limit are throttled and these requests are stored in memory by the Mesos master.

### 4.6 *Container QoS management*

This aspect covers features of CO frameworks that an application manger must understand in order to efficiently use the resources of a user group while also achieving the intended QoS level of its applications.

Supporting high utilization of allocated resources while also maintaining desired QoS levels of applications, during either normal execution or resource contention and failures, is a complex goal. To support this complex goal, CO frameworks are designed with the following two goals in mind:

- Resource allocation models have been developed that support QoS differentiation between containers while also allowing for over-subscription of resources to improve server consolidation.
- CO frameworks offer various mechanisms to application managers for controlling scheduling decisions that influence the performance of the application. These decisions include the placement of inter-dependent containers and data, and prioritization of containers during resource contention.

Note that the offered features do not provide strong SLA guarantees at the level of application-specific metrics (e.g. latency or throughput) but include general mechanisms that can be used to balance the competing goals of improved resource utilization and controllable performance of the application.

### 4.6.1 Common features

**Container CPU and memory allocation with support for oversubscription.** This sub-aspect covers common features of CO frameworks that an application manager must understand to (i) allocate sufficient resources to a container to achieve its intended performance level, but also to (ii) allow flexible reallocation of idle resources to improve resource utilization.

In general, the allocation of computational resources to a container is governed by means of resource allocation policies. Container orchestration frameworks differ in their support for resource allocation policies and also differ in the type of resources that can be limited. In the following, we set out the available support for the different types of resources.

*Minimum guarantees and maximum limits for CPU and memory.* Kubernetes and Docker Swarm provide support for minimum guarantees and maximum limits for CPU and memory, while Mesos-based frameworks supports minimum guarantees for CPU and maximum limits for both CPU and memory:

- Kubernetes manages a <request, limit> [366] pair for CPU and memory for each container and Pod. A Request defines the resource quantity that is always guaranteed to the container (e.g. a requests of 1.5 CPU imply that 1 CPU core and 50% of another CPU core is fully assigned to the container), while a Limit specifies the maximum resource quantity that can be used by this container (e.g. a request of 1.5 CPU and a Limit of 2 CPU specifies that the container is guaranteed 1.5 CPU cores, but it can take up until 2 full CPU cores if the processing power is not used by other containers). When a CPU limit is crossed [367] by a container, the container will be throttled. When a memory limit is crossed [368], the process using the most memory in the container is killed. Note that when the Request is set lower than the Limit, the container is guaranteed the Request but can opportunistically consume the difference between Request and Limit if some resources are not being used by other containers. It has been shown in Borg, the predecessor of Kubernetes, that setting Requests and Limits in the above ways increases resource utilization [52]. Logically, when Requests and Limits of the enclosing Pod are defined, the sum of the Requests and the sum of the Limits of its containers must always be lower than the Request and Limit of the Pod. The current implementation [369] of Requests and Limits for Docker containers uses specific options of the docker run command.
- Docker Swarm integrated mode implements a similar model as Kubernetes called a <reservation, limit> [370] with the same semantics as a <request, limit> pair in Kubernetes.
- Docker Swarm stand-alone supports all resource allocation options of the docker run command; Since Docker 1.13+ it is possible to model minimal guarantees (i.e. reservations) as well as maximum limits for CPU and memory. Minimal guarantees for CPU involves however complex configuration [371] of the Linux kernel's CFS scheduler [372] either through --cpu-shares or a combination of --cpu-quota and --cpu-period options.
- In Mesos, different isolator modules [193] exists for enforcing resource allocation modules for CPU and memory:
    - Various cgroups-based isolator modules are used for enforcing among others: CPU guarantees and limits [373] and memory limits. The isolator for CPU supports minimal guarantees (but based on complex configuration of CFS-based CPU shares [372]) and maximum limits (by means of CFS-based bandwidth control [374]).
    - Mesos v1.1.0+ [375] provides support POSIX rlimits [375], which consist of a soft and hard limit for CPU and memory. The soft limit does not imply a guarantee however, it implies an effective limit which is set by the application manager; this limit may however be increased until the hard limit, which is set by the cluster administrator. The advantage of POSIX rlimits is that it does not only support limiting CPU and mem but also other POSIX resources such as nproc and memlock. POSIX rlimits are currently not used by any Mesos-based CO framework however.
    - Mesos also offers support for another kind of over-subscription in the form of revocable resources [376], which are resources that are already reserved for other processes but currently not used. However these resources are best-effort and can be revoked anytime by Mesos (see also Section 4.6). Only Aurora [377] uses this Mesos feature.
- Aurora [378] and Marathon [379] support only minimal guarantees for CPU (based on CPU-shares) and maximum limits for memory.

*Abstraction of cpu-shares for enforcing CPU guarantees.* Note that Mesos, Aurora, Marathon and Docker Swarm stand-alone rely on CPU-shares of the CFS Scheduler for implementing minimal guarantees. CPU-shares are however difficult to configure because cpu-shares are always defined as weights that are relative to the cpu-shares of other co-located containers: for example, if a new container is started, then the cpu-shares declared by that new container reduce the weights of the already running containers. Kubernetes [366] and Docker Swarm integrated mode [370], on the other hand, offer higher-level abstractions for expressing minimal guarantees that hide the complexity of cpu-shares.

**Allocation of other resources.**

*Limits for NVIDIA GPU* are supported by Mesos [380], Aurora [378] and Marathon [381] (and DC/OS [382]). Kubernetes [383] offers partial support for GPU allocation. First, containers cannot requests fractions of a GPU, only a whole GPU. A single GPU device can neither be shared between containers.

*Limits for disk resources.* Mesos offers support for hard [241] and soft limits for disk usage. Hard limits for disk usage are adopted by Aurora [384], Marathon [381] (and DC/OS). Kubernetes [385] offers support for setting a <request, limit> pair for usage of a node's local root partition (ephemeral storage).

**Controlling scheduling behavior by means of placement constraints.** All CO frameworks allow to restrict the placement decision of the default scheduling algorithm by means of various user-specified constraints in order to improve the QoS level of applications. These user-specified constraints support placing inter-dependent application containers and data close or far from each other in the network topology. Different types of constraints are supported:

*Restrict the set of nodes by evaluating over node labels.* CO frameworks allow to restrict the set of nodes on which a specific container can be scheduled by means of evaluating over node labels or attributes (see Section 4.8.2). A label is defined as a <key, value> pair. A number of such labels are predefined like the hostname of the node.

*Evaluate over custom labels.* Custom labels can also be defined: in Docker Swarm integrated mode [316] and Kubernetes [386], custom labels can be dynamically added or removed, whereas in Marathon [202] and DC/OS [387] custom attributes [388] can only be changed by (re)starting the Mesos agent with the desired list of attributes.

*More expressive constraints.* The CO frameworks differ in the expressiveness of the constraints. Docker Swarm integrated mode [389] offers set-based inclusion operators for both label keys and label values.

Kubernetes [390] does not only offer the same set-based inclusion operator, but also more expressive affinity and anti-affinity constraints [391] between pods and nodes, and inter-pod affinity and anti-affinity constraints [391]. Finally, taints and tolerations [392] work together to ensure that pods are not scheduled onto inappropriate nodes such as nodes with specific hardware such as GPUs. As such pods that don't need GPU resources are kept off those nodes.

Aurora [106] and Marathon [107] also offer different kinds of operators for evaluating over attributes.

**Controlling preemptive and re-scheduling behavior.** This sub-aspect covers common features of CO frameworks that an application manager must understand in order to customize the pre-emptive scheduling and rescheduling logic of CO frameworks such that an intended QoS level for a containerized application is achieved during several exceptional conditions: (i) resource contention at the scheduler level, (ii) out-of-resource node conditions, (iii) node failures, (iv) container start failures and (v) unbalanced services of which the containers are not spread across different nodes..

*Pre-emptive scheduling.* Kubernetes [393] and Aurora [394] use priorities between containers for killing low-priority containers in case the scheduler cannot find a node with enough available resources for scheduling a new container.

*Container eviction when a node runs out of resources.* A fully packed node will likely run out of resources in Kubernetes and Docker Swarm when the maximum resource limits of multiple containers on that node are set higher than their minimum guarantees. After all, the default scheduling algorithm of Kubernetes [395] and Docker Swarm[5] will allocate containers to a node so that, for each resource type, the sum of the containers' minimum guaranteed resources does not exceed the capacity of that node.

To handle such out-of-resource conditions, Kubernetes [396] distinguishes between different QoS classes of containers. Kubernetes orders Pods in the best-effort, burstable and guaranteed QoS classes depending on the containers' Request and Limits. Kubernetes provides supports for proactive and reactive out-of-resource handling:

- Proactive handling by means of Pod eviction [397]: When a node is about to run out of CPU, memory or disk resources, the local kubelet agent can trigger the eviction of a Pod consuming the most resources from that node. Moreover when CPU and memory resources are affected, Pods are also ranked according to the QoS class they belong:
    o The best-effort class will be evicted first.
    o A Pod of the bursty class that consumes the greatest amount of the starved resource relative to their request are evicted thereafter.
    o A pod of the guaranteed class will never be evicted because of another Pod's resource consumption, unless a system daemon (docker, kubelet, journald) is consuming more resources than were reserved (cfr Section 4.5, *Reserving resources for the CO framework)*.

    The scheduler will try to place the evicted Pod on another node in case the Pod is controlled by a Replicaset or Deployment.

- Reactive handling by means of killing a container [398]: When a node runs out of memory resources because the local kubelet agent did not evict a pod on time, container processes that are consuming the largest amount of memory relative to their request will be killed first. When a node runs out of CPU resources, it will not be killed but the CPU will be throttled

Aurora [358] also orders tasks in the revocable, pre-emptible, and preferred classes. Revocable tasks have the lowest priority as they can only use revocable resources [377], which are resources that are currently not needed but can always be reclaimed by other running Mesos tasks (see Section 4.5).

*Container eviction on node failures.* Node failure detection is performed by means of different kinds of health checks by the master. When a node is considered failed, the master reschedules containers on that node to healthy nodes. Mesos [399] offers more advanced support in the fact that it distinguishes between multiple failure scenarios (failed or partitioned agents) and multiple recovery tactics depending on whether the frameworks on the failed agent have enabled checkpointing [400].

*Container lifecycle handling.* All CO frameworks, except Docker Swarm stand-alone, manage *the life cycle of a container as a state machine* from the moment a request to create a new container arrives at the Master API. This means that when containers are placed on a node, but they are stuck in a pending or staging state, application managers are automatically informed via their CLI, web GUI or via an event (see Section 4.8.2). In Kubernetes [401], container developers can use a container life cycle hook framework to run code triggered by events during their management lifecycle.

*Re-distributing unbalanced services.* As container clusters are very dynamic, the distribution of containers over nodes may become unbalanced over time, for example when adding new nodes to the cluster. Docker Swarm integrated mode [402] will not automatically redistribute unbalanced containers of a service to idle nodes in order to avoid temporary service disruptions. Instead the application manager can force by means of the docker service update command that the containers of a particular service are re-distributed.

In Kubernetes, Pods may also arrive in an unbalanced state. For example, we experienced that the Kubernetes scheduler does not redistribute pods of a service across multiple nodes after suspended or shutdown VMs come back on-line. This isn't a problem with clusters with high number of nodes but in small clusters with a few nodes, it is a problem in the sense that all pods of a service may get stuck on the same node [403]. However, a future version of Kubernetes [404] will contain a Descheduler component, which automatically evicts Pods of unbalanced services based on a policy. When these pods belong to a replication controller, the replication controller will ensure that a replacement Pod will be placed on an appropriate node by the default scheduler.

---

[5] There is no documentation on this. Run-time tests indeed showed that only reservations are taken into account by Docker Swarm's scheduling algorithm.

**Table 7**. Commonly supported features for the "container QoS management" aspect.

Cell legend:
- *partially supported*: The feature is partially supported by the CO framework. A URL to relevant documentation has been included. The name of the URL explains the essence of what is supported.
- future: The feature is not yet part of the open-source distribution of the CO framework. It has however been planned according to the documentation, or there is a separate incubation project. A URL to relevant roadmap documentation is included.

| Container QoS management sub-aspects | Features | Swarm stand-alone | Swarm integrated | Kubernetes | Mesos | Mesos + Aurora | Mesos + Marathon | DC/OS |
|---|---|---|---|---|---|---|---|---|
| | | Sa | Si | Ku | Me | Au | Ma | Dc |
| Container CPU and memory allocation with support for oversubscription | *Minimum guarantees for CPU* | ✓ | ✓ | ✓ | ✓ | ✓ | ✓ | Dlgt |
| | *Abstraction of cpu-shares for CPU guarantees* | | ✓ | ✓ | | | | |
| | *Minimum guarantees for memory* | ✓ | ✓ | ✓ | | | | |
| | *Maximum limits for CPU* | ✓ | ✓ | ✓ | ✓ | | | |
| | *Maximum limits for memory* | ✓ | ✓ | ✓ | ✓ | ✓ | ✓ | Dlgt |
| Allocation of other resources | *Limits for NVIDIA GPU* | | | no gpu sharing | ✓ | ✓ | ✓ | Dlgt |
| | *Limits for disk resources* | | | local storage | ✓ | ✓ | ✓ | Dlgt |
| Controlling scheduling behavior by means of placement constraints | *Evaluate over node labels/attributes* | ✓ | ✓ | ✓ | n/a | ✓ | ✓ | Dlgt |
| | *Define custom node labels/attributes* | ✓ | ✓ | ✓ | ✓ | ✓ | ✓ | Dlgt |
| | *More expressive constraints* | ✓ | ✓ | ✓ | n/a | ✓ | ✓ | Dlgt |
| Controlling preemptive scheduling and re-scheduling behavior | *Preemptive scheduling* | | | ✓ | | ✓ | | |
| | *Container eviction when out-of-resource* | | | ✓ | | ✓ | | |
| | *Container eviction on node failure* | ✓ | ✓ | ✓ | ✓ | ✓ | ✓ | Dlgt |
| | *Container lifecycle handling* | | ✓ | ✓ | ✓ | ✓ | ✓ | Dlgt |
| | *Re-distributing unbalanced services* | | ✓ | future | | | | |

## 4.6.2 Unique features

Docker Swarm stand-alone offers the following unique feature for the sub-aspect "Container CPU and memory allocation with support for oversubscription":

- The Remote Docker API supports updating resource reservations and limits without recreating the container [330]. This is a dangerous operation because human users can set resource reservations higher than the actual available resources of the local node and therefore threaten the QoS of other containers on that node. Therefore, such operations should be managed automatically, for example by a vertical scaler approach [128].

Mesos and Kubernetes offers the following unique features for the sub-aspect "allocation of other resources":

Kubernetes:

- Extended Resources [406] allow cluster administrators to add new node-level resources of random kind. Extended resource quantities must be integers and cannot be overcommitted. As such, a pod's request and limit for an extended resources must be equal if both are declared.
- Kubernetes supports the allocation of pre-allocated huge pages [407] by applications in a Pod. Contemporary computer architectures support bigger pages [408] for virtual memory so that CPU and OS need less memory address lookups for retrieving a piece of data, thereby speeding up performance.

Mesos contributes to improved network performance isolation between containers:

- When containers are interconnected via the routing mesh network, the port mapping isolator [165] includes extensive support for network isolation between containers: port range limits , rate limits for container egress traffic [409].
- The Mesos containerizer also supports network isolation between containers for virtual networks [410]. The cgroups/net_cls [410] Isolator module enables cluster operators to implement network performance isolation and segmentation by means of the Linux kernel's network classifier cgroup [411] (net_cls), which provides an interface to tag network packets with a class identifier. These class identifiers can be used by kernel modules such as qdisc (for traffic engineering) and net-filter (for firewall rules) to enforce network performance and security policies. These policies can be specified by a cluster administrator through tools such as tc [412] and iptables.
  Unfortunately, little of these features is currently used by Aurora and Marathon. Only container port ranges of the port mapping network isolator are managed in Marathon.

Kubernetes also offers the following unique feature for the sub-aspect "controlling scheduling behavior by means of placement constraints":

- CPU management policies [413] allow application managers to exclusively reserve a set of CPUs for a specific Pod and once the Pod has been allocated to a set of CPUs of a particular node, the Pod cannot be migrated to another node. This feature is important for workloads where CPU cache affinity and scheduling latency significantly affect workload performance.

### 4.7 Securing clusters

This aspect covers features that a cluster administrator must understand in order to setup a secure cluster. Note this aspect focuses on the security at the level of the container orchestration framework only as it does not entails features related to the security of the applications running inside containers.

#### 4.7.1 Common features

**User identity and access management.** All CO frameworks provide *secure access to their Master API by means of authentication and authorization of users*. The CO frameworks differ in the range of supported authentication and access control models, as well as the plug-ability of the solutions:

- Both Docker Swarm stand-alone [414] and integrated mode [415] rely on TLS-based authentication of the Docker daemon. With respect to authorization, the default authorization model is all or nothing: any user with permission to access the Docker daemon can run any Docker client command. However, it is also possible to start the Docker daemon with external authorization plugins. Docker EE's Universal Control Plane [416] (UCP) also supports role-based access control [347].

- Kubernetes supports various authentication strategies [417] and various mainstream access control models [418]. Kubernetes 1.11+ [419] also supports improved plugin framework for supporting third-party credential providers. Cloud providers, vendors, and other platform developers can now release binary plugins to handle authentication for specific cloud-provider IAM services, or that integrate with in-house authentication frameworks that aren't supported by the open-source distribution, such as Active Directory.
- Mesos supports CRAM-MD5-based authentication [420] of cluster administrators and frameworks, which are both identified through principals. A framework's principals is the person responsible for that specific frameworks.
- Aurora supports authentication and role-based authorization of users [421] by means of Apache Shiro [422].
- Marathon supports authentication [423] and authorization [424] of users with respect to the master API of the container orchestration framework.
- The DC/OS Enterprise distribution of Marathon supports provider-based authentication for single sign-on [425] via either SAML or OpenID Connect. DC/OS Enterprise also provides directory-based authentication [426] based on LDAP. Finally authorization is permission [427]-based.

*Tenant-aware access control.* All CO frameworks, except Docker Swarm stand-alone, support *tenant-aware access-control* that grants users, teams or organizations specific access permissions to a particular user group (see user groups and quota in Section 4.5).

Docker EE's UCP uses role-based access control [347] for granting users and teams access to Collections [348]. Kubernetes [418] associates by default authorization rules to a specific Namespace [349].

Mesos [428] supports authorization of frameworks, which means that cluster administrators can configure which principals can register frameworks under which roles [352] (i.e. resource quota reserved for a particular user group – see Section 4.5).

Aurora [429]'s authorization documentation unfortunately uses the same term for managing access permissions and user groups, i.e. roles: on the one hand, Aurora's basic authorization module [430] allows associating users to roles and permissions to these roles. On the other hand, the latter permissions may include access to the resources of a particular user group, which is named Job Role [394] in Aurora.

The DC/OS Enterprise distribution of Marathon [431] grants users and organizations access to Service Groups [351].

**Cluster network security.**

*Authentication of worker nodes with the master API* is supported by Docker Swarm stand-alone [432], Docker Swarm integrated mode [433], Kubernetes [434], Mesos [420] and Aurora [422]. Docker Swarm and Kubernetes rely both on TLS-based public key certificates. Worker nodes use a client certificate in order to join the cluster in a secure manner.

*Automated bootstrap of authentication tokens for worker nodes* is also supported by Docker Swarm integrated mode [435] and Kubernetes [436] . An authentication token is essentially a symmetric key that enables worker nodes to more easily register with the master node to join the cluster. For Kubernetes, this feature is only out-of-the-box supported in the kubeadm [437] deployment tool.

*Authorization of CO agents on worker nodes towards the master API* is additionally supported by Kubernetes , Mesos [438], and Aurora [422]. In Kubernetes, this feature allows to grant master API access permissions to the Kubelet agent of any node based on the containers which are currently running on that node. Mesos can be configured with an ACL to allow or deny worker nodes to (re)-register with the master. Aurora relies on Zookeeper's ACL mechanisms [439] for controlling Aurora-specific actions of the worker nodes.

**Table 8**. Commonly supported features for the "securing clusters" sub-aspect.

Cell legend:
- $..$: Support for the feature is not included in the open-source distribution of the CO framework, but is included in a commercial product or cloud service. The URL to the corresponding documentation is included. The name of the URL refers to the name of the product or service.
- externalComponent: Support for the feature is not included in the open-source distribution of the CO framework, but the feature is supported by a third party component or platform. The name of the URL refers to the name of the component. The URL to the corresponding documentation is included.

| Securing clusters sub-aspects | Features | Swarm stand-alone | Swarm integrated | Kubernetes | Mesos | Mesos + Aurora | Mesos + Marathon | DC/OS |
|---|---|---|---|---|---|---|---|---|
| | | Sa | Si | Ku | Me | Au | Ma | Dc |
| User identity and access management | *Authentication of users with master API* | ✓ | ✓ | ✓ | ✓ | ✓ | ✓ | Extnd |
| | *Authorization of users with master API* | | ✓ | ✓ | ✓ | ✓ | ✓ | $Extnd$ |
| | *Tenant- aware ACLs* | | $Docker EE$ | ✓ | ✓ | ✓ | | $Add$ |
| Cluster network security | *Authent. of worker nodes with master API* | ✓ | ✓ | ✓ | ✓ | ✓ | | Dlgt |
| | *Automated bootstrap of worker tokens* | | ✓ | kube adm | | | | |
| | *Authorization of CO agents on workers* | | | ✓ | ✓ | ✓ | | |
| | *Encryption of control messages* | | ✓ | $GKE$ | | | | $Add$ |
| | *Restricting external access to service ports* | | ✓ | ✓ | | | | Add |
| | *Encryption of application messages* | | ✓ | weave NET | | | | $Add$ |

*Encryption of control messages between masters and workers* is supported by Docker Swarm integrated mode [440] and Kubernetes Container-as-a-Service offering Google Container Engine [441]. Moreover, DC/OS [442] can be configured to startup in a strict or permissive security mode that respectively enables or enforces TLS encryption of communications between masters and agents.

*Encryption of application messages* is supported by Docker Swarm integrated mode and DC/OS as an optional feature. In Docker Swarm integrated mode [443], it is possible to turn on IPSEC encryption of application messages per overlay network. In DC/OS [442] permissive or security mode, encryption of application messages is also enabled/enforced but only for user services that offer a TLS certificate of their own. Finally, the Weave NET plugin of Kubernetes also supports encryption of application messages.

*Restricting external access to service ports*. As stated in Section 4.3 on container networking – sub-aspect External access to services – containers of which the services are exposed via a service port can be accessed from outside the cluster if there exists a cluster node with a load balancer which has an IP address that is routable from outside the cluster. However, a security risk thus ensues that any container with a service port is susceptible to outside malicious attacks, especially if the load balancer is deployed distributed on every nodes. In order to manage this security risk, one needs a way to segregate the nodes of a cluster into those attached to a private network only and those attached to a private and public network.

With this end in view, Docker Swarm integrated mode [444] allows starting master and worker nodes with a specific IP address or network interface to which it should listen so that service ports on that node are only accessible from the network to which the IP address or network interface is attached.

Kubernetes v1.10+ [445] added a similar feature but also allows to specify a range of IP addresses instead of a single IP address when starting a master or worker node.

Finally, DC/OS [446] distinguishes directly between private and public node types. Public nodes support inbound connections from outside the cluster and are thus primarily meant for externally facing load balancers like marathon-lb or edge-lb. Private nodes cannot be directly accessed from outside the cluster.

### 4.7.2 Unique features

Kubernetes offers the following unique feature for the sub-aspect "user identity and access management"**:**

- Auditing [447] provides a security-relevant chronological set of records documenting the sequence of activities that have affected the overall state of the Kubernetes cluster. These activities can be performed by individual application managers, the cluster administrators or Kubernetes-specific software components.

Docker Swarm integrated mode, Kubernetes and DC/OS offer the following unique features for the sub-aspect "cluster network security":

Docker Swarm integrated mode:

- Encryption of Swarm manager logs [448] is automatically performed to protect data from attackers.

Kubernetes:

- Kubelet authentication and authorization [449] allows to protect the kubelet's HTTP endpoint on every worker node.
- Network policies [450] are specifications of how groups of pods are allowed to communicate with each other and other network endpoints.

### 4.8 *Securing containers*

This aspect covers features that an application manager must understand in order to manage sensitive-information, manage passwords for getting access to private Docker repositories, and limiting the attack interface of containers by limiting the access of containers towards the underlying kernel.

### 4.8.1 Common features

**Protection of sensitive data and proprietary software.** Docker Swarm integrated mode [451], Kubernetes [452], Mesos [453], Marathon [454] and the Enterprise distribution of DC/OS [455] offer concepts for *storing sensitive information such as private keys in Secret API objects* which encompass one or more encrypted data fields.

*Pulling images from a private Docker registry.* Docker Swarm, Kubernetes, Mesos and Marathon offer support for automated docker login to a private Docker registry using a Docker username and password. The Docker daemon requires that a user must log in using the docker login command, which asks for a username and password of the DockerHub. A successful login creates or updates then a config.json file that holds an authorization token.

Docker Swarm [302] allows to set the --with-registry-auth option of the docker service create command in order to pass the stored authorization token when creating a service.

Kubernetes [456] allows generating a secret which automatically includes the authorization token that results from the docker login command. A Pod's configuration file must then include this secret for pulling images from a private registry.

Similarly, Marathon [457] requires to store the config.json file as a secret that must then also be included in a Pod's configuration file.

Finally, Mesos' DockerContainerizer runtime [458] supports passing the Docker config.json file as a flag to the Mesos agent such that the authorization token is passed by default for all container images . This Mesos feature can thus be used in any container orchestration framework such as Aurora.

**Improved security isolation**: One of the weaknesses of containers is that they have a broader security attack surface than virtual machines: containers run on the same host operating system and thereby enlarge the attack surface in comparison to virtual machines that run on a more compact hypervisor. For this reason, all major cloud providers continue to use a virtual machine as a key abstraction for representing a node in order to protect their assets.

Therefore, besides the basic isolation mechanisms at the level of linux containers, i.e. cgroups and namespaces [129], container orchestration frameworks additionally leverage existing security modules in the Linux kernel in order to configure on a per container basis what a container is allowed to do in terms of linux system calls, file permissions, etc.

These modules include SELinux, AppArmor, seccomp and Linux capabilities. SELinux [460] and AppArmor [461] are security modules that can limit by means of access control policies what a process can do. Seccomp-bpf [462] allows filtering of framework calls using a configurable policy implemented using Berkeley Packet Filter [463] rules. Linux capabilities [464] allows to give a user-level process specific root-level permissions. As such a process can be granted access to what it needs without running the process as root.

All container orchestration frameworks have just began to integrate with these different security modules. Note that the following security features are supported by Docker-engine and therefore also for Docker Swarm stand-alone. However, these security features are not yet available for Services in Docker Swarm integrated mode.

*Setting Linux capabilities per container* is supported by Docker Swarm stand-alone [465], Kubernetes [466] and Mesos .

*Setting SELinux labels per container* is supported by the Fedora Atomic distributions of Docker-engine [467] and by Kubernetes [468].

*Setting custom AppArmor profiles per container* is supported by Docker Swarm stand-alone [469] and is a beta feature of Kubernetes [470].

*Setting custom seccomp profiles per container* is supported by Docker Swarm stand-alone [471] and there is also alpha support for seccomp profiles in Kubernetes [472].

*Higher-level aggregate objects for storing multiple security profiles.* Kubernetes offers a g*eneric aggregate object* in the Kubernetes API, named SecurityContext [473], which manages per container and per Pod which Linux capabilities, custom profiles, SELinux labels and other privileges must be applied. Docker [474] has launched a design proposal and work-in-progress library for supporting a similar generic object, called entitlements [475]. Such entitlements are actually envisioned as higher-level abstractions for encompassing security profiles of Services in Docker Swarm integrated mode as well as Pods in Kubernetes. Moreover, a possible future direction entails that image publishers can already store a preconfigured entitlement with the container image and sign it as part of a trusted bundle.

**Table 9**. Commonly supported features for the "securing containers" aspect.

Cell legend:
- future: The feature is not yet part of the open-source distribution of the CO framework. It has however been planned according to the documentation, or there is a separate incubation project. The URL to relevant roadmap documentation is included.
- externalComponent: Support for the feature is not included in the open-source distribution of the CO framework, but the feature is supported by a third party component or platform. The name of the URL refers to the name of the component. The URL to the corresponding documentation is included.

| Securing containers sub-aspects | Features | Swarm stand-alone | Swarm integrated | Kubernetes | Mesos | Mesos + Aurora | Mesos + Marathon | DC/OS |
|---|---|---|---|---|---|---|---|---|
| | | Sa | Si | Ku | Me | Au | Ma | Dc |
| Protection of sensitive data and software | *Storage of sensitive-data as secrets* | | ✓ | ✓ | ✓ | | ✓ | Extnd |
| | *Pull image from a private Docker registry* | | ✓ | ✓ | ✓ | | ✓ | Extnd |
| Improved security isolation | *Setting Linux capabilities per container* | ✓ | future | ✓ | ✓ | | | future |
| | *Setting SELinux labels per container* | Red Hat | future | ✓ | | | | |
| | *Setting AppArmor profiles per container* | ✓ | future | ✓ | | | | |
| | *Setting seccomp profiles per container* | ✓ | future | ✓ | | | | |
| | *Higher-level aggregate objects* | future | future | ✓ | | | | |

### 4.8.2 Unique features

Kubernetes offers the following unique features for the sub-aspect "improved security isolation":

Kubernetes:

- A run-time verification of SecurityContexts by means of a PodSecurityPolicy [476] API object. Kubernetes validates that the SecurityContext of each container is set with the appropriate profiles, capabilities and privileges. If the validation check fails, the container is not permitted to start. A PodSecurityPolicy also sets default profiles, capabilities and privileges for containers without an explicit SecurityContext object. PodSecurityPolicy objects also control a broader set of restrictions [476] for containers including the range or process ids under which a container must run, and the allowed type of volumes. PodSecurityPolicy objects are specified and managed per namespace. It can be controlled by means of role-based authorization [477] which users are allowed to perform what actions on a PodSecurityPolicy object for each namespace.
- Support for configuring the Linux sysctl interface [478]. Linux sysctl interface allows an administrator to modify kernel parameters at runtime. Sysctl parameters are either considered safe or unsafe by Kubernetes. Kubernetes enables all safe sysctl parameters by default. All unsafe sysctl parameters must be enabled manually per node. A pod that is set with an unsafe sysctl feature must be scheduled using the taints and toleration feature (see Section 4.6).

### 4.9 *Application and cluster management*

This aspect covers features of CO framework that a cluster administrator or application manager must understand in order to manage various non-functional requirements of respectively the cluster or the containerized applications. These management services rely on the Identity and Access Management functionality (see Section 4.7) in order to support customized instances of their functionality to cluster administrators and application managers.

### 4.9.1 Common features

**Creation, management and inspection of cluster and applications.** To support user-friendly usage of the Master API, *a Command Line Interface (CLI)* with a well-defined command structure is provided in all CO frameworks.

*Web UI.* Docker [479], Kubernetes [480] and DC/OS [481] offer beside their HTTP-based Master API and Command-Line Interface (CLI) also a graphical user interface for inspecting and managing the state of all objects that can be managed via the Master API, e.g. nodes, services, containers, replication levels of containers, volumes, user groups, multi-tenant authorization controls etc. Erroneous states such as unhealthy containers or failed nodes can also be inspected. Docker's dashboard also offers a tab for managing the deployed networks. Marathon [482] also offers a dashboard, which is still supported in the DC/OS distribution but no longer developed in favor of DC/OS's dashboard. DC/OS can be configured to control user access to the different tabs of its dashboard [483] and to Marathon's dashboard [484]. Finally, Aurora [485] also offers a Web UI, but this only supports limited functionality of the scheduler API, such as finding running jobs[6].

*Labels for organizing and selecting subsets of API objects.* Finally, the CLI and/or dashboard of Docker, Kubernetes, Mesos and DC/OS also use labels for organizing and selecting subsets of containers, services and nodes according to multiple dimensions (e.g. development vs production environments, front-end vs database tiers). As already stated in Section 4.6, labels are <key, value> pairs where the key represents a dimension and the value a particular subset of objects in that dimension. Objects can of course belong to multiple dimensions and therefore be associated with multiple labels. Subsets of particular objects can be selected by means of the set-based inclusion operator over their (key, value) pairs as well as their key. Docker Swarm supports service labels [486] and node labels [487]. In Kubernetes, labels can be attached to various API objects [248] and the Kubernetes CLI and dashboard allows to select objects by their labels. Mesos supports task labels [488], while the DC/OS distribution of Marathon allows to attach labels to Marathon applications and Mesos tasks [489].

*Inspection of cluster-wide resource usage.* These GUIs and associated CLIs also support inspection of (aggregate statistics of) cluster-wide resource usage in order to inspect global cluster state and health. Docker EE's Universal Control Plane [479] shows CPU and memory resource consumption of nodes. Kubernetes' dashboard [480] shows CPU and memory usage at different levels: cluster-wide, per node, and for each separate pod. Aurora's Observer component [490] enables browser-based access of disk usage metrics per task. DC/OS [481]'s dashboard shows CPU, memory and disk usage at similar levels: cluster-wide, per node, per service.

**Monitoring resource usage and health.** Kubernetes, Mesos and the DC/OS distribution of Mesos and Marathon offer *central monitoring of resource usage of services and containers*. Kubernetes [491] provided in the past the Heapster add-on service that allows monitoring of CPU, mem, storage and network resources [492] at different levels: Pods, Nodes, etc. As Heapster stores the collected metrics in a time-series database, it is possible to consult the metrics of the past.

However Heapster has been deprecated since Kubernetes v.11. [493] Heapster has been replaced by two resource metrics pipelines: (i) a core Metrics API [494] that is used to support monitoring of Pods and auto-scaling of Pods and only stores metrics for the short term and (ii) several independent full metrics pipelines [495] of which the most prominent are the Prometheus open-source project with built-in support for Kubernetes and the Google Cloud Monitoring.

---

[6] A new Web UI is created in release 0.19.0 which provides the ability to inject your own custom UI components.

Mesos [496] exposes at every agent an HTTP endpoint with aggregate resource consumption metrics for containers running under that agent. When using marathon-lb [162] for exposing the service of a container, statistics for the network interface of that container [497] can be monitored at the same HTTP endpoint. When using a CNI network [498], network statistics of a container can also be queried.

DC/OS supports a central Metrics API [499] for monitoring containers and Marathon applications. This also involves monitoring network usage per container.

*Central monitoring of resource usage by CO framework components.* Besides monitoring the resource usage of services and containers, Docker Swarm, Kubernetes, Mesos, Aurora and DC/OS also support central monitoring of (aggregated statistics) of resource usage by CO framework components.

- For Docker Swarm [500], it is possible to use Prometheus for monitoring the resource usage of multiple cluster nodes.
- Kubernetes [501] offers support for monitoring performance and health metrics of the ControllerManager component and health metrics can also be monitored for persistent volume operations in GCE, AWS, Vsphere and OpenStack.
- Mesos [502] offers two monitoring concepts: Counters keep track of discrete events and are monotonically increasing, e.g. the number of failed tasks cluster-wide. Gauges represent a sample from a continuously monitored metric such as the uptime of a master and whether the master is elected.
- Aurora [503] does not only support these two Mesos concepts, but also offers threshold-based alerts.
- Marathon [504] offers gauges, timers and meters. DC/OS' metrics endpoint combines Mesos and Marathon monitoring data [505].
- Mesos [506]-based frameworks also support monitoring GPU usage. Kubernetes v1.9 [507] has also just released support for GPU monitoring.

*Reusable and configurable framework for checking the health of containers*. All CO frameworks also offer a framework for developing custom health checks per container. Different health check methods are possible including HTTP check and check via a shell command. Relevant configuration parameters include the timeout period, the interval between two checks and the minimum number of consecutive failed checks for the health check to be considered failed. Docker Swarm uses the HEALTHCHECK [508] instruction of a container image or a customized health check as part of a ComposeV2/V3 file [509]. Kubernetes [312] additionally lets the kubelet agent on every node restart containers that have failed the health check. Mesos v1.2.0+ [510] introduces a general framework for task health checking, whereas in Mesos v1.4.0+ [511] the interpretation of the health check can also be delegated to the framework. Marathon v1.4.0+ [512] has deprecated its original health checking mechanism and adopted Mesos' health checking framework instead. Finally, Aurora [513] still uses its own health checking mechanism.

*Central monitoring of distributed events.* Docker Swarm integrated mode, Kubernetes, Mesos, Aurora and Marathon also support an API for monitoring of events about new requests for creating services, container, container state changes and errors. In Docker Swarm integrated mode [514], events can be monitored using the docker events command line interface. In Kubernetes [515], events are Kubernetes objects that are accessible via the Master API. To avoid filling up master's disk, a retention policy is enforced: events are removed one hour after the last occurrence. To provide longer history and event aggregation a third party solution such as Stackdriver [516] must be used. In Mesos [517], the Operator API supports subscribing to an event stream. In Aurora [518], it is possible to configure a simple HTTP webhook to receive task state change events. In Marathon [519], all API requests and scaling events are captured by an event bus to which can be subscribed via the REST API of Marathon.

**Logging and debugging of CO framework and containers.**

*Logging of containers* is supported via the CLI and/or dashboard of Docker Swarm [520], Kubernetes [521], Mesos [522] and DC/OS' Marathon [523].

*Internal logging of CO framework components* is supported by all CO frameworks. Which specific logging tool is used, depends on the used deployment method: when the CO framework is deployed as a set of containers, container logging can be used (see Logging and debugging in Section 4.4); when the CO framework is installed as Linux package, the journald [524] service is used; Mesos uses the glog logging library [525]. Marathon exposes framework logs as part of its REST API [526].

*Integration with external log aggregation frameworks* is documented in Docker Swarm [527], Kubernetes [528] and DC/OS [529].

**Cluster maintenance.**

*Cluster state backup and recovery* is a built-in feature of Docker Swarm integrated mode [530], Mesos [531], Aurora [532] and Marathon [533]. For Kubernetes an external project for cluster state management operations [534] such as backup and restore exists. Note that Mesos uses state machine replication (SMR) for storing the state of the entire cluster, including the state of the running frameworks. Aurora uses Mesos' SMR while Marathon does not.

*Documentation about how to upgrade a running cluster to a next release* is provided by Kubernetes [535], DC/OS distribution of Mesos [536], Aurora [537] and Marathon [538].

*The effect of an upgrade on running containers.* Docker [539] enables that when the Docker daemon crashes or is shut down for upgrade on a node, the containers on that node can continue running. In Kubernetes [535], the effect of an upgrade on running Pods depends on the used deployment tool or cloud platform. For example the kubeadm [540]deployment tool supports upgrades without affecting running Pods, while in Google Compute Engine [541], Instance Groups [542] are used to sequentially destroy and recreate each node with new software; and any pods that are running on a destroyed node are either automatically recreated when associated with a ReplicaSet object, or must be manually re-created after the upgrade is finished. In Mesos [400], a framework's running tasks can reconnect to the new Mesos agent after an upgrade if the framework checkpointing flag is turned on for the framework. Aurora [537], Marathon [538] and DC/OS [536] have this flag turned on by default.

*A CLI command for draining all containers from a node for maintenance* is supported by Docker Swarm integrated mode [543], Kubernetes [544], Mesos [545], Marathon [546] and DC/OS [547]. Note that Mesos offers higher-level support to cluster administrator for announcing maintenance time windows to frameworks before the actual draining of nodes.

*Garbage collection of containers and/or images* is differently supported by different CO frameworks. Docker [548] supports manual garbage collection of images at the level of the local registry; Kubernetes' kubelet agent [549] supports automated garbage collection of container images as well as containers. Mesos v1.5 [550] supports automated garbage collection of Docker images for the Unified Container Runtime, but not containers. Finally, DC/OS extends Mesos with support for garbage collection of container images for both the Unified Container Runtime as well as the Docker containerizer. Moreover, the architecture of DC/OS [551] also includes support for garbage collection of Docker containers.

**Multi-cloud support.**

*One cluster across multiple availability zones or regions.* The design of Mesos [552]-based frameworks, in particular DC/OS allows that *one cluster can be more easily deployed across multiple availability zones* or regions because these CO frameworks have generic and automated support for setting up replicated masters (see Highly-Available Master/Manager architecture in Section 4.1). Docker Swarm stand-alone [553] as well as integrated mode [554] also allow deploying multiple master nodes across multiple availability zones. Kubernetes [555] provides limited support for multi-zone deployments as generic support for automated HA master setups is not provided. However, Kubernetes-as-a-Service platforms such as Google Kubernetes Engine (GKE) and Amazon Elastic Container Service for Kubernetes (Amazon EKS) [42] offer scalable and highly-available Kubernetes clusters where multiple masters can be deployed across different availability zones.

*Recovering from network partitions.* Mesos [399] has good support for dealing and recovering from network partitions. Aurora v0.20.0 [130] has added an optional and experimental feature for using the Mesos partition-aware APIs in order to customize the job or service recovery strategy. Users of Aurora can set partition policies [556] per job of whether or not to reschedule and how long to wait for the partition to heal.

*Management of multiple clusters across multiple clouds.* Docker's Docker Cloud [557], Kubernetes' kubefed [558], and DC/OS' multi-cluster CLI [559] also offer CLI commands for managing multiple clusters across one or more cloud providers.

*Federated authentication:* Kubernetes's federated API [560] and DC/OS' single-sign-on across clusters [561] capability support federated authentication of users.

*Multi-zone/multi-region workloads*: All CO frameworks, except Docker Swarm stand-alone, allow to control the availability of a service by spreading its containers across multiple fault domains (i.e., availability zones, regions or datacenters). Docker Swarm integrated mode [389], Mesos [562], Aurora [563], Marathon and DC/OS [564] require that nodes are in advance labeled with their zone, region or datacenter and offer a placement preference operator that ensures that containers of a service are spread across these different fault domains. DC/OS [565] also offers fault domain detect scripts for AWS EC2 and Azure nodes that automatically start Mesos agents with the detected zones and regions. Kubernetes [566] uses another approach: It uses its extensive support for federating multiple container clusters across different fault domains. Kubernetes' kubefed command line interface can then be used to deploy federated instances of all Kubernetes API objects such as Deployments, ReplicaSets, StatefulSets, Jobs, Services, Secrets, ConfigMaps, etc. There is also alpha support for federated autoscalers [567]. Moreover, cross-cluster service discovery is supported as well.

**Table 10**. Commonly supported features for the "application and cluster management" aspect.

Cell legend:
- **future**: The feature is not yet part of the open-source distribution of the CO framework. It has however been planned according to the documentation, or there is a separate incubation project. The URL to relevant roadmap documentation is included.
- **externalComponent**: Support for the feature is not included in the open-source distribution of the CO framework, but the feature is supported by a third party component or platform. The URL to the corresponding documentation is included. The name of the URL refers to the name of the component.
- **$..$**: Support for the feature is not included in the open-source distribution of the CO framework, but is included in a commercial product or cloud service. The URL to the corresponding documentation is included. The name of the URL refers to the name of the product or service.
- *partial support*: the CO framework offers partial support for the feature. The URL to a relevant documentation page is included. The name of the URL refers to the essence of what is being supported.

| Application and cluster management sub-aspects | Features | Swarm stand-alone | Swarm integrated | Kubernetes | Mesos | Mesos + Aurora | Mesos + Marathon | DC/OS |
|---|---|---|---|---|---|---|---|---|
| | | Sa | Si | Ku | Me | Au | Ma | Dc |
| Creation, management and inspection of cluster and applications | *Command-line interface (CLI)* | ✓ | ✓ | ✓ | ✓ | ✓ | ✓ | Sprsd |
| | *Web UI* | | $Docker EE$ | ✓ | ✓ | ✓ | ✓ | Sprsd |
| | *Labels for organizing API objects* | ✓ | ✓ | ✓ | ✓ | | ✓ | Dlgt |
| | *Inspection of resource usage graphs* | | $Docker EE$ | ✓ | | *disk usage* | | Add |
| Monitoring resource usage and health | *Monitoring container resource usage* | | | ✓ | ✓ | | | Extnd |
| | *Monitoring CO framework resource usage* | | Prometheus | ✓ | ✓ | ✓ | ✓ | Dlgt |
| | *Framework for container health checks* | ✓ | ✓ | ✓ | ✓ | ✓ | ✓ | Extnd |
| | *Distributed events monitoring* | | ✓ | ✓ | ✓ | ✓ | ✓ | Dlgt |
| Logging and debugging of CO framework and containers | *Logging of containers* | ✓ | ✓ | ✓ | ✓ | | | Extnd |
| | *Logging of CO framework components* | ✓ | ✓ | ✓ | ✓ | | ✓ | Extnd |
| | *Integration with log aggregator systems* | | $Docker EE$ | ✓ | | | | Add |
| Cluster maintenance | *Cluster state backup and recovery* | | ✓ | *future* | ✓ | ✓ | ✓ | Dlgt |
| | *Official cluster upgrade documentation* | | | ✓ | ✓ | ✓ | ✓ | Extnd |
| | *Upgrade does not affect active containers* | ✓ | ✓ | *kube adm* | ✓ | ✓ | | Dlgt |
| | *Draining a node for maintenance* | | ✓ | ✓ | ✓ | | ✓ | Dlgt |
| | *Garbage collection of containers/images* | *images* | *images* | ✓ | *images* | | | Extnd |
| Multi-cloud support | *A cluster across availability zones/regions* | ✓ | ✓ | *$GKE$ $AWS$* | ✓ | ✓ | ✓ | Extnd |
| | *Recovering from network partitions* | | | | ✓ | ✓ | | |
| | *Management of multiple clusters* | | ✓ | ✓ | | | | Add |
| | *Federated authentication across clusters* | | | ✓ | | | | Add |
| | *Multi-zone/multi-region workloads* | | ✓ | ✓ | ✓ | ✓ | ✓ | Extnd |

4.9.2 Unique features

**Creation, management and inspection of cluster and applications.** Docker offers the following unique feature for the sub-aspect "creation, management and inspection of cluster and applications":

- Docker's CLI [568] comes with command-line completion for Docker Swarm integrated mode.

**Monitoring resource usage and health.** Kubernetes, Aurora and DC/OS offer the following unique features for the sub-aspect "monitoring resource usage and health":

Kubernetes:

- The Cluster autoscaler [569] is a tool that automatically adjusts the size of the Kubernetes cluster by adding or removing nodes, e.g. when all nodes are running out of resources or nodes are idle.

Aurora:

- SLA metrics [570] of running and recently completed jobs (e.g. Median Time to Start) are reported in different scopes: per cluster, per job, or per node size (in terms of CPU, memory and disk resource sizes).

DC/OS:

- Custom node and cluster health checks [571] can be configured during installation.

**Logging and debugging of CO framework and containers.** Kubernetes offers the following unique feature for the sub-aspect "logging and debugging of CO framework and containers":

- Port forwarding [572] allows a developer to connect his local workstation to a running Pod for debugging

**Cluster maintenance.** Kubernetes offers the following unique features for the sub-aspect "cluster maintenance":

- A disruption budget [573] enables an application manager to limit the number of concurrent voluntary disruptions that his application experiences due to cluster maintenance operations. A request to drain a node for maintenance will be denied if that request would violate the disruption budget of any Pod on that node.
- The hosted Kubernetes Engine [574] provides automated support for upgrades of Kubernetes. Upgrading the etcd key-value store [575] of the master is always a manual operation however.

**Multi-cloud support.** Kubernetes offers the following unique features for the sub-aspect "multi-cloud support":

- Support for the Open Service Broker API [576] in order for containers to use services that are offered by other cloud providers.
- Kubernetes offers a separate federated API with federated instantiations of several single-cluster API objects [577] such as deployments, daemon sets, ingress, etc.
- Multi-cluster service discovery and management [578] : a federated service consists of different service shards that are deployed across different Kubernetes clusters in different cloud availability zones. Service discovery using the federated DNS name of the service will return the service shard that is closed and still healthy.

**5. Quantitative analysis with respect to genericity**

This section presents the results of quantitative analysis of the collected data in Section 4 to determine evidence of significant differences in genericity between aspects and CO frameworks. We structure the presentation of these results in accordance with the research questions RQ4-RQ6 (see Section 1.2).

A CO framework is more generic than another CO framework when it supports a higher number of common features. After all, the more features are supported, the broader the set of application and cluster configurations that can be supported and managed by a CO framework. The same measure can also be used to quantify differences in genericity between (sub)-aspects.

We also take into account the number of unique features for quantifying the differences in genericity because Kubernetes has a relatively large number of unique features. Since Kubernetes is already supported by many public cloud providers and Docker EE and DC/OS also offer support for Kubernetes as an alternative orchestrator, these unique features are widely available at a large set of private and public cloud platforms.

**RQ4: How are functional (sub)-aspects ranked in terms of number of common and unique features?** Table 11 presents an overview of the number of common and unique features found for the 9 aspects of container orchestrations. The table ranks the aspects according to the number of common feature implementation strategies by CO frameworks. We see that the functional aspects of "application configuration and deployment", "application and cluster management", "container networking" and "container QoS management" count the most common feature implementation strategies. On the other hand, the aspects of "securing containers", and "resource quota management" counts the lowest number of common feature implementation strategies.

**Table 11.** Functional aspects ranked according to the number of common feature implementation strategies by CO frameworks. If a common feature is partially supported by or only supported in the commercial version of a CO framework, the implementation strategy is counted as ½. Finally, the number of common and unique features of each functional aspect are also presented.

| Aspects | #common features | #implementation strategies | #unique features |
|---|---|---|---|
| Application configuration and deployment | 29 | 130.5 | 10 |
| App and cluster management | 21 | 104 | 10 |
| Container networking | 20 | 82 | 8 |
| Container QoS Management | 15 | 69 | 6 |
| Cluster architecture and setup | 13 | 63 | 2 |
| Securing clusters | 9 | 36 | 4 |
| CO framework customization | 6 | 32 | 9 |
| Securing containers | 7 | 19.5 | 3 |
| Resource quota management | 4 | 12.5 | 1 |
| Total | 124 | 548.5 | 53 |

These numbers should only be used as a measure for ranking sub-aspects in terms of genericity: the more common features are identified in a specific aspect, the larger the set of concerns that are covered by this aspect. Of course, real genericity entails to the actual number of common feature implementation strategies by the different CO frameworks. This number of common feature implementation strategies is, in turn, a metric for the size of the set of all possible application and cluster configurations that can be managed by a particular CO framework.

Table 12 ranks the functional sub-aspects according to the number of common feature implementation strategies. Again, this metric is a measure for ranking sub-aspects in terms of genericity. For example, we see that the sub-aspect "persistent volumes" counts the most common features and the most common feature implementation strategies. This is because of two reasons:

- Besides the main functional requirement of persistent storage, various orthogonal orchestration features for management of persistent volumes can be distinguished. Moreover most of these features are supported by almost all CO frameworks.
- The adoption of the Docker volume plugin architecture by Mesos-based systems as well as the CSI specification by Kubernetes and Mesos has also been recorded as an additional feature.

Secondly the sub-aspect "services networking" counts also a high number of common features because of again two similar reasons:

- No less than 3 alternative approaches to services networking can be distinguished that are all supported by multiple CO frameworks and within each alternative approach one can distinguish at a lower nested level between different alternative load balancing strategies.

- There are again two standardization initiatives related to this sub-aspect: Docker's libnetwork architecture and the CNI specification.

**Table 12.** Functional sub-aspects ranked according to the number of common feature implementation strategies by CO frameworks.

| Sub-aspects | #common features | #implementation strategies | #unique features |
|---|---|---|---|
| Persistent volumes | 9 | 47 | 6 |
| Services networking | 8 | 35 | 2 |
| Service upgrades | 8 | 32 | 1 |
| Architectural patterns | 5 | 31.5 | 0 |
| Reusable container configuration | 5 | 26 | 2 |
| Installation methods and deployment tools | 7 | 25.5 | 2 |
| Supported workload types | 7 | 25.5 | 1 |
| Cluster maintenance | 5 | 25 | 2 |
| Container CPU and mem allocation with support for over-subscription | 5 | 23 | 1 |
| Creation, management and inspection of cluster and applications | 4 | 22.5 | 1 |
| Service discovery and external access | 6 | 22 | 6 |
| Monitoring resource usage and health | 4 | 22 | 3 |
| Multi-cloud deployments | 5 | 19.5 | 3 |
| Controlling scheduling behavior by means of placement constraints | 3 | 19 | 0 |
| Cluster network security | 6 | 18.5 | 3 |
| Controlling preemptive scheduling and re-scheduling behavior | 5 | 18 | 1 |
| Plugin architecture for network services | 4 | 17.5 | 0 |
| User identity and access management | 3 | 17.5 | 1 |
| Unified container runtime architecture | 3 | 17 | 0 |
| Framework design of orchestration engine | 3 | 15 | 9 |
| Logging and debugging of CO framework and containers | 3 | 15 | 1 |
| Resource quota management | 4 | 12.5 | 1 |
| Protection of sensitive data and proprietary software | 2 | 10 | 0 |
| Improved security isolation | 5 | 9.5 | 3 |
| Allocation of other resources | 2 | 9 | 4 |
| Host ports conflict management | 2 | 7.5 | 0 |
| Configuration management approach | 1 | 6 | 0 |
| Total | 124 | 548.5 | 53 |

An interesting question is whether there is a linear association between the number of common feature implementation strategies and the number of unique features across sub-aspects. We ranked these two vectors using R's rank function with the parameter ties.method set to "average", i.e. when two sub-aspects have the same number of strategies or unique features, their absolute ranks are replaced by the mean of these absolute ranks.

According to several existing linear association measures (see Table 13) there is a very weak association. As such, there is no relation between the number of common feature implementation strategies and the number of unique features. The independence between these variables is confirmed by the chi-square test (p = 0.3264) and the linear-by-linear association test (p-value = 0.4199) using the coin R package.

The weak linear association implies that when determining the overall risk of feature lock-in for a specific sub-aspect, one should study unique features and the number of common feature implementation strategies for a specific sub-aspect independently in order to come to an accurate estimation.

**Table 13**. Application of existing association measures for ordinal data using the DescTools R package.

| Statistic | Value | 95% confidence interval |
|---|---|---|
| Kendall's Tau-b | 0.1730347 | (-0.1424626; 0.4885320) |
| Stuart's Tau-c | 0.1760402 | (-0.1430893; 0.4951698) |
| Somers' D C⏐R | 0.1589595 | (-0.1288827; 0.4468018) |
| Goodman Kruskal's Gamma | 0.1903114 | (-0.1601409; 0.5407637) |

**RQ5: How are CO frameworks ranked in terms of number of supported common features?** As shown in Figure 4, Kubernetes implements the highest number of common features (but also supports the highest number of unique features).

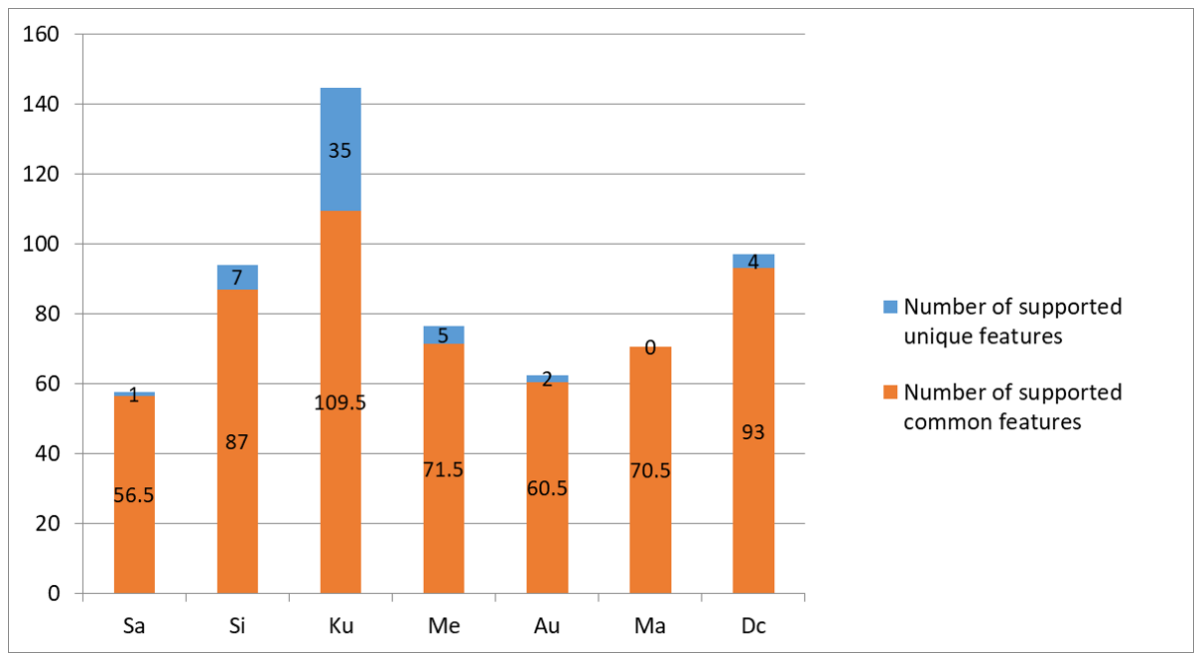

**Figure 4**. Comparison of CO frameworks according to the total number of supported features.

**RQ6a. Which functional (sub)-aspects are best supported by a CO framework in terms of common features?**

As shown in Figure 5, Kubernetes implements the highest number of common features for 6 aspects. Docker Swarm integrated mode supports the most common features for the aspects "container networking" and "securing clusters". Finally, DC/OS supports the most common features for the aspect "application and cluster management".

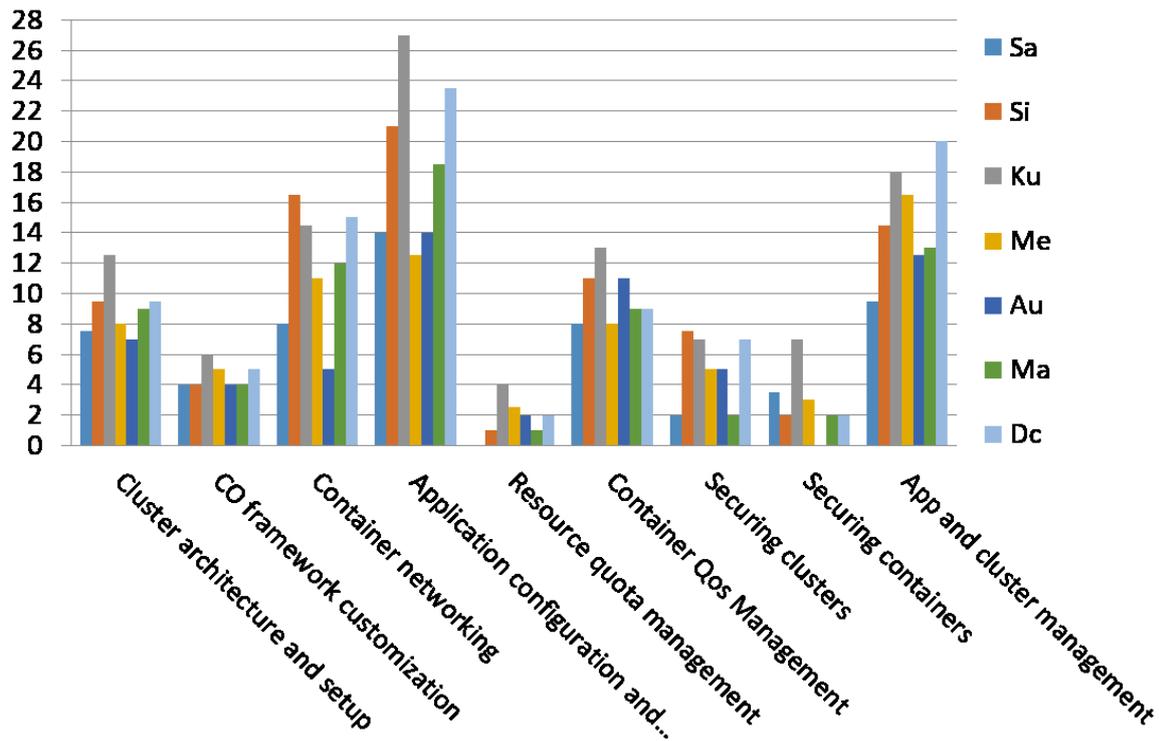

**Figure 5.** The number of common feature implementation strategies supported by each CO framework is shown for each of the 7 CO frameworks.

Table 27 presents an overview of the number of common feature implementation strategies per CO framework and per (sub)-aspect. We find significant differences in ranking between the frameworks when applying the Friedman test for unreplicated designs [135] (p-value=2.668e-08). To deal with tied observations in this test, we again compute ranks using R's rank() method where ranks for tied observations are replaced by their mean.

We also performed during post-hoc analysis a pairwise comparison between CO frameworks using the Nemenyi multiple comparison test with q approximation for unreplicated blocked data [135] (see Figure 6).

```
      Sa        Si        Ku        Me        Au        Ma
Si  0.06877    –         –         –         –         –
Ku  3.5e-05   0.44516    –         –         –         –
Me  0.63667   0.90711   0.03010    –         –         –
Au  0.99977   0.17533   0.00021   0.85620    –         –
Ma  0.89562   0.65762   0.00609   0.99918   0.98311    –
Dc  0.00182   0.93666   0.97533   0.27595   0.00765   0.09599
```

Figure 6. Resulting p-values of the Nemenyi multiple comparison test. For p-values <= 0.05, we can reject the null hypothesis, i.e. there is no significant difference in overall ranking between a pair of CO frameworks).

Based on the p-values of the Nemenyi test, we find that Docker Swarm stand-alone and Aurora differ significantly from both Kubernetes and DC/OS. Moreover there is a significant difference between Kubernetes on the one hand and Mesos and Marathon on the other hand:

- Docker Swarm stand-alone and Aurora are indeed clearly less generic in terms of offered features than the other CO frameworks. After all, Aurora is specifically designed for running long-running jobs and cron jobs, while Docker Swarm stand-alone is also a more simplified framework with substantial less automated management.

**Table 14**. For each functional (sub)-aspect, the number of common feature implementation strategies by each CO framework are shown and the framework(s) with the highest number is/are also shown.

| Aspects | Sub-aspects | CO frameworks | | | | | | | FW(s) with most common features |
|---|---|---|---|---|---|---|---|---|---|
| | | Sa | Si | Ku | Me | Au | Ma | Dc | |
| **cluster architecture and setup** | | **7.5** | **9.5** | **12.5** | **8** | **7** | **9** | **9.5** | **Ku** |
| | Configuration management approach | 1 | 1 | 1 | 0 | 1 | 1 | 1 | All but Me |
| | Architectural patterns | 5 | 5 | 4.5 | 3 | 4 | 5 | 5 | Sa/Si/Ma/Dc |
| | Installation methods and deployment tools | 1.5 | 3.5 | 7 | 5 | 2 | 3 | 3.5 | Ku |
| **CO system customization** | | **4** | **4** | **6** | **5** | **4** | **4** | **5** | **Ku** |
| | Unified container runtime architecture | 3 | 3 | 3 | 2 | 2 | 2 | 2 | Sa/Si/Ku |
| | Framework design of orchestration engine | 1 | 1 | 3 | 3 | 2 | 2 | 3 | Ku/Me/Dc |
| **Container networks** | | **8** | **16.5** | **14.5** | **11** | **5** | **12** | **15** | **Si** |
| | Services networking | 3 | 7.5 | 6 | 4.5 | 2 | 5 | 7 | Si |
| | Host ports conflict management | 1 | 2 | 1 | 0.5 | 1 | 1 | 1 | Si |
| | Plugin architecture for network services | 3 | 3 | 2.5 | 3 | 0 | 3 | 3 | Sa/Si/Me/Ma/Dc |
| | Service discovery and external access | 1 | 4 | 5 | 3 | 2 | 3 | 4 | Ku |
| **Application configuration and deployment** | | **14** | **21** | **27** | **12.5** | **14** | **18.5** | **23.5** | **Ku** |
| | Supported workload types | 2 | 4 | 6.5 | 1 | 3 | 4 | 5 | Ku |
| | Persistent volumes | 7 | 7 | 7.5 | 8.5 | 3 | 6.5 | 7.5 | Me |
| | Reusable container configuration | 3 | 4 | 5 | 3 | 3 | 4 | 4 | Ku |
| | Service upgrades | 2 | 6 | 8 | 0 | 5 | 4 | 7 | Ku |
| **Resource quota management** | | **0** | **1** | **4** | **2.5** | **2** | **1** | **2** | **Ku** |
| **Container QoS Management** | | **8** | **11** | **13** | **8** | **11** | **9** | **9** | **Ku** |
| | Container CPU and mem allocation with support for over-subscription | 4 | 5 | 5 | 3 | 2 | 2 | 2 | Si/Ku |
| | Allocation of other resources | 0 | 0 | 1 | 2 | 2 | 2 | 2 | Me/Au/Ma/Dc |
| | Controlling scheduling behavior by means of placement constraints | 3 | 3 | 3 | 1 | 3 | 3 | 3 | All but Me |
| | Controlling preemptive scheduling and re-scheduling behavior | 1 | 3 | 4 | 2 | 4 | 2 | 2 | Ku/Au |
| **Securing clusters** | | **2** | **7.5** | **7.5** | **5** | **5** | **2** | **7** | **Si** |
| | User identity and access management | 1 | 2.5 | 3 | 3 | 3 | 2 | 3 | Ku/Me/Au/Dc |
| | Cluster network security | 1 | 5 | 4.5 | 2 | 2 | 0 | 4 | Si |
| **Securing containers** | | **3.5** | **2** | **7** | **3** | **0** | **2** | **2** | **Ku** |
| | Protection of sensitive data and proprietary software | 0 | 2 | 2 | 2 | 0 | 2 | 2 | All but Sa/Au |
| | Improved security isolation | 3.5 | 0 | 5 | 1 | 0 | 0 | 0 | Ku |
| **App and cluster management** | | **9.5** | **14.5** | **18** | **16.5** | **12.5** | **13** | **20** | **Dc** |
| | Creation, management and inspection of cluster and applications | 3 | 3 | 4 | 3 | 2.5 | 3 | 4 | Ku/Dc |
| | Monitoring resource usage and health | 1.5 | 2.5 | 4 | 4 | 3 | 3 | 4 | Ku/Me/Dc |
| | Logging and debugging of CO framework and containers | 2.5 | 2.5 | 3 | 2 | 1 | 1 | 3 | Ku/Dc |
| | Cluster maintenance | 1.5 | 3.5 | 3.5 | 4.5 | 3 | 4 | 5 | Dc |
| | Multi-cloud deployments | 1 | 3 | 3.5 | 3 | 3 | 2 | 4 | Dc |
| Total # common feature implementation strategies | | 56.5 | 87 | 109.5 | 71.5 | 60.5 | 70.5 | 93 | 548.6 |

We only recommend Docker Swarm stand-alone as a possible starting point for developing one's own CO framework. This is a relevant direction because 28% of surveyed users in the most recent OpenStack survey [4], responded that they have built their own CO framework instead of using existing CO frameworks (see also Figure 1). We make such recommendation because the API of Docker Swarm stand-alone is the least restrictive in terms of the range of offered options for common commands such as creating, updating and stopping a container. For example, Docker Swarm stand-alone is the only framework that allows to dynamically change resource limits without restarting containers. Such less restrictive API is a more flexible starting point for implementing a custom developed CO framework.

- The significant difference between Kubernetes and Mesos can be partially explained by the fact that Mesos by itself is not a complete CO framework as Mesos enables fine-grained sharing of resources across different CO frameworks such as Marathon, Aurora and DC/OS. It is moreover self-explaining that there are no significant differences between Mesos on the one hand and Aurora, Marathon and DC/OS on the other hand, because many feature implementation strategies of the latter three CO frameworks rely on Mesos.

The significant difference between Kubernetes and Marathon can be explained by the fact that very few new features have been added to Marathon since the start of DC/OS. After all DC/OS is the extended Mesos+Marathon distribution that has also an enterprise edition.

There are no significant differences between the other CO frameworks. However, for 13 sub-aspects, a specific CO framework distinguishes itself by offering the most common features in that sub-aspect. In particular, Kubernetes, Docker Swarm integrated mode, DC/OS and Mesos are the most distinguishing frameworks:

- Kubernetes has the absolutely most features for 7 sub-aspects:
    1. Installation methods and deployment tools
    2. Service discovery and external access
    3. Supported workloads
    4. Reusable container configuration
    5. Service upgrades
    6. Resource quota management
    7. Improved security isolation

    For all 7 sub-aspects, the open-source distribution of Kubernetes supports all common features of these sub-aspects. As such Kubernetes is very generic with respect to these sub-aspects.

- Docker Swarm integrated mode has the most features for 3 sub-aspects:
    1. Services networking
    2. Host ports conflict management
    3. Cluster network security

    For the first two sub-aspects, Docker Swarm integrated mode offers support for all common features, while for the last sub-aspect, the open-source distribution of Docker Swarm integrated mode offers support for all common features except *authorization of CO agents on worker nodes*.

- DC/OS has the most features for 2-sub-aspects:
    1. Cluster maintenance
    2. Multi-cloud deployments

    For the first sub-aspect, DC/OS offers support for all common features of this sub-aspect by building upon Mesos and Marathon and providing detailed manual instructions for upgrading DC/OS. For the second sub-aspect, DC/OS offers support for all common features except *recovery from network partitions*.

- Mesos has the most features for 1 sub-aspect:
    1. Persistent volumes

    After all, Mesos offers support for both Docker volumes as well as CSI-based volumes.

There are furthermore tied observations for 14 sub-aspects (see Table 15):

1. Configuration management approach. All CO frameworks except Mesos offer support for *declarative configuration management*.
2. Architectural patterns. The open-source distributions of Docker Swarm stand-alone, Docker Swarm integrated mode, Marathon and DC/OS all support *automated setup of highly* *available clusters*, where Kubernetes only provides support for this feature in particular commercial versions.

3. Unified container runtime architecture. Docker Swarm stand-alone, Docker Swarm integrated mode and Kubernetes support the *OCI standard*, while Mesos-based frameworks do not yet.

4. Framework design of orchestration engine. Kubernetes, Mesos and DC/OS are the only 3 frameworks that support all common features of this sub-aspect.

5. Plugin architecture for network services. Mesos and DC/OS are the most generic frameworks as they offer support for both the *CNI specification* and *Docker's libnetwork*. On the other hand, Docker Swarm stand-alone and Docker Swarm integrated mode offer support for *separation of data and control traffic*.

6. Container CPU and mem allocation with support for over-subscription. Docker Swarm integrated mode and Kubernetes are the only CO frameworks that support over-subscription of resources. Moreover for CPU, these frameworks offer higher-level abstractions that hide the complexities of using concepts of the Linux scheduler. Kubernetes also offer concepts for oversubscription of local ephemeral storage resources.

7. Allocation of other resources. All Mesos-based frameworks offer support for *disk limits* (in terms of storage size of persistent volumes) ánd *GPU limits* (in terms of milliseconds of GPU).

**Table 15**. Tied observation for 14 sub-aspects.

| Aspects | Sub-aspects | CO frameworks | | | | | | | #ties between FWs |
|---|---|---|---|---|---|---|---|---|---|
| | | Sa | Si | Ku | Me | Au | Ma | Dc | |
| **Cluster architecture and setup** | | | | | | | | | **10** |
| | Configuration management approach | ✓ | ✓ | ✓ | | ✓ | ✓ | ✓ | 6 |
| | Architectural patterns | ✓ | ✓ | | | | ✓ | ✓ | 4 |
| **CO system customization** | | | | | | | | | **6** |
| | Unified container runtime architecture | ✓ | ✓ | ✓ | | | | | 3 |
| | Framework design of orchestration engine | | | ✓ | ✓ | | | ✓ | 3 |
| **Container networking** | | | | | | | | | **4** |
| | Plugin architecture for network services | ✓ | ✓ | | ✓ | | | ✓ | 4 |
| **Container QoS Management** | | | | | | | | | **14** |
| | Container CPU and mem allocation with support for over-subscription | | ✓ | ✓ | | | | | 2 |
| | Allocation of other resources | | | | ✓ | ✓ | ✓ | ✓ | 4 |
| | Controlling scheduling behavior by means of placement constraints | ✓ | ✓ | ✓ | | ✓ | ✓ | ✓ | 6 |
| | Controlling preemptive scheduling and re-scheduling behaviour | | | ✓ | | ✓ | | | 2 |
| **Securing clusters** | | | | | | | | | **4** |
| | User identity and access management | | | ✓ | ✓ | ✓ | | ✓ | 4 |
| **Securing containers** | | | | | | | | | **5** |
| | Protection of sensitive data and proprietary software | | ✓ | ✓ | ✓ | | ✓ | ✓ | 5 |
| **Application and cluster management** | | | | | | | | | **7** |
| | Creation, management and inspection of cluster and applications | | | ✓ | | | | ✓ | 2 |
| | Monitoring resource usage and health | | | ✓ | ✓ | | | ✓ | 3 |
| | Logging and debugging of CO framework and containers | | | ✓ | | | | ✓ | 2 |
| Total tied observations per CO framework | | 5 | 7 | 11 | 6 | 5 | 5 | 11 | 50 |

8. Controlling scheduling behavior by means of placement constraints. All CO frameworks provide similar support for placement constraints although Mesos-based frameworks offer complex support for concisely expressing that no two containers of the same service are deployed on the same node.

9. Controlling preemptive scheduling and re-scheduling behaviour. Kubernetes and Aurora offer the most extensive support for different prioritization schemes in order to prevent that higher-priority containers do not get scheduled or suffer from resource contention at the node level because of lower-priority containers.

10. User identity and access management. Kubernetes, Mesos and Aurora offer the most extensive support for authentication and authorization of cluster administrators and application managers because the open-source distributions of these frameworks offer support for *tenant-aware access control lists*. The commercial versions of Docker and DC/OS also offer support for this feature, though.

11. Protection of sensitive data and proprietary software. All CO frameworks, except Docker Swarm stand-alone and Aurora, offer support for *secrets* as well as *pulling container images from a private Docker registry*. Docker Swarm stand-alone and Aurora do not offer support for any of these two features.

12. Creation, management and inspection of cluster and applications. The open-source distributions of Kubernetes and DC/OS offer the most extensive *command-line interfaces* and *web-based user interfaces* with support for common features such as *labels for organizing API objects* and *visual inspection of resource usage graphs*. The commercial version of Docker also includes a web-based UI with the same set of features, though.

13. Monitoring resource usage and health. Kubernetes, Mesos and DC/OS all offer support for *monitoring container resource usage, monitoring CO framework resource usage, a framework for container health checks* and a *distributed events monitoring system.* As main difference, Docker Swarm and Aurora lack support for monitoring container resource usage.

14. Logging and debugging of containers and CO framework. The open-source distribution of Kubernetes and DC/OS and the commercial version of Docker Swarm offer support for integrating existing log aggregation systems.

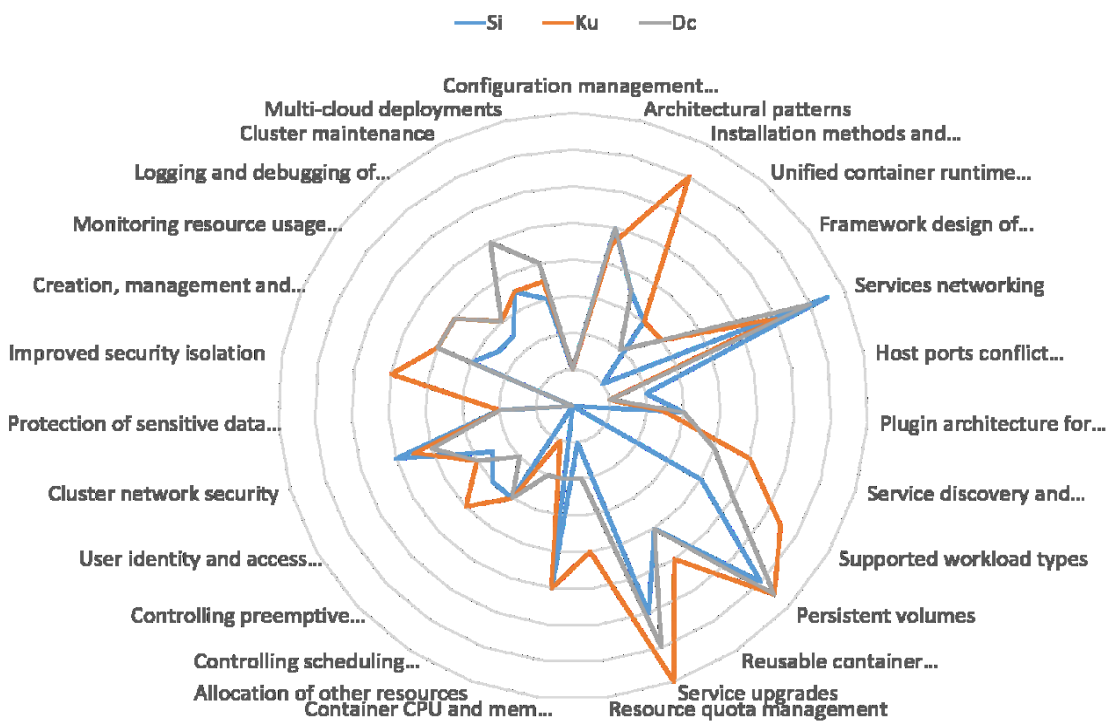

**Figure 7**. Radar chart of Docker Swarm integrated mode, Kubernetes and DC/OS to graphically present in which sub-aspects these CO frameworks support the highest number of common features.

If we would rank CO frameworks in terms of counting those sub-aspects for which they offer the most common features as well as those aspects with tied observations where they share the 1st position with other CO frameworks in terms of highest number of common features, then Kubernetes ranks highest with 18 sub-aspects, DC/OS with 13 sub-aspects, Docker Swarm integrated mode with 9 sub-aspects, Mesos with 7 sub-aspects and finally Marathon, Docker Swarm stand-alone and Aurora with 5 sub-aspects. We graphically represent the top 3 CO frameworks using a radar diagram (see Figure 7).

**RQ6. Which functional sub-aspects are best supported by a CO framework in terms of common features ánd unique features?** Kubernetes clearly offers the highest number of unique features (see Figure 4). When adding up common and unique features, Kubernetes even supports the highest number of features for all 9 aspects (see Figure 8).

We argue that it is fair to take the large number of unique features of Kubernetes into account when ranking CO frameworks with respect to genericity. After all, as already stated in Section 1.1, both Docker EE and DC/OS also offer support for Kubernetes as an alternative orchestrator. Moreover, the stability assessment of Section 7 will show that only a few unique features of Kubernetes incur a higher risk of feature deprecation.

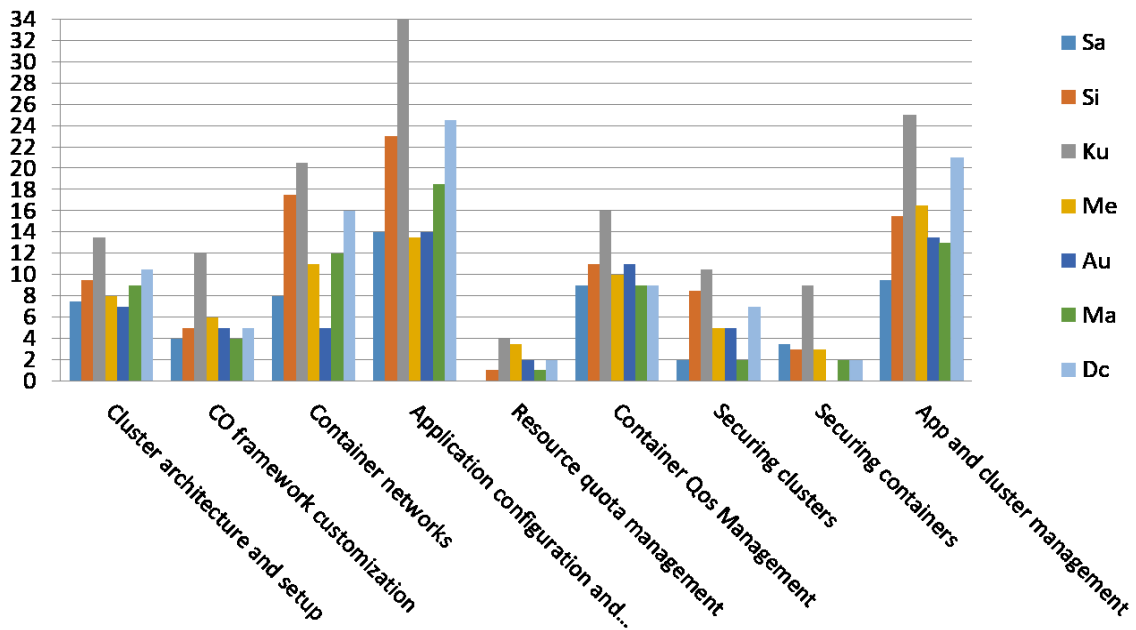

**Figure 8.** The total number of features supported by each of the CO frameworks is shown for the 9 aspects.

Table 16 provides an overview of the total number of features per (sub-)aspect and per CO framework. We find a more significant differences in ranking between the frameworks when re-applying the Friedman test for unreplicated designs (p-value=1.729e-10). We also performed a pairwise comparison between CO frameworks using the Nemenyi multiple comparison test with q approximation for unreplicated blocked data [135]. Based on the p-values of the Nemenyi test, we find that Docker Swarm stand-alone and Aurora still differ significantly from both Kubernetes and DC/OS and these differences are more replicated for Kubernetes but not for DC/OS. Similarly, the differences between Kubernetes on the one hand and Mesos and Marathon on the other hand has become also more significant. This can be explained by the fact that Kubernetes introduces the most unique features of all CO frameworks whereas DC/OS, Mesos and Marathon introduce little to no unique features.

Table 16. Overview of the number of total number of features (i.e. common + unique features) per (sub)-aspect and CO framework. The last column also shows which framework(s) support(s) the highest number of features per sub-aspect. Unique features of which the development has been explicitly announced as halted are not included.

| Aspects | Sub-aspects | CO frameworks | | | | | | | FW(s) with most features |
|---|---|---|---|---|---|---|---|---|---|
| | | Sa | Si | Ku | Me | Au | Ma | Dc | |
| **Cluster architecture and setup** | | **7.5** | **9.5** | **13.5** | **8** | **7** | **9** | **10.5** | **Ku** |
| | Configuration management approach | 1 | 1 | 1 | 0 | 1 | 1 | 1 | All but Me |
| | Architectural patterns | 5 | 5 | 4.5 | 3 | 4 | 5 | 5 | Sa/Si/Ma/Dc |
| | Installation methods and deployment tools | 1.5 | 3.5 | 8 | 5 | 2 | 3 | 4.5 | Ku |
| **CO framework customization** | | **4** | **5** | **12** | **6** | **5** | **4** | **5** | **Ku** |
| | Unified container runtime architecture | 3 | 3 | 3 | 2 | 2 | 2 | 2 | Sa/Si/Ku |
| | Framework design of orchestration engine | 1 | 2 | 9 | 4 | 3 | 2 | 3 | Ku |
| **Container networks** | | **8** | **17.5** | **20.5** | **11** | **5** | **12** | **16** | **Ku** |
| | Services networking | 3 | 8.5 | 6 | 4.5 | 2 | 5 | 8 | Si |
| | Host ports conflict management | 1 | 2 | 1 | 0.5 | 1 | 1 | 1 | Si |
| | Plugin architecture for network services | 3 | 3 | 2.5 | 3 | 0 | 3 | 3 | Sa/Si/Me/Ma/Dc |
| | Service discovery and external access | 1 | 4 | 11 | 3 | 2 | 3 | 4 | Ku |
| **Application configuration and deployment** | | **14** | **23** | **34** | **13.5** | **14** | **18.5** | **24.5** | **Ku** |
| | Supported workload types | 2 | 4 | 8.5 | 1 | 3 | 4 | 5 | Ku |
| | Persistent volumes | 7 | 7 | 11.5 | 9.5 | 3 | 6.5 | 8.5 | Ku |
| | Reusable container configuration | 3 | 5 | 6 | 3 | 3 | 4 | 4 | Ku |
| | Service upgrades | 2 | 7 | 8 | 0 | 5 | 4 | 7 | Ku |
| **Resource quota management** | | **0** | **1** | **4** | **3.5** | **2** | **1** | **2** | **Ku** |
| **Container QoS Management** | | **9** | **11** | **16** | **10** | **11** | **9** | **9** | **Ku** |
| | Container CPU and mem allocation with support for over-subscription | 5 | 5 | 5 | 3 | 2 | 2 | 2 | Sa/Si/Ku |
| | Allocation of other resources | 0 | 0 | 3 | 4 | 2 | 2 | 2 | Me |
| | Controlling scheduling behavior by means of placement constraints | 3 | 3 | 3 | 1 | 3 | 3 | 3 | All but Me |
| | Controlling preemptive scheduling and re-scheduling behavior | 1 | 3 | 5 | 2 | 4 | 2 | 2 | Ku |
| **Securing clusters** | | **2** | **8.5** | **10.5** | **5** | **5** | **2** | **7** | **Ku** |
| | User identity and access management | 1 | 2.5 | 4 | 3 | 3 | 2 | 3 | Ku |
| | Cluster network security | 1 | 6 | 6.5 | 2 | 2 | 0 | 4 | Si/Ku |
| **Securing containers** | | **3.5** | **3** | **9** | **3** | **0** | **2** | **2** | **Ku** |
| | Protection of sensitive data and proprietary software | 0 | 2 | 2 | 2 | 0 | 2 | 2 | Si/Ku/Me/Ma/Dc |
| | Improved security isolation | 3.5 | 1 | 7 | 1 | 0 | 0 | 0 | Ku |
| **App and cluster management** | | **9.5** | **15.5** | **25** | **16.5** | **13.5** | **13** | **21** | **Ku** |
| | Creation, management and inspection | 3 | 4 | 4 | 3 | 2.5 | 3 | 4 | Si/Ku/Dc |
| | Monitoring resource usage and health | 1.5 | 2.5 | 5 | 4 | 4 | 3 | 5 | Ku/Dc |
| | Logging and debugging | 2.5 | 2.5 | 4 | 2 | 1 | 1 | 3 | Ku |
| | Cluster maintenance | 1.5 | 3.5 | 5.5 | 4.5 | 3 | 4 | 5 | Ku |
| | Multi-cloud deployments | 1 | 3 | 6.5 | 3 | 3 | 2 | 4 | Ku |
| Total number of feature implementation strategies | | 57.5 | 94 | 144.5 | 76.5 | 62.5 | 70.5 | 97 | 602.5 |

In addition to these existing differences, we observe an additional, significant difference between Docker Swarm integrated mode and Docker Swarm stand-alone. This can be explained by the fact that the former introduces more unique features than the latter. There are no significant differences between other CO frameworks.

However, for 17 sub-aspects there is a specific CO framework that supports the highest number of common and unique features:

- Kubernetes offers the most features for 15 sub-aspects:
    1. Installation methods and deployment tools (1 unique feature)
    2. Framework design of orchestration engine (6 unique features)
    3. Service discovery and external access (6 unique features)
    4. Supported workload types (2 unique feature)
    5. Persistent volumes (4 unique features)
    6. Reusable container configuration (1 unique feature)
    7. Service upgrades (0 unique features)
    8. Resource quota management (0 unique features)
    9. Controlling preemptive scheduling and re-scheduling behaviour (1 unique feature)
    10. User identity and access management (1 unique feature)
    11. Cluster network security (2 unique features)
    12. Improved security isolation (2 unique features)
    13. Logging and debugging (1 unique feature)
    14. Cluster maintenance (2 unique features)
    15. Multi-cloud deployments (3 unique features minus the halted feature = 2 unique features)

    With respect to the first three sub-aspects, Kubernetes is the only framework that offers Kubernetes-as-a-service on top of major public cloud providers. It also offers 6 unique features that are relevant for service discovery and external access on top of public cloud providers as well as 6 unique features for the customizability of the orchestration engine and the master API and offers.

- Docker Swarm integrated mode loses the 1$^{st}$ rank for the sub-aspect "cluster network security" to Kubernetes, but still offers the most features for the sub-aspects "services networking" and "host port conflict management".

- Mesos does not offer anymore the most features for the sub-aspect "persistent volumes", which is now more elaborately supported by Kubernetes. Instead it offers the most features for the sub-aspect "allocation of other resources". In particular, Mesos supports network isolation between containers.

- DC/OS does not anymore offer the absolute most features in any sub-aspect.

```
        Sa       Si       Ku       Me       Au       Ma
  Si  0.03010   –        –        –        –        –
  Ku  4.3e-07  0.16340   –        –        –        –
  Me  0.77501  0.63667  0.00064   –        –        –
  Au  0.99992  0.07489  2.5e-06  0.91777   –        –
  Ma  0.97949  0.25975  4.1e-05  0.99688  0.99832   –
  Dc  0.01069  0.99992  0.31007  0.42464  0.03010  0.13117
```

**Figure. 9** Resulting p-values of the Nemenyi multiple comparison test when counting common and unique features (see Table 16). For p-values <= 0.05, we can reject the null hypothesis (i.e., there is no significant difference in overall ranking between a pair of CO frameworks).

Finally, there are tied observations for 9 remaining sub-aspects (see Table 17). If we would rank CO frameworks in terms of counting those sub-aspects for which they offer the most features as well as those aspects with tied observations where they share the highest number of features, then Kubernetes ranks highest with 22 sub-aspects, then Docker Swarm integrated mode with 10 sub-aspects, then DC/OS with 7 sub-aspects, then Docker Swarm stand-alone with 6 aspects, then Marathon with 5 aspects, and finally Mesos and Aurora with both 3 sub-aspects. We graphically present the top 3 CO frameworks using a radar chart (see Figure 10).

**Table 17**. Tied observation for 9 sub-aspects when counting common and unique features.

| Aspects | Sub-aspects | CO frameworks | | | | | | | #ties between FWs |
|---|---|---|---|---|---|---|---|---|---|
| | | Sa | Si | Ku | Me | Au | Ma | Dc | |
| **Cluster architecture and setup** | | | | | | | | | **10** |
| | Configuration management approach | ✓ | ✓ | ✓ | | ✓ | ✓ | ✓ | 6 |
| | Architectural patterns | ✓ | ✓ | | | | ✓ | ✓ | 4 |
| **CO system customization** | | | | | | | | | **3** |
| | Unified container runtime architecture | ✓ | ✓ | ✓ | | | | | 3 |
| **Container networking** | | | | | | | | | **5** |
| | Plugin architecture for network services | ✓ | ✓ | | ✓ | | ✓ | ✓ | 5 |
| **Container QoS Management** | | | | | | | | | **9** |
| | Container CPU and mem allocation with support for over-subscription | ✓ | ✓ | ✓ | | | | | 3 |
| | Controlling scheduling behavior by means of placement constraints | ✓ | ✓ | ✓ | | ✓ | ✓ | ✓ | 6 |
| **Securing containers** | | | | | | | | | **5** |
| | Protection of sensitive data and proprietary software | | ✓ | ✓ | ✓ | | ✓ | ✓ | 5 |
| **Application and cluster management** | | | | | | | | | **5** |
| | Creation, management and inspection of cluster and applications | | ✓ | ✓ | | | | ✓ | 3 |
| | Monitoring resource usage and health | | | ✓ | | | | ✓ | 2 |
| Total tied observations per CO framework | | 6 | 8 | 7 | 2 | 2 | 5 | 7 | 37 |

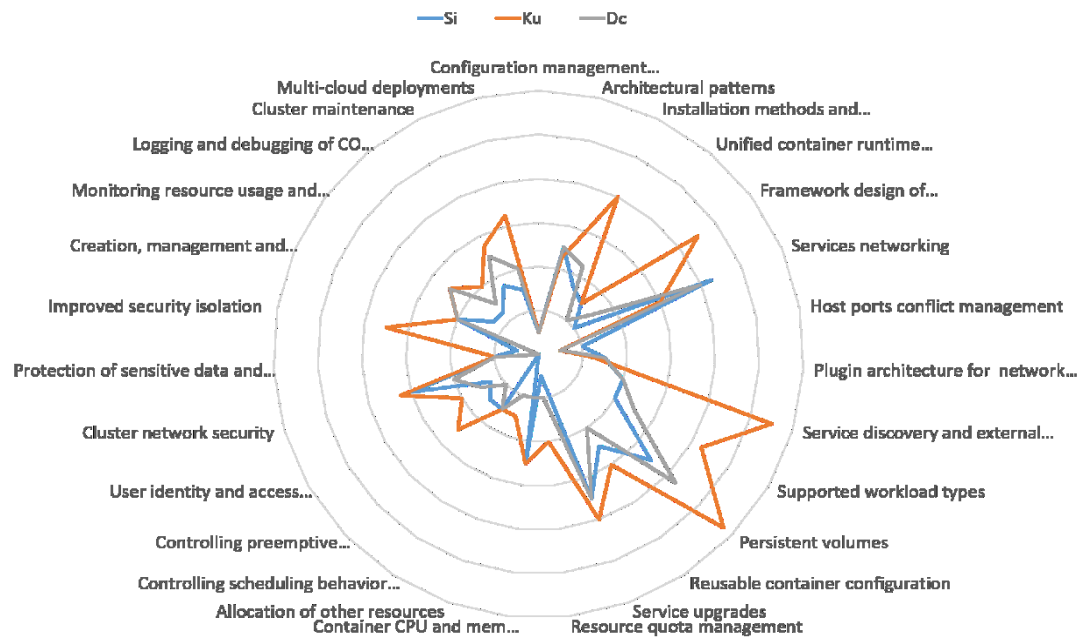

**Figure 10**. Radar chart of Docker Swarm integrated mode, Kubernetes and DC/OS for common + unique features.

## 6. Assessment of Maturity

This section answers research questions RQ7 and RQ8.

**RQ7. What is the maturity of a CO framework with respect to a common feature or a functional (sub)-aspect?** For each of the 9 functional aspects, we present a table that maps each common feature to a timeline that orders CO frameworks according to the time they have released the alpha version of the common feature. In the supplementary material of this article, we provide tables with direct hyperlinks to the documentation about these alpha versions of the features. Sections 5.1-5.9 present the main findings that can be drawn from these tables.

## 6.1 Cluster architecture and setup

Table 18 presents the timeline of when the features of the "cluster architecture and setup" aspect have been introduced by the different CO frameworks. Firstly, it shows that these are basic features that are released as part of the first versions of the frameworks. Secondly, the implementation strategy for these features has also rarely been changed except in the "installation methods and tools" sub-aspect. Finally, as Mesos, Aurora and Marathon have been created earlier, they pioneered in all sub-aspects of the cluster architecture and setup aspect. One notable exception is in the sub-aspect *installation methods and tools*, where Kubernetes is the first CO framework that is offered as a hosted solution by cloud providers.

## 6.2 Container orchestration framework customization

Table 19 shows the historical timeline of the features of the "CO framework customization" aspect. Firstly, with respect to the "unified container runtime" sub-aspect, it shows that despite the popularity of Docker, it didn't last long before other container runtimes have been offered by CO frameworks. However, Docker's *containerd* initiative for creating a unified runtime architecture has been timely and has been pushed by the Cloud Native Computing foundation as de-facto standard for unified container runtime architecture [36]. Secondly, Mesos clearly pioneered with its highly modular software architecture of the core orchestration engine, but Kubernetes has also been highly extensible from the start of the project and this extensibility has been continuously improved (see Section 4.2.2).

## 6.3 Container networking

With respect to the "services networking" sub-aspect (see Table 20), networking with global service ports and host ports found its roots in Mesos v0.20 that added support for network isolation between containers without relying on a virtual bridge. The network isolation module prevents a single container from exhausting the available network ports, consuming an unfair share of the network bandwidth or significantly delaying packet transmission for others. Subsequently, Aurora v0.8 pioneered with a fully-functioning host port networking approach using the central Mesos-DNS service as service proxy. Aurora also pioneered in the "host port conflict management" sub-aspect by supporting dynamic allocation of host ports.

Kubernetes v0.6 pioneered in the integrated design of a routing mesh for service ports and virtual IP network support for containers with a distributed L4 load balancer. Docker Swarm integrated mode v1.12 fully adopted this integrated design of two networking approaches with a distributed load balancer. Marathon v1.0.0 introduced then a centralized L4-L7 load balancer for global service ports, while DC/OS v1.10 introduced a similar L4-L7 load balancer with support for load balancing both container-orchestrated services and non-container orchestrated services.

Mesos v0.25 has introduced support for virtual networks as part of its Mesos containerizer runtime. However, this Mesos feature has never been used by Aurora or Marathon. Later, Mesos v1.0.0 deprecated this initial support in favor of CNI-based networking.

Marathon v0.14 initially supported an IP-per-container feature, but this feature has been deprecated and all efforts were spent in adding support for virtual IP addresses for containers to DC/OS. DC/OS v1.8 introduced separate components for implementing respectively: a virtual container network, a distributed DNS server, name-based VIPs and a distributed L4 load balancer. Later in DC/OS v1.11, these different component have been aggregated into a composite dcos-net component that runs in an Erlang VM. Finally Marathon v1.5.0 again provided support for virtual IP networks for containers.

**Table 18.** Timeline of when each CO framework introduced support for features of the "cluster architecture and setup" aspect. Rows are features, aggregated by sub-aspect, while columns are semesters.

---

*Cell Legend for Tables 18- 27*

- **Sa**: Docker Swarm stand-alone
- **Si**: Docker Swarm integrated
- **Ku**: Kubernetes
- **Me**: Mesos
- **Au**: Mesos+Aurora
- **Ma**: Mesos+Marathon
- **Dc**: DC/OS
- **Xx1**: 1st version of corresponding feature by CO framework Xx
- **Xx2**: 2nd version of the corresponding feature by Co framework Xx
- **Xx**: The CO framework Xx currently has deprecated the corresponding feature in the mean time
- **Xx1**: The CO framework Xx has superseded the 1st version of the corresponding feature with a later 2nd version

---

| Cluster architecture and setup sub-aspects | Features | before Jun 13 | Jul 13 – Dec 13 | Jan 14 – Jun 14 | Jul 14 – Dec 14 | Jan 15 – Jun 15 | Jul 15 – Dec 15 | Jan 16 – Jun 16 | Jul 16 – Dec 16 | Jan 17 – Jun 17 | Jul 17 – Dec 17 | Jan 18 - Jun 18 |
|---|---|---|---|---|---|---|---|---|---|---|---|---|
| Configuration management approach | *Declarative configuration management* |  | Ma<br>Au |  | Ku |  | Dc<br>Sa | Si |  |  |  |  |
| Architectural patterns | *Master-Worker architecture* | Me | Ma<br>Au |  | Ku | Sa |  | Dc | Si |  |  |  |
|  | *Highly-available (HA) master design* |  | Me<br>Au |  | Ma |  | Sa<br>Ku | Dc | Si |  |  |  |
|  | *Generic, automated setup of HA masters* |  | Au |  | Ma |  | Sa | Dc | Si<br>Ku |  |  |  |
|  | *Versioned HTTP API + client libraries* |  | Ma |  |  |  | Ku |  | Me | Dc | Sa<br>Si |  |
|  | *Simple, policy-rich scheduling algorithm* |  | Ma<br>Au |  | Ku<br>Sa |  |  | Dc | Si |  |  |  |
| Installation methods and tools for setting up a cluster | *Dockerized CO software* |  |  |  | Me<br>Ma | Sa1<br>Ku1 |  |  | Ku2<br>Sa2 |  |  |  |
|  | *VM images with CO software for local dev* |  |  | Me | Ku1 | Au | Ma |  | Ku2 Dc |  |  |  |
|  | *Linux packages + CLI for cluster setup* |  |  |  | Me<br>Ma |  |  | Au<br>Dc | Si<br>Ku |  |  |  |
|  | *Configuration management tools* |  |  | Me | Ku |  |  |  |  |  |  |  |
|  | *Cloud-provider tool or platform* |  |  |  | Ku |  |  | Sa | Si<br>Dc |  |  |  |
|  | *Cloud-provider independent tools* |  |  |  |  |  |  | Dc Ku | Si |  |  |  |
|  | *Microsoft Windows or Windows Server* |  |  |  |  |  |  | Me | Si<br>Ku |  |  |  |

With respect to the "network plugin architecture" sub-aspect, Kubernetes v1.0.0 pioneered with different implementations of its innovating virtual container network model. Very soon thereafter, Docker Swarm stand-alone v1.0.0 and Docker v1.9 released an innovating network plugin architecture, libnetwork, for creating and removing virtual container networks at runtime and installing new network plugins at run-time. Kubernetes v1.2 adopted subsequently the CNI specification. Then, Mesos v1.0.0 supported both CNI-based and Docker-based network plugins and this Mesos feature has been made available in DC/OS v1.9 and Marathon v1.5.0.

With respect to the sub-aspect "service discovery and external access to services", Kubernetes v0.8 pioneered with an internal DNS service and Kubernetes v0.18 pioneered in external access to services and added later many other unique features. Docker v1.10 and DC/OS v1.7 pioneered with a distributed DNS for Docker Swarm stand-alone.

**Table 19.** Timeline of when each CO framework introduced support for features of the "CO framework customization" aspect.

| CO framework customization sub-aspects | Features | before Jun 13 | Jul 13 – Dec 13 | Jan 14 – Jun 14 | Jul 14 – Dec 14 | Jan 15 – Jun 15 | Jul 15 – Dec 15 | Jan 16 – Jun 16 | Jul 16 – Dec 16 | Jan 17 – Jun 17 | Jul 17 – Dec 17 | Jan 18- Jun 18 |
|---|---|---|---|---|---|---|---|---|---|---|---|---|
| Unified container runtime architecture | *Unified container runtime architecture* | | | | | | | Me Sa | Si Au Ku | Ma Dc | | |
| | *Support for OCI specifications* | | | | | | | Sa | Si | | | Ku |
| | *Other supported container runtimes* | | | | Me | Ku | Sa | | Au Si | Ma Dc | | |
| Framework design of orchestration engine | *External plugin architecture* | | | | Me | Sa1 | Ku Ma | Dc | Si1 Sa2 Si2 | | | |
| | *Plugin-architecture for schedulers* | | | | Me | | | Ku Dc | | Ma | Au | |
| | *Modular interceptors* | | | | Me Au1 | Ku1 | | Dc | | Ku2 | Au2 | |

### 6.4 *Application configuration and deployment*

With respect to the "supported workload types" sub-aspect (see Table 21), concepts for configuring different workload types have been defined quite early during the start of the Kubernetes, Aurora and Marathon projects and these concepts were also quite similar across the different frameworks. Exceptions to this are the concept of Pod, which has been introduced first by Kubernetes, and support for composite applications which has been introduced by Marathon.

The "persistent volumes" sub-aspect counts the highest number of feature implementation strategies when accumulating the effort done for all frameworks. Kubernetes v0.6 introduced support for external persistent volumes. During the period of Jul 2015-Dec 2015, Docker v1.7 also documented support for persistent volumes and Docker v1.8 introduced a plugin architecture for different volume plugins. A distinguishing feature of this new plugin framework is that plugins could be installed at any time in a running cluster, while Kubernetes' volume plugin framework required a rebuild of the framework software in order to add a new implementation.

By August 2016, Docker's plugin architecture for volumes has also been supported by Mesos v1.0.0, Marathon v1.3.0 and DC/OS v1.8. However, in February 2017, Docker v1.13+ redesigned its plugin framework completely; it's not clear if this new plugin framework is supported in Mesos-based framework. In May 2017, Google and Mesosphere, the company behind Mesos, Marathon and DC/OS, initiated an attempt to define a common specification, named CSI, for exposing container storage providers to containers at run-time. Alpha support for CSI has been added to Kubernetes v1.9 and DC/OS 1.11 around the end of 2017.

With respect to "the reusable container configuration" and "service upgrades" sub-aspect, features have been added gradually over the lifetime of the CO frameworks.

**Table 20.** Historical timeline of "container networking" features.

| Container networking sub-aspects | Features | | before Jun 13 | Jul 13 – Dec 13 | Jan 14 – Jun 14 | Jul 14 – Dec 14 | Jan 15 – Jun 15 | Jul 15 – Dec 15 | Jan 16 – Jun 16 | Jul 16 – Dec 16 | Jan 17 – Jun 17 | Jul 17 – Dec 17 | Jan 18 – Jun 18 |
|---|---|---|---|---|---|---|---|---|---|---|---|---|---|
| Services networking | Routing mesh for stable global service ports | distributed Layer 4 load balancer( based on ipvs) | | | | | | Ku1 (no ipvs) | | Si (with ipvs) | | Ku2 (with ipvs) | |
| | | With centralized L4-L7 LB without ipvs | | | | | | Me | Ma1 Ku | Si1 Dc1 | | Ma2 Dc2 | Si2 |
| | Virtual IP network for containers | L4 distributed LB (with ipvs support) | | | | | | Ku1 (no ipvs) | | Si (with ipvs) Dc1 (no ipvs) | Dc2 (with ipvs) | Ku2 (with ipvs) | |
| | | with stable DNS name for service | | | | Ku | | | | Si | Dc | | |
| | | IP per container | | | | Ku | | Sa Me1 | Ma1 | Si Me2 Dc | | Ma2 | |
| | Host port networking | mapping container port to host port | | | | Me Ku | Au | Sa Ma1 | | Dc1 | Si | Ma2 | Dc2 |
| | | with stable DNS name for service | | | | | | Me Au | | Ma Dc1 | Dc2 | Si | |
| | | host mode networking | | | | | | Sa | | Me | Si | Ma | Dc |
| Host ports conflict management | Dynamic allocation of host ports | | | | | | Au | | Ma1 Dc | Sa | Si | Ma2 | |
| | Management of host port conflicts | | | | | | | Ku | | | Si | | |
| Plugin architecture for network services | Network plugin architecture | | | | | | | Ku1 Sa | Ku2 | Me Si | Dc | Ma | |
| | Support for CNI specification | | | | | | | | Ku | Me | Dc | Ma | |
| | Support for Docker's network architecture | | | | | | | Sa | | Me Si | Dc | Ma | |
| | Separation of data and control traffic | | | | | | | | | | | Ku Sa Si | |
| Service discovery and external access | Internal DNS for service discovery | distributed DNS server on every node | | | | | | Sa Dc | | Si | | | |
| | | centralized DNS server | | | | Ku | | Me | | Ma Dc Au | | | |
| | DNS SRV records (only supported by centralized DNS server) | | | | | Ku | | Me | Ma | Dc Au | | | |
| | Bypassing the L4 service load balancer | | | | | Ku | | | | Si | | | |
| | Exposing services to external clients outside the cluster via routing mesh | | | | | Ku | | | Ma | Si Dc | | Dc2 | |
| | Co-existence of service IPs and global service ports for a single service | | | | | Ku | | | | Si | | | |

**Table 21.** Historical timeline of the "application configuration and deployment" features.

| Application configuration and deployment sub-aspects | Features | before Jun 13 | Jul 13 – Dec 13 | Jan 14 – Jun 14 | Jul 14 – Dec 14 | Jan 15 – Jun 15 | Jul 15 – Dec 15 | Jan 16 – Jun 16 | Jul 16 – Dec 16 | Jan 17 – Jun 17 | Jul 17 – Dec 17 | Jan 18 - Jun 18 |
|---|---|---|---|---|---|---|---|---|---|---|---|---|
| Supported workload types | Pods | | | | Ku | | | | Me Dc | Ma | | |
| | Container-based jobs | | Au | | | | Ku | | Dc | | | |
| | Container-based services | | | | Ma Ku | | Au | Dc1 | Dc2 Si | | | |
| | Elastic scaling of services | | | | Ma Ku1 | | | Au Ku2 Sa | Dc Si | | | |
| | Auto-scaling of services | | | | | | | Ku | | | | Dc |
| | Global containers | | | | Ku1 | | | Ku2 | Si | | | |
| | Composite applications | | | | Ma | | | Dc | Ku1 | Sa Si | Ku2 | |
| Persistent volumes | Local volumes | | | | Ku1 | | Me Sa Au | Ma Dc | Si | | | Ku2 |
| | Automatic (re)scheduling | | | | | | Me Au | Ma Dc | | | | Ku |
| | Shareable volumes between containers | | | | Ku | | Sa | | Si Me | | | |
| | External volumes | | | | Ku | | Au Sa | | Si Me Ma Dc | | | |
| | Volume plugin architecture | | | | Ku | | Sa | | Si Me Ma Dc | | | |
| | Run-time installation of volume plugins | | | | | | Sa | | Si Me1 Ma Dc1 | | Ku Me2 | Dc2 |
| | Docker Engine Plugin framework support | | | | | | Sa | | Si Me Ma Dc | | | |
| | Common Storage Interface (CSI) support | | | | | | | | | | Ku Me | Dc |
| | Dynamic provisioning of volumes | | | | | | Me Au | Ma Dc | Si Ku | Sa | | |
| Reusable container configuration | Pass environment variable to container | | | | Me Ma | | Sa | Ku Me2 Au | Si | Ma2 Dc2 | | |
| | Self-inspection API | | | | | | Ku | | | Ma Dc | | |
| | Separate configuration data from image | | | | | | | | | Ku Si | | |
| | Custom ENTRYPOINT | | | | Me Ma | | Ku Sa | Au Me2 | Si Dc | Ma2 Dc2 | | |
| | Custom CMD | | | | Me Ma | | Ku Sa | Au Me2 | Si Dc | Ma2 Dc2 | | |
| Service upgrades | Rolling upgrades of services | | | | Au Ma | | Ku1 | Ku2 | Dc | Si | | |
| | Monitoring of a rolling upgrade | | | | Ma | | Au | Ku | Dc | Si | | |
| | Roll back | | | | | | Au | | Dc | Ku | Si | |
| | Configuration of custom readiness checks | | | | Ku | | Au | | Ma Dc | | | |
| | Customizing the rolling upgrade process | | | | Ma | | | Ku | Dc | Si | | |

|  |  |  |  |  |  |  |  |  |  |  |  |
|---|---|---|---|---|---|---|---|---|---|---|---|
| Support for performing canary deployments |  |  |  |  |  | Ku |  | Dc Au |  |  |  |
| In-place updates of app configurations |  |  |  |  | Ku Dc | Sa | Si |  |  |  |  |
| Non-disruptive, in-place updates |  |  |  |  |  | Sa | Ku | Si |  |  |  |

## 6.5 Resource quota management

As shown in Table 22, Aurora and Kubernetes pioneered with the features related to managing API objects for different user groups and quota limits on computing resources and amount of API objects. Kubernetes is the only CO framework that offers support for all 5 features of this aspect. Mesos offers support for partitioning computing resources but across different scheduler frameworks running on top of a Mesos cluster.

**Table 22**. Historical time line of the "resource quota management" features.

| Resource quota management aspect | Features | before Jun 13 | Jul 13 – Dec 13 | Jan 14– Jun 14 | Jul 14 – Dec 14 | Jan 15 – Jun 15 | Jul 15– Dec 15 | Jan 16 – Jun 16 | Jul 16 – Dec 16 | Jan 17 – Jun 17 | Jul 17 – Dec 17 | Jan 18– Jun 18 |
|---|---|---|---|---|---|---|---|---|---|---|---|---|
| Resource quota management | Partitioning API objects in user groups |  |  |  |  | Au Ku |  | Me | Dc | Si |  |  |
|  | CPU and memory quota per user group |  |  |  |  | Ku Au |  | Me1 |  |  |  | Me2 |
|  | Disk quota per user group |  |  |  |  | Au |  | Me1 | Ku |  |  | Me2 Me3 |
|  | Object count quota limits per user group |  |  |  |  | Ku |  | Me |  |  |  |  |
|  | Reserving resources for the CO framework |  |  |  |  |  |  | Ku |  |  | Ma Dc |  |

## 6.6 Container QoS management

With respect to the sub-aspect "container CPU and memory allocation with support for oversubscription" (see Table 23), Kubernetes and later Docker Swarm integrated mode offer resource allocation policies for CPU and memory that support oversubscription and that hide the complexity of using cpu-shares, which are relative weights. So these two frameworks are the preferred choice when optimal server consolidation is important.

With respect to the "allocation of other resources" sub-aspect, Mesos-based frameworks pioneered with both disk limits and GPU limits.

With respect to the sub-aspect "controlling scheduling behavior by means of placement constraints", Mesos-based frameworks also pioneered in supporting various types of expressive placement preferences. Note that Kubernetes has been the most actively developed framework with respect to supporting various types of expressive placement preferences.

Finally, with respect to the sub-aspect "controlling preemptive scheduling and re-scheduling behavior", Mesos-based frameworks, in particular Aurora, has pioneered in supporting preemptive scheduling while Docker Swarm integrated mode has pioneered in redistributing un-balanced services.

## 6.7 Securing clusters

With respect to the "user identity and access management" sub-aspect (see Table 24), Mesos-based frameworks and Kubernetes took the forefront in developing user authentication and authorization with respect to their respective master API. Kubernetes pioneered with tenant-aware access control.

With respect to the "cluster network security" sub-aspect, Mesos and Aurora pioneered respectively with authentication of worker nodes with the Mesos master and authentication of Executors with the scheduler of Aurora. The first release of Docker Swarm integrated mode contained several innovating features related to automated bootstrap of a secure cluster when installing the cluster and adding new nodes to the cluster.

### 6.8 Securing containers

With respect to the "protection of sensitive data and proprietary software" sub-aspect, Kubernetes v0.20 is the first framework to provide support for all features of this aspect (see Table 25). Support for secrets has been added much later by the other frameworks.

**Table 23.** Historical timeline of the "Container QoS management" features.

| Container QoS management sub-aspects | Features | before Jun 13 | Jul 13 – Dec 13 | Jan 14 – Jun 14 | Jul 14 – Dec 14 | Jan 15 – Jun 15 | Jul 15 – Dec 15 | Jan 16 – Jun 16 | Jul 16 – Dec 16 | Jan 17 – Jun 17 | Jul 17 – Dec 17 | Jan 18 – Jun 18 |
|---|---|---|---|---|---|---|---|---|---|---|---|---|
| Container CPU and memory allocation with support for oversubscription | Minimum guarantees for CPU | | | Me | Sa | Ma Ku | Au | | Si Dc | | | |
| | Abstraction of CPU-shares for CPU guarantees | | | | | Ku | | | Si | | | |
| | Minimum guarantees for memory | | | | | Ku | | | Si | Sa | | |
| | Maximum limits for CPU | | | Me1 | | Ku | | | Si Me2 | Sa | | |
| | Maximum limits for memory | | | Me1 | Sa | Ma Ku | Au | | Si Dc Me2 | | | |
| Allocation of other resources | Limits for NVIDIA GPU | | | | | | | | Me Au Ma | Dc Ku | | |
| | Limits for disk resources | | | | | | Me | Ma Au | | | Dc | Ku |
| Controlling scheduling behavior by means of placement constraints | Evaluate over node labels/attributes | | Ma | | | Ku Au | Sa | | Si Dc | | | |
| | Define custom node labels/attributes | | Me | Ma | | Ku Au | Sa | | Si | Dc | | |
| | More expressive constraints | | | Ma | | Au | Sa | Ku1 | Ku2 Ku3 Dc | Si | | |
| Controlling preemptive scheduling and re-scheduling behavior | Preemptive scheduling | | | | | Au | | Ku1 | | | Ku2 | |
| | Container eviction when out-of-resource | | | | | | Au | | Ku1 | Ku2 | | |
| | Container eviction on node failure | | Me | Ma | | Au | | Ku | Dc Sa | Si | | |
| | Container lifecycle handling | Me | Ma | | | Au Ku | | | Dc | Si | | |

Also, with respect to the sub-aspect "Improved security isolation", Kubernetes and Docker pioneered by adding support for different access control mechanism of the Linux kernel. Kubernetes pioneered also by adding policy-based management for aggregating sets of access control rules and applying them at different levels of granularity: both at the level of individual containers as well as at the level of user groups (see Section 4.8.2).

### 6.9 Application and cluster management

With respect to the "creation, management and inspection of cluster and applications" sub-aspect (see Table 26), basic CLI and Web UI features are part of the first release of each CO framework. Kubernetes pioneered with the features for organizing API objects by means of labels and visualization of resource usage graphs.

With respect to the "monitoring resource usage and health" sub-aspect, Kubernetes pioneered in support for monitoring container resource usage, while Aurora pioneered in monitoring resource usage by the CO framework itself. Marathon pioneered in a framework for health checks and distributed event monitoring.

**Table 24.** Historical timeline of the "securing clusters" features.

| Securing clusters sub-aspects | Features | before Jun 13 | Jul 13 – Dec 13 | Jan 14– Jun 14 | Jul 14 – Dec 14 | Jan 15 – Jun 15 | Jul 15– Dec 15 | Jan 16 – Jun 16 | Jul 16 – Dec 16 | Jan 17 – Jun 17 | Jul 17 – Dec 17 | Jan 18– Jun 18 |
|---|---|---|---|---|---|---|---|---|---|---|---|---|
| User identity and access management | *Authentication of users with master API* | | Me1 | Me2 | Ma Ku1 | Au | | Sa Dc | Si | | | Ku2 |
| | *Authorization of users with master API* | | | | Me Ku | Au | Ma | Dc | Si | | | |
| | *Tenant-aware ACLs* | | | | Ku | Au | | | Me | Dc Si | | |
| Cluster network security | *Authentication of worker nodes with master API* | | | Me | | | | Sa Au Dc | Si Ku | | | |
| | *Automated bootstrap of worker tokens* | | | | | | | | Si | Ku | | |
| | *Authorization of CO agents on workers* | | | | | | | Au | Me | Ku | | |
| | *Encryption of control messages* | | | | | | | | Si Ku | Dc | | |
| | *Encryption of application messages* | | | | | | Ku | | Si | Dc | | |
| | *Restricting access to service ports* | | | | | | | | Dc | Si | | Ku |

With respect to the "logging and debugging of CO framework and containers" sub-aspect, logging of containers and logging of CO framework components are part of the first release of each CO framework, except Marathon.

**Table 25.** Historical timeline of the "securing containers" features.

| Securing containers sub-aspects | Features | before Jun 13 | Jul 13 – Dec 13 | Jan 14– Jun 14 | Jul 14 – Dec 14 | Jan 15 – Jun 15 | Jul 15– Dec 15 | Jan 16 – Jun 16 | Jul 16 – Dec 16 | Jan 17 – Jun 17 | Jul 17 – Dec 17 | Jan 18– Jun 18 |
|---|---|---|---|---|---|---|---|---|---|---|---|---|
| Protection of sensitive data and software | *Storage of sensitive-data as secrets* | | | | | Ku | | | | | Si Dc | Me Ma |
| | *Pull image from a private Docker registry* | | | | Me Ku | Ma | | | | Si | Dc | |
| Improved security isolation | *Setting Linux capabilities per container* | | | | | | | Ku | Sa | Me | | Si |
| | *Setting SELinux labels per container* | | | | | | | Ku | Sa | | | Si |
| | *Setting AppArmor profiles per container* | | | | | | | | Sa | Ku | | Si |
| | *Setting seccomp profiles per container* | | | | | | | | Sa | Ku | | Si |
| | *Higher-level aggregate objects* | | | | | | | Ku | | | | Sa Si |

With respect to the "cluster maintenance" sub-aspects, Mesos-based frameworks have pioneered in all features of the sub-aspect.

Finally, with respect to the "multi-cloud support" aspect, Mesos-based frameworks have pioneered in installing a single cluster across multiple availability zones that can handle and recover from network partitions as well as multi-zone deployments of services of which the services are spread across different availability zones. Kubernetes has pioneered installing and managing multiple clusters across different availability zones and federating these clusters using a separate authentication and control plane.

**Table 26**. Historical timeline of the "application and cluster management" features.

| Application and cluster management sub-aspects | Features | before Jun 13 | Jul 13 – Dec 13 | Jan 14– Jun 14 | Jul 14 – Dec 14 | Jan 15 – Jun 15 | Jul 15– Dec 15 | Jan 16 – Jun 16 | Jul 16 – Dec 16 | Jan 17 –Jun 17 | Jul 17 – Dec 17 | Jan 18– Jun 18 |
|---|---|---|---|---|---|---|---|---|---|---|---|---|
| Creation, management and inspection of cluster and applications | Command-line interface (CLI) | | Ma | Au | Me Ku | Sa | | | Dc | Si | | |
| | Web UI | | | | Au | Me | Ma | Ku Dc | Sa Si | | | |
| | Labels for organizing API objects | | | | | Ku | Sa Me | Ma | Dc | Si | | |
| | Inspection of resource usage graphs | | | | | | | Ku Dc | | Sa Si | Au | |
| Monitoring resource usage and health | Monitoring container resource usage | | | | | | Ku1 | | Me | Dc | Ku2 | |
| | Monitoring CO framework resource usage | | | | Au | | | Me | Ma Dc | Ku | | Sa Si |
| | Framework for container health checks | | | Ma | | Au | | Ku | Me | Sa Si Dc Ku | | |
| | Distributed event monitoring | | | | | Ma | | | Dc | Me Au Si Dc | | |
| Logging and debugging of CO framework and containers | Logging of containers | | | | | | Sa | Ku | Me | | | |
| | Logging of CO framework components | Me | | | | Ku | Sa Ma1 | Au | | Si | Dc Ma2 | |
| | Integration with log aggregator frameworks | | | | | | | | Sa Si Ku | Dc | | |
| Cluster maintenance | Cluster state backup and recovery | | | | Me | Au | | | Si | | Ma Dc | Ku |
| | Official cluster upgrade documentation | Me | | | | Ku | Ma | | | Au Dc | | |
| | Upgrade does not affect running containers | | | Me | | | | Ma | Sa Si | Ku Au Dc | | |
| | Draining a node for maintenance | | | | | | | Me | Si Ku | Dc | | Ma |
| | Garbage collection of containers and images | | | | | | | | | Dc1 Ku | | Me Dc2 |
| Multi-cloud support | One cluster across availability zones | | Me Au | | Ma | | Ma | Sa | Au Si | Ku Dc | | |
| | Recovering from network partitions | | | | | | | | Me | | | Au |
| | Management of multiple clusters | | | | | | | Ku | | Si | Dc | |
| | Federated authentication | | | | | | | | Ku | | | Dc |

**RQ8. Which functional sub-aspects are mature enough to consider them as part of the stable foundation of the overall domain? Which CO frameworks have pioneered in what sub-aspect?** In

this section we aim to rank different sub-aspects by their overall maturity by determining the time when support for a sub-aspect has been consolidated[7] by a pioneering CO framework for the first time. A sub-aspect is considered to be consolidated when a comprehensive subset of the common features from this sub-aspect has been implemented by a pioneering framework.

Figure 11 shows an overall timeline that ranks sub-aspects with respect to their maturity. For each sub-aspect, the figure shows which CO framework has pioneered in consolidating the sub-aspect. We define a sub-aspect as being consolidated when a coherent subset of the common features of that sub-aspect has been established by the pioneering framework.

With respect to identifying those sub-aspects that are considered mature and well-understood, we are guided by the criteria that (i) the sub-aspect has been consolidated by the pioneering framework at least two traditional release cycles of 18 months [130] ago[8], (ii) the corresponding feature implementation strategies of the pioneering framework have at least reached beta-stage in the meantime and (iii) there are no deprecation or removal events of important features in the latest traditional release cycle.

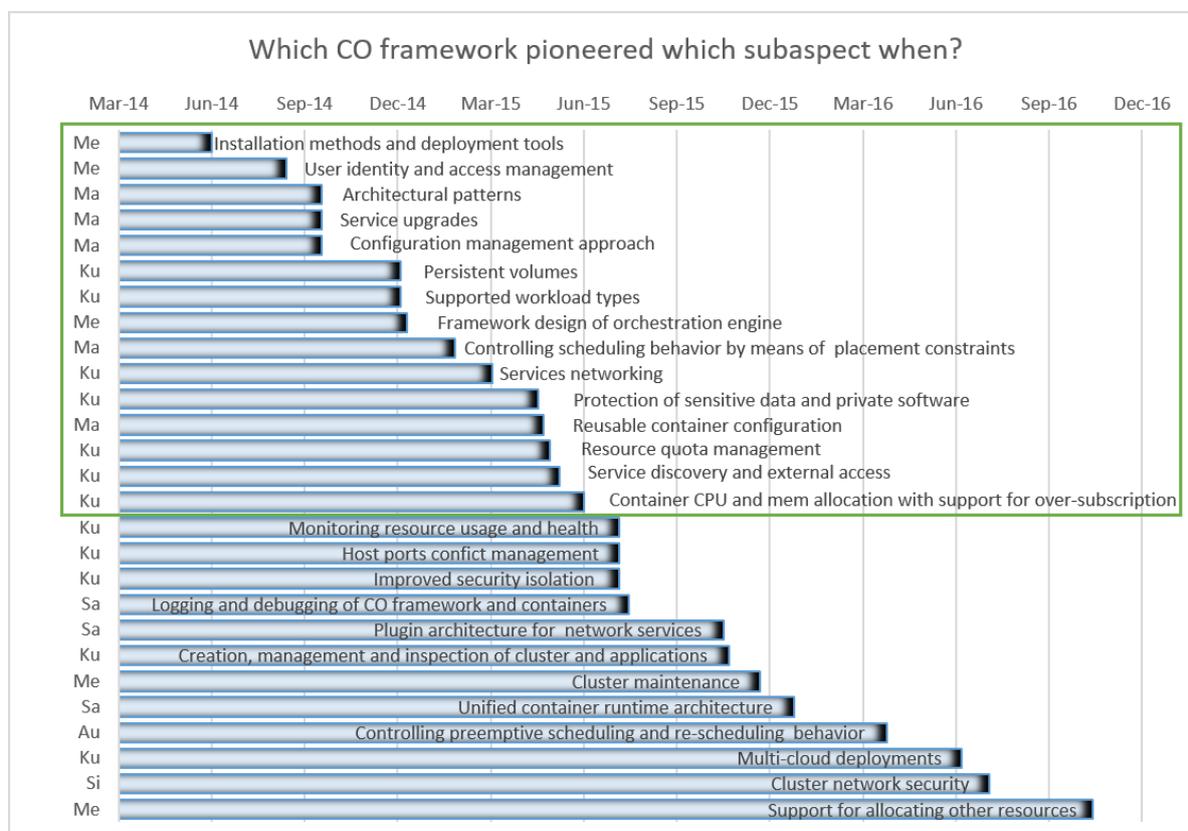

**Figure 11**. Timeline of when support for a sub-aspect have been consolidated by a CO framework.

This leads us to the observation that 15 out of 27 sub-aspects can be considered mature and well-understood (see green rectangle in Figure 11). Some sub-aspects that have been consolidated at least 36 months ago are not yet considered mature because they fail to meet the other two criteria:

- The sub-aspect "monitoring resource usage and health" is still in flux as Kubernetes' monitoring service (Heapster) has recently been completely replaced by two new monitoring services.
- Host port conflict management is expected to evolve due to the growing importance of supporting service networking in true host mode.

---

[7] i.e., a comprehensive subset of the common features from this sub-aspect has been implemented.

[8] These two release cycles are needed for letting other CO framework adopt and develop similar features.

- Improved security isolation support by Kubernetes has not been substantially adopted by other orchestration frameworks; instead security isolation is becoming a customizable property of container runtimes themselves.
- Logging support has remained very basic in all frameworks. Instead many third-party companies have already offered commercial solutions for centralized log management.
- The network plugin architecture of Kubernetes has remained in alpha-stage, while Docker's network plugin architecture is also expected to evolve because Docker EE supports Kubernetes as an alternative orchestrator.
- Inspection of cluster applications is expected to evolve towards a fully reflective interface so that it becomes possible to support application-specific instrumentation of different types of container orchestration functionality. This evolution already has happened at the level of container runtimes (e.g. crictl[9]) but is expected to extend towards orchestration framework functionality as well. Examples of relevant instrumentation scenarios include customizations to service load balancing, fault tolerance extensions to service identity to support global checkpointing of services and enactment customization of rolling upgrades.
- Cluster maintenance, especially cluster upgrades, remains poorly automated.

Figure 11 also presents the creativity of CO frameworks by showing on the left which CO frameworks pioneered in consolidating a sub-aspect, i.e. establishing a coherent subset of the common features of that sub-aspect. Kubernetes has pioneered in 12 of the 27 sub-aspects. Mesos+Marathon in 10 of these sub-aspects, Docker Swarm in 4 sub-aspects, and Aurora in 1 sub-aspect. As such, the Kubernetes project has been the most creative in terms of pioneering new features despite being a younger project than Mesos, Marathon and Aurora.

## 7. Qualitative assessment with respect to stability

This sections answers research questions R9 and R10.

**RQ9. What are the relevant standardization initiatives and which CO frameworks align with these initiatives?** The stability of a CO framework software depends among other factors on its alignment with standardization initiatives. Increased openness to such standardization initiatives also creates more potential for researchers and entrepreneurs to contribute innovating technology that can be integrated in multiple CO frameworks.

In Section 4, we have identified several standardization initiatives towards common specifications to improve the plug-ability of various components including container runtimes, container networking services and storage drivers for external persistent volumes. Table 27 gives an overview of these standardization initiatives and by which CO frameworks they are adopted.

**Table 27**. Overview of existing standardization initiatives and their support.

---

| Sub-aspects with features that relate to standardization initiatives | Standardization initiatives | Swarm stand-alone | Swarm integrated | Kubernetes | Mesos | Mesos+ Aurora | Mesos + Marathon | DC/OS |
|---|---|---|---|---|---|---|---|---|
| | | Sa | Si | Ku | Me | Au | Ma | Dc |
| Unified container runtime architecture | Open Container Initiative (OCI) spec | ✓ | ✓ | ✓ | future | | | |
| | Containerd container runtime architecture | ✓ | ✓ | ✓ | | | | |
| Plugin architecture for network services | Container Network Interface (CNI) | | | ✓ | ✓ | | ✓ | ✓ |
| | Docker's libnetwork (aka CNM) | ✓ | ✓ | | ✓ | | ✓ | ✓ |
| Persistent volumes | Docker's volume plugin system | ✓ | ✓ | | ✓ | | ✓ | ✓ |
| | Common Storage Interface (CSI) | | | ✓ | ✓ | | | ✓ |

The OCI specification for pluggable container runtimes has been accepted by Docker EE and Kubernetes, although Mesos has announced to add support for OCI soon.

Different standards for container networking (CNI, libnetwork) and persistent storage (CSI, Docker volumes) are not compatible across respectively Kubernetes and Docker Swarm. In opposition, DC/OS, provides encompassing support for all initiatives:

- DC/OS supports both CNI-based network plugins and Docker's libnetwork architecture.
- Moreover it supports both Docker volumes as well as the CSI specification for persistent volumes.

As such with respect to networking and storage plugins, DC/OS and Mesos-based frameworks in general are the most open frameworks. With respect to container runtimes, Kubernetes and Docker Swarm are the most open frameworks.

In general we can state that DC/OS is the most interesting platform for prototyping novel techniques for container networking and persistent volumes because DC/OS' adherence to all relevant specifications in these two areas maximizes the potential to deploy these techniques in Docker Swarm and Kubernetes as well. Docker or Kubernetes are best fit for prototyping innovating container runtimes.

However, a widespread adoption of Kubernetes by cloud providers and cloud orchestration platforms[10] has also occurred after he Cloud Native Computing Foundation pushed Kubernetes as de-facto standard in container orchestration and launched a certification programme for production-grade commercial Kubernetes offerings [37]. As a result, Docker volumes and Docker's libnetwork architecture, which are not supported by Kubernetes, may face the risk of not being further developed or halted. We estimate this risk to be low however because Docker offers its volume and networking architecture as separate building blocks that are relatively loosely coupled from its orchestrator Swarm.

**RQ10. What is the risk that common or unique features might become deprecated in the future?** If a particular CO framework halts the development of a particular feature or even deprecates the feature without offering a replacing feature update, then the development of company products or research prototypes that heavily rely on those features might also get compromised. Also t

---

[10] Cloud orchestration platforms such as Rancher [625] and Juju [626] that in the past allowed to manage different CO frameworks on multiple cloud providers, nowadays only support Kubernetes.

In this section we will assess the risk that development of features will be halted in the future or features are even deprecated. With respect to common features, we have studied the volatility of features in the past by counting the number of feature additions versus the number of feature deprecations in Section 4.9.2. Surprisingly, we have found very little volatility in terms of feature being deprecated without a replacing feature update. We recorded in total 626 feature additions; 48 out of these 626 additions comprised an update of an existing feature without deprecating the existing implementation strategy of the feature; finally only 9 out of 626 feature additions comprised a feature update with deprecation or removal of the old implementation strategy of the feature. As such, if we assume that the past is good indicator for the future, the risk that a common feature will be deprecated by a CO framework without being replaced with an alternative new feature implementation strategy is less than 2%.

With respect to unique features, we assume that the risk may be higher. After all, if the team developing a specific unique features faces even small problems, there is less incentive to resolve these problems in comparison to common features that are supported by other CO frameworks as well. This risk should be taken into account by research and development projects that consider relying on those unique features.

Table 28 summarizes the 54 unique features found across the 7 CO frameworks, as presented in Section 4. These unique features are again organized according to the 27 sub-aspects.

Table **28**. Unique features of Docker Swarm, Kubernetes, Mesos, Aurora, Marathon and DC/OS.

Column Legend:
- **Sa**: Docker Swarm stand-alone
- **Si**: Docker Swarm integrated
- **Ku**: Kubernetes
- **Me**: Mesos
- **Au**: Mesos+Aurora
- **Ma**: Mesos+Marathon
- **Dc**: DC/OS

| Container orchestration aspects and sub-aspects | Swarm stan'd-alone | Swarm integrated | Kubernetes | Mesos | Mesos + Aurora | Mesos + Marathon | DC/OS |
|---|---|---|---|---|---|---|---|
| *Cluster architecture and setup* | Sa | Si | Ku | Me | Au | Ma | Dc |
| Configuration management approach | | | | | | | |
| Architectural patterns | | | | | | | |
| Installation methods and tools for setting up a cluster | | | Kubernetes-as-a-Service | | | | GUI-based installation wizard |
| *CO framework customization* | Sa | Si | Ku | Me | Au | Ma | Dc |
| Unified container runtime architecture | | | | | | | |
| Framework design of orchestration engine | | install plugins as global Swarm services | cloud-provider plugin; custom API objects | Resource provider abstraction to customize how Mesos Agent synchronizes | custom worker agent software | | |

| | Sa | Si | Ku | Me | Au | Ma | Dc |
|---|---|---|---|---|---|---|---|
| | | | aggregation of additional APIs<br><br>annotations to API objects<br><br>discovery of a node's hardware features<br><br>dynamic worker agent reconfiguration | with the Mesos Master about available resources and operations on those resources | | | |
| **Container networking** | **Sa** | **Si** | **Ku** | **Me** | **Au** | **Ma** | **Dc** |
| Services networking | | SCTP protocol support | | | | | load balancing of non-container-based services |
| Host ports conflict management | | | | | | | |
| Plugin architecture for network services | | | | | | | |
| Service discovery and external access | | | Exposing service via LB of cloud provider<br><br>synchronize exposed services with external DNS providers<br><br>hide Pod's virtual IP behind Node IP<br><br>override DNS lookup with custom /etc/hosts entries in Pod<br><br>override name server with custom /etc/resolv in Pod<br><br>install another DNS server in cluster | | | | |
| **App configuration/deployment** | **Sa** | **Si** | **Ku** | **Me** | **Au** | **Ma** | **Dc** |
| Supported workload types | | | initialization containers<br><br>vertical pod auto-scaler | | | | |
| Persistent volumes | | | deploying and managing stateful services<br><br>raw block volumes<br><br>dynamically grow volume size<br><br>dynamic maximum volume count | local volume can be shared between tasks from different frameworks | | | tools and libraries for integration with and deployment of stateful services |
| | | | | | | | |

| | Sa | Si | Ku | Me | Au | Ma | Dc |
|---|---|---|---|---|---|---|---|
| Reusable container configuration | | Run a simple service initiation system inside a container | injection of configs at Pod creation time | | | | |
| Service upgrades | | Customizing the enactment of the rollback of a service | | | | | |
| **Resource quota management** | **Sa** | **Si** | | **Me** | **Au** | **Ma** | **Dc** |
| | | | | request rate limiting of Mesos frameworks | | | |
| **Container QoS management** | **Sa** | **Si** | **Ku** | **Me** | **Au** | **Ma** | **Dc** |
| Container CPU and `memory allocation with support for oversubscription | updating resource reservations and limits without restarting the container | | | | | | |
| Allocation of other resources | | | define custom node resources of random kind<br><br>scheduling of huge pages | network performance isolation between containers for routing mesh networks<br><br>network performance isolation between containers for virtual networks | | | |
| Controlling scheduling behavior | | | | | | | |
| Controlling preemptive scheduling and re-scheduling | | | Pods with cpu-cache affinity cannot be evicted from a node once CPUs have been allocated | | | | |
| **Securing clusters** | **Sa** | **Si** | **Ku** | **Me** | **Au** | **Ma** | **Dc** |
| User identity and access management | | | audit of master API requests | | | | |
| Cluster network security | | encryption of master/manager logs | access control of the kubelet worker agent's HTTP endpoint on each node<br><br>network policies for regulating communication of Pods | | | | |
| **Securing containers** | **Sa** | **Si** | **Ku** | **Me** | **Au** | **Ma** | **Dc** |
| Protection of sensitive data and proprietary software | | | | | | | |

| Improved security isolation | | Customize a service isolation mode in Windows | Run-time verification and enforcement of system-wide Pod security policies for governing privileges, capabilities and access control profiles of containers<br><br>support for configuring Linux kernel parameters at run-time[ | | | | |
|---|---|---|---|---|---|---|---|
| *App and cluster management* | **Sa** | **Si** | **Ku** | **Me** | **Au** | **Ma** | **Dc** |
| Creation, management and inspection of cluster and applications | | command-line auto-completion | | | | | |
| Monitoring resource usage and health | | | auto-scaling of cluster | | SLA metrics on Aurora's performance | | custom node and cluster health checks |
| Logging and debugging of CO framework and containers | | | debug running Pod from local work station | | | | |
| Cluster maintenance | | | disruption budget to minimize the number of disruptions due to maintenance<br><br>automated upgrade of the Kubernetes Engine on Google Cloud | | | | |
| Multi-cloud support | | | API for using externally managed services<br><br>federated API with federated instantiations of Kubernetes API objects<br><br>service discovery of the closest healthy service shards | | | | |

For some unique features, it is fairly obvious that the risk of being halted or deprecated is low:

- As Mesos is an underlying framework for multiple scheduler frameworks, all common and unique features of Mesos stem from requirements of multiple scheduler frameworks. Therefore, the unique features of Mesos are assets that have a low risk of becoming halted or deprecated without a replacing feature update because multiple scheduler framework depend on them.
- In the "resource quota management" aspect there is only 1 unique feature from Mesos, which that has a low risk of being halted or deprecated as noted above.

- The "container QoS Management" aspect counts just 3 unique features. Kubernetes introduces 2 unique features for improving performance management for memory- and CPU-bound workloads. Docker Swarm stand-alone allows adjusting resource allocation policies of containers at run-time. These are all useful in their own right.

- All unique features in aspects "securing clusters" and "securing containers" are useful additions. Moreover, as these aspects are not well supported by many CO frameworks, we expect that improving security is an important future work that still needs to be done. Unique features in these CO frameworks will certainly not be deprecated without introducing a replacing feature update with similar, but improved functionality.

For the remaining unique features, we discuss whether or not they run the risk of being halted or deprecated.

### 7.1 Cluster architecture and setup

There are only 2 unique features in the sub-aspect "installation methods and deployment tools".

**Installation methods and deployments tools.** Kubernetes is the only framework with certified commercial Kubernetes-as-a-Service [111] offerings that fully automate the setup and management of Kubernetes clusters. At the moment, at least a dozen of public cloud providers provide uch certified offerings. This certification programme is created by the Cloud Native Computing Foundation that pushes Docker engine and Kubernetes as de-facto standards for container runtimes, container orchestration, respectively. As such we believe this feature will certainly be further developed and strengthened to consolidation the position of Kubernetes across public cloud providers.

### 7.2 CO framework customization

There are no unique features for the sub-aspect "unified container runtime architecture". As such we only discuss the other sub-aspect.

**Framework design of the orchestration engine.** Kubernetes supports several novel types of extension points that are non-existent in other CO frameworks.

1. Extensibility of the API which includes support for extending existing API objects with annotations, adding custom API objects, and even adding entire new APIs.
2. Cloud controller management concept that enables cloud provider specific code and the Kubernetes core to evolve independently
3. Support for custom computing resources (see the "container QoS management" aspect) and corresponding plugins for automated detection of the existence of that hardware on a node.

Clearly, the first two features are assets of Kubernetes because the extensibility of the API is a major enabler for portability because customers can create specific APIs for themselves that abstract Kubernetes-specific APIs. The third feature on the other hand is too limited at the moment as it only allows to specify resource quantities as integers. This implies that a single instance of a custom resource cannot be shared among containers. For example, GPUs can only be allocated as a whole, which means that a GPU cannot be shared by multiple Pods. We expect that this third extension point to be further improved or halted.

### 7.3 Container networking

There are only unique features in the sub-aspects "services networking" and "service discovery and external access".

**Services networking.** Docker Swarm offers support for applications of cellular networks. As shown in the quantitative analysis, Docker Swarm offers the most common features for the services networking aspect where it is possible to dynamically add multiple networking plugins that can co-exist. As such, giving this strong foundation for service networking, we might see Docker Swarm being used in specific technology segments such as cellular networks, cyber-physical systems, and connected and autonomous vehicles. Of course, performance overhead introduced by the service networking approach [58] is the main obstacle that needs to be tackled first.

**Service discovery and external access.** Kubernetes is clearly positioned as the best framework to expose container orchestrated services that run on public cloud providers such as AWS, Google Cloud and Microsoft Azure. We don't expect that other frameworks can compete here. Indeed Kubernetes offers a huge number of features for enabling external access to container orchestrated services such as automated integration with the load balancing service of a cloud provider and automated synchronization with external DNS providers. So indeed, we believe these unique features are assets of Kubernetes that will be further developed to further strengthen the position of Kubernetes as main CO framework for public cloud providers.

### 7.4 *Application configuration and deployment*

The unique features of the sub-aspects "reusable container configuration" and "service upgrades" are all very useful additions. We believe these unique features may be adopted by other CO frameworks.

However, the risk of being halted or deprecated is less clear for the unique features of the sub-aspects "persistent volumes" and "supported workload types".

**Persistent volumes.** With respect to persistent volumes, Kubernetes v.1.10 [617] and Kubernetes v1.11 [618] has added several additional unique features so that its *StatefulSet* concept for automated management of database clusters meets the requirements for production environments.

First, performance improvements have been made. Kubernetes has added support for *raw block storage* that is often required by databases to attain their full performance capacity. Moreover it has added support for *dynamic volume count limits* that can be configured on a per node basis. Second, to ensure that fluctuations in actual disk usage versus expected disk usage can be efficiently handled, Kubernetes has added support for *resizing existing volumes*.

DC/OS takes a completely different approach on automated deployment of stateful applications such as database clusters. Namely it runs stateful services in a separate scheduler framework that interacts with the central Mesos scheduler to place instances of stateful services across nodes. DC/OS also offers a library and associated SDK [342] for user-friendly development and performance tuning of such scheduler framework. An on-line service catalog [619] with default available services, e.g. various databases such as Cassandra, streaming frameworks such as Kafka, continuous integration frameworks such as Jenkins, and machine learning frameworks such as TensorFlow. Note that the scheduler frameworks for database clusters do not have to use containers for installing the stateful services, but instead rely on traditional configuration management tools that directly install the services from Linux packages. An interesting question is whether the aforementioned performance overheads of CO frameworks for running databases can be avoided in the non-containerized approach of DC/OS.

Kubernetes and DC/OS are definitively two camps of opposite approaches. We believe that when high-performance database workloads must be targeted where database Pods must run close to the physical data storage location in the data center, DC/OS' database services might be the preferred choice because they have native performance and Mesos' protocol for allocation and reservation of local disk resources is very mature.

On the other hand, DC/OS' strategy to offer a separate Mesos framework for running databases increases the risk of vendor lock-in. A relevant remark here is that DC/OS' Edge-lb load balancer offers integrated support for load balancing container-orchestrated and non-container-orchestrated workloads which includes the abovementioned stateful services.

**Supported workload types.** With respect to auto-scaling concepts, Kubernetes provides besides the Horizontal Pod Autoscaler [259] (HPA) also the Vertical Pod Autoscaler [334] (VPA). These autoscalers are primarily meant to dynamically optimize the required resources for an application in accordance with fluctuations in the workload of customer requests. These autoscalers are generic in the sense that the offered configuration concepts and mechanisms can support autoscaling of different types of applications such as ReplicaSets as well as StatefulSets.

With the increasing focus of recent Kubernetes releases to improve QoS management, the question arises if these auto-scaling concepts can also be configured to meet service-level objectives (SLOs). However, we have demonstrated in previous research that the Horizontal Pod Autoscaler is too simplistic for meeting service level objectives (SL0s) of database clusters. We handled this problem by developing a tailored auto-scaler component that is customized to the type of database cluster [58]. Unless the HPA for StatefulSets can be tailored via Kubernetes' *annotations* and *modular interceptors*, the HPA for StatefulSets will need to be redeveloped by relying on a framework or library where custom auto-scaling policies and complex event monitoring policies can be specified and enforced.

DC/OS takes another approach to horizontal auto-scaling: it only offers third party tutorials [260] for building various types of auto-scalers. As stated above, DC/OS already offers a library and SDK [342] for configuring and deploying stateful services. Logically, this is the right layer for adding dedicated auto-scaling features for databases.

The Kubernetes' VPA concept is promising but there is one big disadvantage with respect to SLO compliance: adjusting resource allocation policies of Pods requires killing these Pods and waiting till the scheduler assigns a new Pod with the adjusted allocation policies. Obviously, this operation needs to be performed at run-time without restarting containers in order to avoid temporary performance degradation with SLO violations. Ironically, although run-time adjustment of container resource allocation policies is by default supported in Docker engine, they are not supported by any CO framework except Docker stand-alone. Indeed recent research presents a middleware for vertical scaling of containers that is implemented on top of Docker engine exactly because the presented middleware requires adjusting resource allocation policies without restarting containers [134].

In summary, existing auto-scalers of Kubernetes are not ready for managing performance SLOs. This lack is also the main reason why we have not grouped these auto-scaling features under the "Container QoS Management" aspect.

*7.5 Application and cluster management*

Most unique features found in this aspect, except those from the sub-aspect "multi-cloud support", are useful additions of functionality that are orthogonal to the core of the CO frameworks. As such we don't see any reason why these features will be deprecated in the long-term future. As such, we assess the features of the multi-cloud support sub-aspect below.

**Multi-cloud support.** Kubernetes has developed an extensive Federation API and associated command line interface for managing and federating multiple container clusters that are possibly located in separate cloud availability zones. A unique feature of Kubernetes is that this Federation API offers many federated instantiations of various API objects such as deployments and namespaces. However, the development of the Federation API has been put on hold and a new effort to build a dedicated federation API apart from the Kubernetes API is planned [621]. This is of course not good news for those companies that have already built their software products on top of the federation API. Note that this does not mean that the other unique features of Kubernetes in that sub-aspect have also a higher risk of being halted.

In opposition, Docker Swarm and all Mesos-based systems have invested most of their effort in building extensive support for running a single container cluster in high availability mode where multiple masters are spread across different cloud availability zones. This kind of multi-zone cluster does not require federated instantiations of existing API objects. Support for such an automated HA cluster across multiple availability zones is not supported by the open-source contribution of Kubernetes; it is only supported by the commercial Kubernetes-as-a-Service offerings on top of AWS and Google cloud.

### 7.6 *Summary of findings*

In this section we have studied to which extent unique features have a risk of being halted or deprecated without a replacing feature update because a competing framework offers a better alternative. The following three features of Kubernetes might incur an increasingly higher risk:

- If the performance overhead of *StatefulSets* for running database clusters cannot be resolved, DC/OS' approach to offer a user-friendly software development kit for generating custom scheduler frameworks for specific database may be the better approach.
- *Horizontal and vertical Pod autoscalers* are not fit to meet SLOs for complex stateful applications like databases. The generic design of these autoscalers will need to be sacrificed so that application managers can develop custom auto-scalers for particular workloads. As such there is a substantial chance that the generic autoscaler will be replaced by different types of auto-scalers.
- The development of the federation API for managing multiple Kubernetes clusters across cloud availability zones has been halted; instead a new API is being planned but there is no consistent effort into this direction. Most likely the federation API will replaced by a simplified API where some *existing federated instantiations* of Kubernetes API objects such as federated namespaces will be deprecated.

## 8. Conclusions

We first discuss in Section 8.1 the threats to validity of our results and the limitations of the overall study. Thereafter, in Section 8.2 we present the main insights that can be drawn from the findings of the study. Finally, in Section 8.3 we outline likely further evolutions in the technology domain in the short term.

### 8.1 *Threats to validity and limitations of study*

In essence, we present in this article a descriptive study based on expert reviews and expert assessments and therefore the main results are qualitative. All quantitative results are based on the identified features in the qualitative part of the study, which is inherently subjective to some extent.

We have thus not used variations of dependent and independent variables with different subject groups. Neither have we used automated metrics such as NLP-based processing of documentation, or amount of code/documentation.

As consequence, a large part of the standard threats to internal and external validity in experiment design are not relevant to this study. As a reminder, threats to internal validity compromise our confidence in stating that the found differences between CO frameworks are correct. Threats to external validity compromise our confidence in stating that the study's results are applicable to other CO frameworks.

As we don't make claims about other CO frameworks, only the following internal validity threats remain relevant:

- Selection bias, i.e. the decision what CO framework to select and the selection of the different features and the overarching (sub)-aspects may be determined subjectively. Thus, we may have missed features or interpreted feature implementation strategies inappropriately.
- Experimenter bias, i.e. unconscious preferences for certain CO frameworks that influence interpretation of documentation; e.g., whether a feature is partially or fully supported by a framework.

### 8.1.1 Selection bias.

We have tried to manage selection bias in our research method by means of three complementary approaches that have been explained in detail in Section 3. Firstly, we have applied a systematic approach and used existing methods if possible; for example, we have applied commonality and variability analysis in feature modelling to find common features (see Section 3.1.2) and we have applied card sorting to group features in usable aspects (see Section 3.1.3).

Secondly, we improved the accuracy of the description of the features and feature implementation strategies by means of an iterative approach. For example, we have first performed a pair-wise comparison of titles of documentation pages and thereafter a detailed review of the documentation pages in full detail. Then, we have asked customers and platform developers to review different versions of this article with respect to the question whether the set of identified features and their comparison makes sense and is complete (see Section 3.5).

Thirdly, we have continuously elaborated our practical experience of CO frameworks by not only testing specific features but also conducting performance evaluation research [58], [59]. This practical experience helps to make better interpretations of documentation.

### 8.1.2 Experimenter bias

It has been challenging to manage experimenter bias because container technology is currently at its peak of inflated expectations according to the Gartner hype cycle, has evolved quickly in the past, and Kubernetes has been adopted by Docker EE and DC/OS and all major public cloud providers.

To stay objective in the mid of such inflated expectations, we have consciously scoped the study to research questions with respect to software qualities that can be objectively measured using simple arithmetic: (i) genericity (in terms of number of supported features) and (ii) maturity (i.e., mapping features to development history on GitHub). To find evidence for overall significant differences between the CO frameworks with respect to genericity, we have used the Friedman and Nemenyi tests due to their effectiveness in un-replicated experimental designs for checking overall ranking of multiple systems with respect to different treatments [135]; in our research, treatments correspond with the 27 sub-aspects and systems with the CO frameworks.

### 8.1.3 Limitations of the study

Besides the above threats to internal validity in experimental design, the study has the following limitations:

* We have only studied the documentation of CO frameworks, not the actual code. We have not used any automated methods for mining features/aspects from code. As such features that can only be extracted from code are not covered in this study.
* Any claims about performance or scalability of a certain CO framework's feature implementation strategy are based on actual performance evaluation of Kubernetes and Docker Swarm integrated mode in the context of the aforementioned publications [58], [59]. Projections of these claims towards performance and scalability of similar feature implementation strategies in Mesos-based frameworks are speculative however.
* The study does not provide findings about the robustness of the CO frameworks such as or the ratio of bugs per line of code, or the number of bug reports per user.

### 8.2 *Lessons learned*

We organize the main conclusions from this study according to the three aforementioned software qualities, and thereafter we summarize the highlights for each of the frameworks

### 8.2.1 Genericity

* The ratio of common features over unique features is relatively large and most common features are supported by at least 50% of the CO frameworks. Such a high ratio of common features allows for direct comparison of the CO frameworks with respect to non-functional requirements such as scalability and performance of feature implementation strategies.
* Features in the sub-aspects "improved security isolation" and "allocation of other resources" are only supported by two or three CO frameworks

- o Although Kubernetes consolidated a full feature set for container isolation policies almost 36 months ago, there is little uptake of these features by the other CO frameworks.
  - o Mesos-based support for allocating GPU and disk resources to co-located containers is only marginally supported by Kubernetes and not supported by Docker Swarm.
- Kubernetes offers the highest number of common features and the highest number of unique features. When adding up both common and unique features, Kubernetes even offers the highest number of features for all 9 aspects and it offers the highest number of features for 15 sub-aspects.
- Significant differences in genericity with Docker EE and DC/OS have however not been found. After all, when taking into account only common features, Kubernetes offers the absolute highest number of common features for 7 sub-aspects, whereas Docker Swarm integrated mode offers the highest number of common features for the sub-aspects "services networking", "host port conflict management" and "cluster network security". Mesos offers the most common features for the sub-aspect "persistent volumes" and DC/OS offers the most common features of the sub-aspects "cluster maintenance" and "multi-cloud deployments".
- In the sub-aspects "services networking" and "host port conflict management", Docker Swarm integrated mode and DC/OS offer support for the features *host mode services networking*, *stable DNS name for services* and *dynamic allocation of host ports*. We have found that the other approaches to services networking such as routing meshes and virtual IP networks introduce quite a substantial performance overhead in comparison to running Docker containers in host mode. As such, a host mode service networking approach with appropriate host port conflict management is a viable alternative for high-performance applications.

### 8.2.2 Maturity

- The 15 sub-aspects identified by the green rectangle in Figure 11 shape a mature foundation for the overall technology domain as these sub-aspects are well-understood by now and little feature deprecations have been found in these sub-aspects.
- Figure 11 further indicates that Kubernetes is the most mature project in terms of pioneering common features despite being a younger project than Mesos, Aurora and Marathon.

### 8.2.3 Stability

- Mesos is the most interesting platforms for prototyping novel techniques for (i) container networking and (ii) persistent volumes because Mesos' adherence to all relevant standardization initiatives in these two areas maximizes the potential to deploy these techniques in Docker Swarm and Kubernetes as well. Docker or Kubernetes are best fit for prototyping innovating techniques for container runtimes.
- The overall rate of feature deprecations among common features in the past is about 2% of the total number of feature updates (i.e., feature additions, feature replacements, and feature deprecations).
- Only one unique feature of Kubernetes, *federated instantiations of the Kubernetes API objects*, has been halted and will probably be deprecated without a replacing feature update.

### 8.2.4 Main insights with respect to Docker Swarm

Although Docker Swarm is the youngest and also least generic framework among the three main vendors, Docker Swarm has clearly contributed an innovative services networking approach and networking plugin architecture.

Docker has actually separated services networking support from Docker Swarm. As such we believe Docker's networking architecture is here to stay. Docker has also recently released an enterprise edition with support for deploying and managing Kubernetes clusters next to Swarm clusters. While the current release does not show any strong integration between Docker and Kubernetes, support for Docker's networking architecture in Kubernetes is a likely future feature request.

### 8.2.5 Main insights with respect to Kubernetes

Kubernetes is the most generic orchestration framework for 7 out of 27 of sub-aspects. Yet, in many sub-aspects the absolute differences in number of supported features is small with respect to the two main other vendors Docker EE and DC/OS.

Kubernetes has also the most unique features. This may be a higher source of vendor lock-in on the one hand, but mainly constitutes a competitive edge. Our analysis of genericity has shown that many unique features of Kubernetes are much stronger a source for increased genericity than a source of vendor lock-in. When taking into account the total of common and unique features of Kubernetes, it counts the highest number of features of 15 sub-aspects.

Kubernetes is also the most mature container orchestration framework as it has pioneered 12 out of the 27 sub-aspects.

Kubernetes is in particular unique by its support for integrating with public cloud platform's load-balancing tier and offering a wide range of external service discovery options. As a result, a large number of public cloud providers have offered a hosted solution or even a Kubernetes-as-a-Service offering.

A weakness of the open source distribution of Kubernetes is that it does not offer support for automated installation of a highly-available cluster with multiple master nodes.

### 8.2.6 Main insights with respect to Mesos and DC/OS

DC/OS, an extended Mesos+Marathon distribution is the second most generic framework. Mesos+Marathon has pioneered also 10 out of the 27 aspects. A strength of Mesos is that it allows fine-grained sharing of cluster resources across multiple scheduler frameworks, which include not only CO frameworks but also non-CO frameworks like Hadoop, Kafka and NoSQL databases. Mesos or DC/OS may also be a viable alternative to companies who seek to setup a highly available cluster in a private cloud with the broadest range of possibilities to integrate container-based applications with non-container based applications. After all, DC/OS offers support for load balancing non-container orchestrated workloads such as databases or high-performance computing applications.

### 8.2.7 Main insights with respect to Docker Swarm alone and Apache Aurora

Docker Swarm stand-alone and Apache Aurora are relatively small CO frameworks that do differ significantly in terms of genericity from DC/OS and Kubernetes. Indeed, Aurora is specifically designed for running long-running jobs and cron jobs, while Docker Swarm stand-alone is also a more simplified framework with substantial less automated management.

We only recommend Docker Swarm stand-alone as a possible starting point for developing one's own CO framework. This is a relevant direction because 28% of surveyed users in the most recent OpenStack survey [4], responded that they have built their own CO framework instead of using existing CO frameworks (see also Figure 1). We make such recommendation because the API of Docker Swarm stand-alone is the least restrictive in terms of the range of offered options for common commands such as creating, updating and stopping a container. For example, Docker Swarm stand-alone is the only framework that allows to dynamically change resource limits without restarting containers. Such less restrictive API is a more flexible starting point for implementing a custom developed CO framework.

### 8.3 *Further evolution in the short term.*

Likely areas for further evolution and innovation include system support for cluster network security and container security, performance isolation of GPU, disk and network resources and network plugin architectures.

As stated above, Kubernetes is the only framework that offers rich support for container security isolation whereas Mesos and DC/OS offer very limited support and Docker EE uses another approach so that security isolation policies in Kubernetes are not easy to migrate to Docker. It is expected that research is needed to better understand this evolution and how lower-level system security guarantees can be designed and verified uniformly.

A weakness of Kubernetes is its limited support for performance isolation of GPU and disk resources and its lack of support for network isolation. Improved support for persistent volumes as part of the Container Storage Interface (CSI) specification effort has been the main focus of the most recent releases of Kubernetes. Network isolation features have also been subject to recent research. It is expected that thus in the near future these features will be considerably improved.

Finally, network plugin architectures themselves will change considerably. Although the Container Network Interface (CNI) specification has been adopted by Kubernetes and Mesos for several years now, the development of such CNI-based network plugin architectures has halted and is still in the alpha stage in Kubernetes. Docker's libnetwork is in particular very dynamic and new features are continuously being added. Better support for high-performance network function virtualization (NFV) without sacrificing automated management is currently also a main focus of current systems research. It is expected that these innovations will trigger similar improvements in virtual networking architectures for containers.

**Author Contributions:** Conceptualization, Eddy Truyen and Dimitri Van Landuyt; Data curation, Eddy Truyen; Funding acquisition, Bert Lagaisse and Wouter Joosen; Investigation, Eddy Truyen; Methodology, Eddy Truyen; Project administration, Bert Lagaisse; Software, Eddy Truyen; Supervision, Bert Lagaisse and Wouter Joosen; Validation, Eddy Truyen and Dimitri Van Landuyt; Visualization, Eddy Truyen; Writing – original draft, Eddy Truyen; Writing – review & editing, Eddy Truyen, Dimitri Van Landuyt and Davy Preuveneers.

**Funding:** "This research was funded by the Agency for Innovation and Entrepreneurship IWT, grant DeCoMAdS, grant number 179K2315, and the Research Fund KU Leuven.

**Acknowledgments:** We thank Bert Robben for his feedback and reviews of drafts of this article. We thank the developers of the CO frameworks, especially the technical writers for the excellent documentation. Finally we thank GitHub for offering the documentation of the CO frameworks in versioned format.

**Conflicts of Interest:** The authors declare no conflict of interest. The funders had no role in the design of the study; in the collection, analyses, or interpretation of data; in the writing of the manuscript, or in the decision to publish the results.